\documentclass[5p,sort,compress]{elsarticle}

\usepackage{fancyhdr}
\usepackage{amsfonts}
\usepackage{url}
\usepackage{amsmath}
\usepackage[labelfont=bf,justification=raggedright,singlelinecheck=false]{caption}
\usepackage{array}
\usepackage{tabu}
\usepackage{multirow}
\usepackage{graphicx}
\usepackage{amsmath}
\usepackage{algorithm}
\usepackage[noend]{algpseudocode}
\usepackage{pdflscape}
\usepackage{rotating}
\usepackage{wrapfig}
\usepackage[inline, shortlabels]{enumitem}
\usepackage{graphbox}
\usepackage{mathtools}
\usepackage{amsmath}
\newcommand*\rfrac[2]{{}^{#1}\!/_{#2}}
\newcolumntype{v}{>{\centering\arraybackslash}m{.18\linewidth} }
\usepackage{hyperref}
\hypersetup{
    colorlinks=true,
    linkcolor=blue,
    filecolor=magenta,      
    urlcolor=cyan,
}
\usepackage[labelfont=bf,   justification=justified,   format=plain]{caption} 
\begin{document}
\title{TRLF: An Effective Semi-fragile Watermarking Method for Tamper Detection and Recovery based on LWT and FNN}
\author[fum]{Behrouz Bolourian Haghighi}
\ead{b.bolourian@mail.um.ac.ir}
\author[fum]{Amir Hossein Taherinia\corref{cor1}}
\ead{taherinia@um.ac.ir}
\author[fum,SCIIP]{Reza Monsefi}
\ead{monsefi@um.ac.ir}
\cortext[cor1]{Corresponding author}
\address[fum]{Computer Engineering Department, Ferdowsi University of Mashhad, Mashhad, Iran }
\address[SCIIP]{Center of Excellence on Soft Computing and Intelligent Information Processing (SCIIP)}
\begin{abstract}
This paper proposes a novel method for tamper detection and recovery using semi-fragile data hiding, based on Lifting Wavelet Transform (LWT) and Feed-Forward Neural Network (FNN). In TRLF, first, the host image is decomposed up to one level using LWT, and the Discrete Cosine Transform (DCT) is applied to each 2$\times$2 blocks of diagonal details. Next, a random binary sequence is embedded in each block as the watermark by correlating $DC$ coefficients. In authentication stage, first, the watermarked image geometry is reconstructed by using Speeded Up Robust Features (SURF) algorithm and extract watermark bits by using FNN. Afterward, logical exclusive-or operation between original and extracted watermark is applied to detect tampered region. Eventually, in the recovery stage, tampered regions are recovered by image digest which is generated by inverse halftoning technique. The performance and efficiency of TRLF and its robustness against various geometric, non-geometric and hybrid attacks are reported. From the experimental results, it can be seen that TRLF is superior in terms of robustness and quality of the digest and watermarked image respectively, compared to the-state-of-the-art fragile and semi-fragile watermarking methods. In addition, imperceptibility has been improved by using different correlation steps as the gain factor for flat (smooth) and texture (rough) blocks.
\end{abstract}
\maketitle
\section{Introduction}
\noindent In the Internet network era, due to the rapid growth of media and signal processing tools, digital content can be spread and conveniently destroyed, tampered, duplicated, or any other manipulations over the Internet. In other words, attacks and threats may challenge the security of digital media, and the content of the media can easily be substituted with fake content. All these modifications are done in the way that it is difficult for the human to see the alteration with naked eyes. Thus, we are not sure about the validity of media that received from the Internet or any other channels. So, it is really a basic and major problem to ensure the integrity and entirety of the received media, and protect the owner’s rights from any manipulations or plagiarism. In the critical cases, such as for proof in a court of law, a small amount of content modification in digital media can change the judgment. So, it is essential to check the authenticity of the media. Therefore, Nowadays the tampered detection and localization are important issues. This problem can be overcome and solved by using digital watermarking techniques \cite{ref51}. 

Digital image watermarking is one of the common topics in information security and data hiding \cite{ref1, ref2}. Watermarking methods hide the secret message, signature or information about the image into the host (cover) image, without attracting attention from of attackers and increasing the size of the file. The information, called a watermark, can be extracted from the watermarked image in order to test the authenticity and integrity of the image or authentication and identification of the owner \cite{ref3, ref4, ref5, ref6, ref7, ref8, ref9}. This process is imperceptible to human observers. In recent years, several watermarking methods have been proposed. These methods can be classified into several categories depending on the requirements, features, and type of applications. 

The main classification is the working domain, that technically divided into spatial and frequency domains \cite{ref2}. Spatial domain methods are the most straight methods by substituting the values in the Least Significant Bits (LSBs) of the host image with information of the watermarks, without applying any transformation. These methods are simple and less complex but not have the ability to resist against different attacks. So, these methods can be considered for fragile watermarking application. Frequency domain technique \cite{ref3, ref4, ref5, ref6, ref7} transforms and embeds the watermark information by modifying the transformed coefficients of the host signal, then apply the inverse transform into the spatial domain \cite{ref7}. Generally, these methods are more robust against different image processing attacks. In recent years, there have been a large amounts of watermarking methods proposed in the frequency domain such as DCT \cite{ref5}, DWT \cite{ref4, ref8}, DFT \cite{ref3}, SVD \cite{ref4, ref5, ref7}, etc. 

On the other hand, watermarking methods can be classified based on their robustness as 1) robust, 2) fragile and 3) semi-fragile watermarking \cite{ref2}. In robust watermarking method \cite{ref3, ref4, ref5, ref6, ref40}, the watermark is able to resist against unintentional or intentional attacks. In other words, robust watermarking is designed to resist any editing operations or attacks in order to copyright protection and ownership verification. Fragile watermarking (called hard authenticate) \cite{ref49, ref48, ref10, ref11, ref12, ref13, ref14, ref15, ref16, ref17, ref18, ref19, ref20} is aimed to be watermark destroyed any unintentional or intentional modification in the image. Therefore, these methods are fragile and sensible to any modifications or forge in order to detect any tampering. In fragile type, a major practical problem appears. For example, its high sensibility even to accidental variation or innocent image editing like compression, etc. The solution is using semi-fragile watermarking instead of fragile type. Finally, in semi-fragile watermarking methods \cite{ref21, ref22, ref23, ref24, ref25, ref26, ref27, ref28, ref29, ref30, ref31, ref32, ref33}, watermark is designed to resist against non-intentional manipulations caused by image processing operations such as compression, geometric, salt and pepper noise, histogram equalization, Gaussian filter, sharpen, and other basic operations of image processing. Furthermore, the watermark is fragile and sensitive against intentional manipulations like content modified by fake information. Generally, fragile and semi-fragile approaches are efficiently employed for image and video content authentication.

Eventually, watermarking techniques can be divided according to the detection process into three categorize as blind, non-blind and semi-blind \cite{ref2}. In blind watermarking methods \cite{ref3, ref7, ref39}, the watermark can be extracted from the cover image without any reference to original images. In other words, these methods do not require watermark, the original image or any extra information during the extraction process. The non-blind watermarking approach, need the original cover image during extraction phase. Totally, these techniques are more robust against image processing attacks. Finally, other extraction methods do not need the original image, but require the watermark or extra information beside the watermarked image, during the extraction process, which is called semi-blind watermarking methods.

Generally, the digital watermarking technique must have the following requirements: 
\begin{enumerate*}[i)]
\item Imperceptible (transparency), \item Difficult to remove the watermark without seriously affecting the quality of an image, \item Robust against various image processing operations and geometrical attacks
\end{enumerate*}. Proposing the method which satisfies these requirements is not an easy work \cite{ref8}. Thus, a trade-off between a number of embedded watermarks and robustness is needed in this field.

In last decade, image authentication and verification has become to the front of image processing \cite{ref50}. The goal for image authentication is to reveal the malicious manipulations and accept valid content in the image. As mentioned, integrity and authenticity of the digital image can be guaranteed by using watermarking technology. Image tamper detection and recovery is a type of watermarking application which maintains the integrity and content protection of digital images. These methods can authenticate and detect tampered regions, using information embedded in the host signal. In other words, integrity and authenticity can be guaranteed by using digital signatures that are the most common technique in protecting media content. If a watermarked image suffers from modification, the watermarking method can localize the tampered region and recover the corrupted content by image digest. Image digest is a compact version of the original image that describes the content of the host image. This process is called tamper detection and recovery.

The rest of the paper is organized as follows. Section 2 reviews the related work of tamper detection and recovery methods. Section 3 presents a brief preliminary about Lifting Wavelet Transform (LWT), Halftoning and WInHD technique, Speeded up robust features (SURF), Feed-forward neural network (FNN). Section 4 describes the proposed semi-fragile watermarking method. Section 5 presents the experimental results and analysis. Finally, the conclusions along with the scope of future works are given in Section 6.
\section{Related Works}

\noindent In this section, first, we will review several tamper detection and recovery techniques, which published in recent years as related works. Then, the contributions of the proposed method are presented in the subsection. 

Nowadays, several watermarking schemes for authenticating and correct tampered regions in digital images have been developed. Authenticating image with fair enough imperceptibility and high detection rate is the challenge of today’s research. Generally, these approaches can be divided into fragile and semi-fragile watermarking. 

Lee and Lin \cite{ref10} proposed an effective dual watermark for image tamper detection and recovery. Dual watermarking techniques can improve the quality of the recovered image. In this scheme, two copies of the watermark are embedded into two different positions in the whole image. Thus, it can provide second chance to correct tampered block in case one copy is destroyed, but this method is not able to detect any content tampering that modifies bits in 5 most significant bits. Results illustrated resilience for covering, removing, cropping and replacing tampering and vulnerable against collage, vector-quantization, and copy-move tampering. In another work \cite{ref11}, probability-based tampering detection scheme for digital images is proposed by Hsu and Tu. This scheme aims to use a probability theory to improve tamper detection rate. In the tamper detection phase, first, the watermarked image identifying through the authentication watermark which embedded in the image. Then, a probability theory is employed to improve previously results and enhance authentication rate. However, the main drawback is, it is not able to restore tampered regions. The experimental results demonstrate that the scheme performs well in terms of authentication accuracy rate. 

Qian et al \cite{ref12} proposed, a fragile watermarking scheme aimed at providing improved restoration capability based on Discrete Cosines Transform. DCT coefficients of 8$\times$8 blocks encoded into different numbers of bits and the authentication and restoration bits are hidden into the three least significant bits planes of the host image. On the receiving side, the authentication bits are extracted to authenticate image, and the restoration bits are used to recover the contents of the tampered regions. Results showed that the accuracy rate of tampered detection has been decreased, due to the usage of a large block size. In \cite{ref13} authors proposed, an effective self-embedding fragile watermarking for image tamper localization and recovery based on Discrete Cosines Transform. This scheme performed an improved tamper localization and recovery algorithm, compared to previous methods. In the proposed scheme for enhancing the security of the algorithm, a non-linear chaotic sequence is been used. In the embedding phase, the watermark is generated by encoding DCT coefficients of each 2$\times$2 blocks and hidden in another block according to the block mapping. The experimental results showed that the tampered region can be successfully detected and has a high detection rate even if the tampered region is up to 80\%. Tong et al. \cite{ref14}, proposed a chaos-based fragile watermarking for image tamper detection and self-recovery. In this scheme, a novel chaos-based method to scramble the blocks is employed, and a sister block embedding scheme is designed to improve the recovery rate. According to experimental results, the scheme is more secure and have a better effect on authentication and recovery, especially for the larger tampered region.

In \cite{ref15}, an effective singular value decomposition (SVD) based image tampering detection and self-recovery is proposed by Dadkhah et al. To improve the tamper detection rate, a mixed block partitioning approach for 4$\times$4 and 2$\times$2 blocks is utilized. The experimental results reveal that the proposed scheme is superior in terms of security, tamper localization, and recovery rate, over the other fragile tamper detection and recovery schemes. In addition, this scheme is able to detect vector-quantization and copy-move tampering. Zhang et al. \cite{ref16} proposed a self-embedding fragile watermarking method based on DCT and fractal coding. In this scheme, three copies of recovery watermark are embedded into different quadrants, that provide two chances for recovery in case of one is destroyed. Similar to \cite{ref12}, in this scheme, the accuracy, and precision of tamper localization decreases due to using a large block size.

In \cite{ref17} authors proposed, an effective self-embedding fragile watermarking scheme for image tampered detection and extreme localization with recovery capability based on DCT. In this scheme, for each 2$\times$2 blocks, two authentication bits, and ten recovery bits are generated from the five most significant bits planes. Authentication and recovery bits are embedded in the three least significant bits of the block itself and corresponding mapped block, respectively. The experimental results illustrate that the proposed method improves the accuracy of tamper detection, because of using small size blocks and also removes the blocking artifacts. In \cite{ref18} self-embedding image authentication scheme using the quick response code features based on SVD is proposed by Wu et al. In this scheme, two self-embedding approaches are presented to involve distinctive singular values, and the quick response code is used to preserve singular values. Both approaches, extract self-characteristics of an image as the authentication information. The experimental results demonstrate that both proposed methods effectively detect the tampered region. 

In \cite{ref19} image tamper detection and recovery based on adaptive embedding rules is proposed. This scheme employs smoothness to distinguish the types of blocks to enhance embedding efficiency, authentication, and recovery effects. In addition, hide more recovery data in a region with major changes to increase recovery quality. The experimental results showed that the proposed scheme yields satisfactory image recovery quality for texture images. In addition, compared with the scheme proposed in \cite{ref10}, this scheme causes less damage to the quality of the original image. In \cite{ref20} authors proposed hierarchical recovery capability for tampered images based on watermark self-embedding. In this scheme, through the hierarchical recovery strategy, the higher most significant bit layers of modified regions have higher priority to be corrected than the lower most significant bit layers. Results showed that the scheme improved the visual quality of the recovered image, even though the tampered area is large.

As previously mentioned, in the fragile scheme, the watermark information may be destroyed by the image processing operations likes compression, histogram equalization, salt and pepper noise, various filtering and geometrical attacks. Recently, in order to achieve more robustness against these attacks, the semi-fragile watermarking schemes have been developed that can simultaneously resist some operations with well localization capability. Now, we briefly review several semi-fragile watermarking schemes which have been proposed in recent years.

In \cite{ref21}, two authentication schemes using random bias and nonuniform quantization based on wavelet transform is proposed. The first method is based on the random bias factor in order to fixed decision boundary, while the second method is employed on nonuniform quantization to improve the detection rate. The Experimental result showed that it provided a good imperceptibility and robustness against JPEG compression, but is not provided any report to evaluate against other types of attacks. In \cite{ref22}, the authors proposed a semi-fragile watermarking scheme for the automatic authentication and recovery of digital image content based on irregular sampling. In this method, when modified blocks are detected, the recovery phase is formulated as an irregular sampling problem. Simulation results demonstrated that the approach minimizes the probability of false positive while maintaining the data integrity of the recovered images.

In \cite{ref23, ref24, ref25}, authors proposed the semi-fragile watermarking scheme for authenticating of the digital image based on wavelet transform. In \cite{ref23}, multiple watermarks are used to focus on both the tamper detection and recovery and perform self-recovery in the case of malicious attack. It also, categorize the modifications and provides a value-added strategy for improving security and efficient authentication rate. Unfortunately, this paper did not report any result to show the resistant of the scheme against various attacks. The proposed scheme in \cite{ref24}, a quantization technique is employed to modify one chosen approximation coefficient of each block to ensure its robustness against various attacks. However, there are two main drawbacks to this scheme. Firstly, the scheme is not able to recover tampered regions. Secondly, this scheme is vulnerable to geometrical attacks. In \cite{ref25} Phadikar et al. proposed a watermarking scheme based on Quantization Index Modulation (QIM) in the Integer Wavelet Transform (IWT). In this scheme, a binary watermark and an image digest are embedded by modulating IWT  coefficients using dither modulation based QIM. The image digest is predefined halftone version of the approximate sub-band generated by two levels of IWT decomposition. On the receiving side, decoder extracts the binary watermark and image digest from a watermarked image for tamper detection and recovery, respectively. Experimental results showed that the scheme provides a superior performance in terms of probability of false positive as well as in tamper correction, compared to previous semi-fragile schemes. However, a low quality of the recovered image and vulnerable to vector quantization, collage, and copy move tampering are main disadvantage of this scheme. In \cite{ref26} authors proposed, two semi-fragile watermarking for image authentication with recovery capability using halftoning technique and Multilayer Perceptron (MLP). In both schemes, a halftone version of the host image is used as image digest. In this scheme, two distinct image authentication methods based on IWT and DCT are proposed. In the authentication phase, the tampered $8\times8$ blocks are detected using the Structural Similarity Index (SSIM) metric. The main novelty in this scheme is employed the MLP neural network to improve the visual quality of the recovered image, compare to \cite{ref25}. The experimental results illustrated the robustness of both algorithms against various attacks. However, two major drawbacks of this scheme are low quality of the watermarked image, and also the fragility of geometrical operations.

Preda \cite{ref27} proposed a semi-fragile watermarking for image authentication with sensitive tamper localization based on wavelet transform. In this scheme, watermark bits embedded in a group of wavelet coefficients by quantization technique. The embedded watermark is robust against JPEG compression. Huo et al. \cite{ref28} proposed a semi-fragile image watermarking algorithm with two-stage detection. In this scheme, the watermark bit is embedded in 8$\times$8 blocks based on DCT. In the authentication phase, the tampered region is detected based on two maps generated from two extracted watermarks. Experimental results showed that the scheme is able to detect tampered region with high probability, but can't recover tampered region.

Al-Otum \cite{ref29} proposed, a semi-fragile scheme for authentication and tamper detection based on an adjusted expanded-bit multiscale quantization technique. In this work, a random watermark bit sequence embedded in the low-frequency subbands of the second level DWT decomposition. On the receiving side, two measures are used to classify the watermarked image as authenticated, incidentally, or maliciously attacked. Experimental results have shown that the suitability of the proposed technique for tamper detection and authentication. However, the proposed scheme was not extended to handle color images, geometric attacks, recover tampered region, and may fail to authenticate tampered region when the watermarked image attacked by a higher JPEG or JPEG2000 compression. In \cite{ref30} the authors proposed, a semi-fragile watermarking technique for content authentication and tamper localization based on singular value decomposition. The aim of this paper is to preserve the image content authentication and localizing the tamper region. In this scheme, the security watermark is obtained by applying the exclusive operation with the singular value. Afterward, a watermark embedded in the 4$\times$4 blocks of the wavelet domain to generate the watermarked image. In authentication step, the watermark has been extracted and rebuilt to detect and localize tampered region. Similarly, this scheme is not able to reconstruct tampered region and is fragile to geometric attacks.

Benrhouma et al. \cite{ref31} presented a tamper detection scheme based on Arnold Cat Map (ACM) and DWT. In this scheme, the approximation coefficient of each block, hidden in the details coefficients of another block as the watermark. In the embedding phase, the blocks pairs are permuted by using an ACM. The experimental results demonstrate the efficiency of the tamper detection and recovery algorithm. A semi-fragile and self-recoverable watermarking scheme based on wavelet group quantization and dual authentication is proposed in \cite{ref32}. In the proposed scheme, a host image is first divided into 16$\times$16 non-overlap blocks. For each block, a five-bit authentication is obtained from the first order statistical moment of the block as the watermark. Then, the watermark is embedded into the middle frequency of another block by a group based wavelet quantization technique. Simulation results showed that the recovered image is a better approximate to the original image compared to the previous methods. The main drawback of \cite{ref31, ref32} is low quality of the watermarked image. Another work in \cite{ref33} proposed an intelligent blind semi-fragile watermarking scheme for authentication and tamper detection of digital images based on curvelet transforms. In this scheme, the watermark embedded in the quantized first level Discrete Curvelet Transform (DCLT) course coefficients. The quantization step of the first level course DCLT coefficients of the cover image is selected intelligently based on the energy contribution of the coefficients. On the receiver side, the similarity between each block of extracted and generated coefficients is compared based on similarity index value. Similarly, this scheme is not able to recover tampered region, too.

\subsection{Our contribution}
\noindent Limitations of the previous semi-fragile methods which reviewed in the literature show that these methods suffer from three disadvantages: 
1) Low robustness to geometrical and non-geometrical attacks, 2) Vulnerable against special tampering include vector-quantization, collage, and copy-move tampering, 3) Low visual quality of the watermarked image, 4) Unable to reconstruct tampered region, 5) Almost entirely focused on gray-scale images.

In this paper, to overcome the disadvantages and problems, an effective semi-fragile watermarking method for tamper detection and recovery based on LWT and FNN is proposed (TRLF). The main novelty of TRLF is that it uses FNN for extracting binary watermark with high correlation, which is not experienced in previous works. The motivation of using FNN is well learning and predicting without requiring feature extraction. Also, it has good generalization ability, even when the watermarked image attacked with the hybrid of various geometric and image processing attacks. In this way, first, one level LWT on cover image is applied, and then divide the diagonal details to 2$\times$2 non-overlapping blocks. After that, Discrete Cosine Transform (DCT) is applied in each block and a binary watermark is embedded by using the adaptive correlation of DC's coefficients of each block. The correlation step (Gain Factor) is optimized by using JPEG compression for detecting texture and flat region. This optimization leads to obtain the highest possible robustness without losing the transparency. TRLF is semi-blind and does not require the original image. In this way, we transfer the compact version of the original image as digest to further utilizes for reconstructing the image geometry and recover tampered region. 

\noindent Key features of TRLF mention below:
\begin{enumerate}[1),leftmargin=*, noitemsep, topsep=0pt]
\item High detection and localization accuracy, and detect copy-move, collage, and vector-quantization tampering.
\item Resist to the most geometric and non-geometric attacks and reconstruct the geometry of the attacked image.
\item High visual quality of the watermarked image, and employ halftoning technique (inverse halftone), in order to effective tamper correction without blocky artifacts.
\item Employ Neural network to authenticate image (without feature extraction), and LWT to improves the processing speed of DWT and construct lossless coefficients.
\item Extract watermark, without transfer host or watermark.
\end{enumerate}

The performance of the proposed tamper detection and recovery method is analyzed against various tampering. Experimental results reveal that it achieved a high level of robustness against the image processing operations and the geometrical attacks, compared with other semi-fragile schemes. Generally, TRLF decreases the probability of mistakes in tamper localization, and enhances the accuracy of authentication, especially when the tampered area is large.
\begin{figure*}[t]
\center
\includegraphics[width=1\textwidth,trim=18cm 20cm 23cm 20cm,clip]{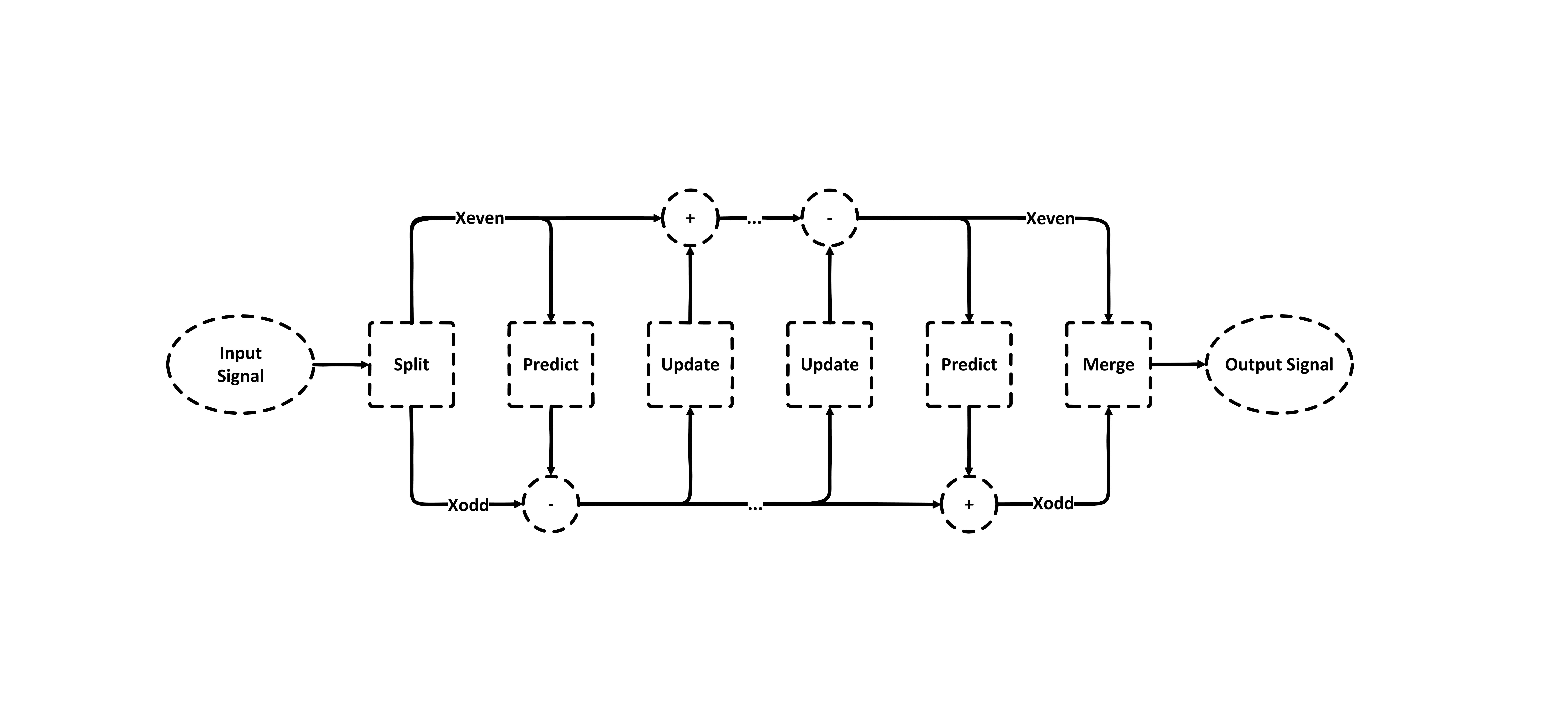}
\caption{Decomposition and reconstruction of LWT.}
\label{fig:LWT}
\end{figure*}

\section{Preliminaries}
\noindent In this section, we present the background requirements for subsequent sections. For this purpose, we begin with a brief introduction to Lifting Wavelet Transform (LWT) and illustrate the Jarvis halftone technique and WInHD, then describe Speeded Up Robust Features (SURF). Finally, review Feed-Forward Neural Network (FNN) technique.

\subsection{Lifting wavelet transform}
\noindent In general, the traditional wavelet transform \cite{ref34} analyzed the signal at different frequency and resolution. This transform decomposed a signal to details and approximate parts. The approximate which contains the low-frequency part of the signal can be then further decomposed to provide more details. This process can continue for several levels. Details contain a high-frequency part of a signal that represents the horizontal, vertical, and diagonal directions. This type of wavelet transform called the first generation. 

LWT was proposed by Sweldens in 1998 and belong to the second generation of wavelet transformation \cite{ref35}. LWT, analyzing signals in time domain, has less computational time and memory requirement than traditional wavelet. Time-domain calculations have two basic advantages. First, there is no need to form a Fourier transformation as a prerequisite and second, this method provides the algorithmic infrastructure that is given the ability to construct signals and easily use of the inverse transform. In addition, it is a flexible transform to create a linear or nonlinear wavelet transformation. Totally, according to the properties described above, LWT conquers and overcomes the limitations and weakness of traditional wavelet. So, the advantages of LWT compare to traditional wavelet, help us to increases the performance of TRLF.

\begin{figure}[h!]
\center
\begin{tabular}{cc}
\includegraphics[width=0.45\columnwidth]{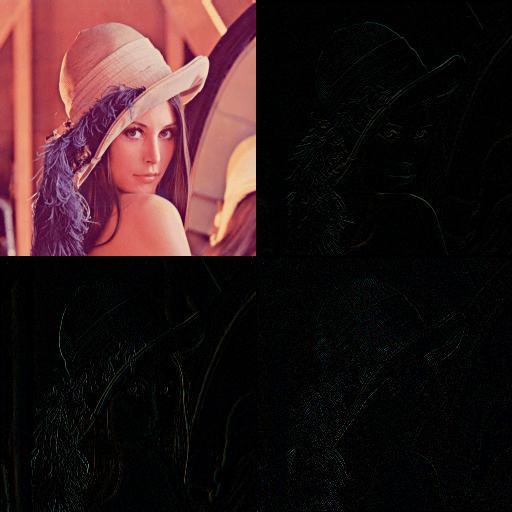}&
\includegraphics[width=0.45\columnwidth]{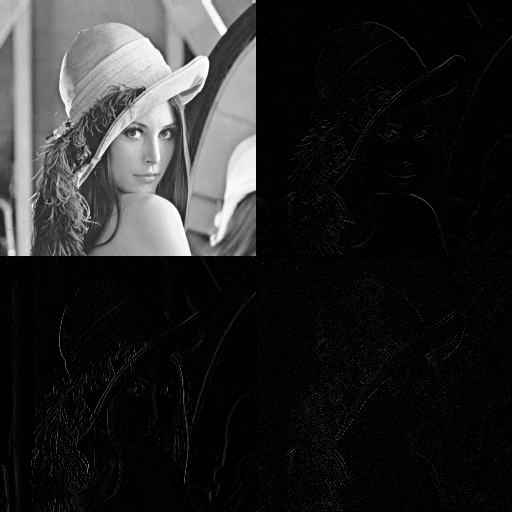}\\ 
(a)&(b)
\end{tabular}
\caption{Applied LWT (Haar) on Lena. (a) Color, (b) Gray.}
\label{fig:LWTLENA}
\end{figure}

Fig. \ref{fig:LWTLENA} shown applied LWT on Lena image. In this implantation, simple filter that call Haar wavelet is used. 

\noindent In fact, the main unique major properties of LWT are: 
\begin{enumerate}[1),leftmargin=*, noitemsep, topsep=0pt]
\item Calculated more efficiently and needs less memory.
\item Analyzing the signal in time domain.
\item Do not have quantization errors.
\item Energy compaction.
\item Solve the selection of basic function problems.
\end{enumerate}

In general, LWT has three main phases (splitting, prediction, and update) which describe as follow. In split step, first separate sample signals to the odd and even sets. Even and odd subsets are scattered in an original set. If an input sequence was random, there would be a correlation between values that are close to each other. Obtaining affiliation or data structures is the main purpose of each compress representation. In prediction step, each odd sample is predicted with a number of even samples that are around. The difference between predicted and actual values of odd samples is considered as component coefficients. If the input current structure was intended to fully comply stipulated, all the details would be zero. Therefore, signal-distortion details of a structure are provided and belong to the current scale level. This means that they should be removed from the signal to obtain a smooth version with lower scale level. This is what has done in lifting wavelet method. In the update step, paired samples reform based on obtained details, as the features of the original signal maintain its approximation. The output of this step approximate coefficients which are the smoother version of the input signal. The decomposition and reconstruction of LWT are illustrated in Fig. \ref{fig:LWT}.
\subsection{ Jarvis halftone technique and WInHD}\label{sec:winhd}
\noindent Digital halftoning is a technique to generate a halftone image by homogeneously distributed black and white dots from a continuous tone image with discrete levels. In other words, digital halftoning is a process which displays an image with only 0 and 255 intensities (black and white colors). Over the last several decades, many halftoning techniques have been developed. These techniques can be classified into various types of technologies such as Error Diffusion (ED), Dot Diffusion (DD), Ordered Dithering (OD), etc \cite{ref36, ref37}.

\begin{figure}[t]
\center
\begin{tabular*}{1\columnwidth} {@{}c@{}c@{} }
\includegraphics[width=0.72\columnwidth,trim= 15cm 10.5cm 15cm 3cm,clip]{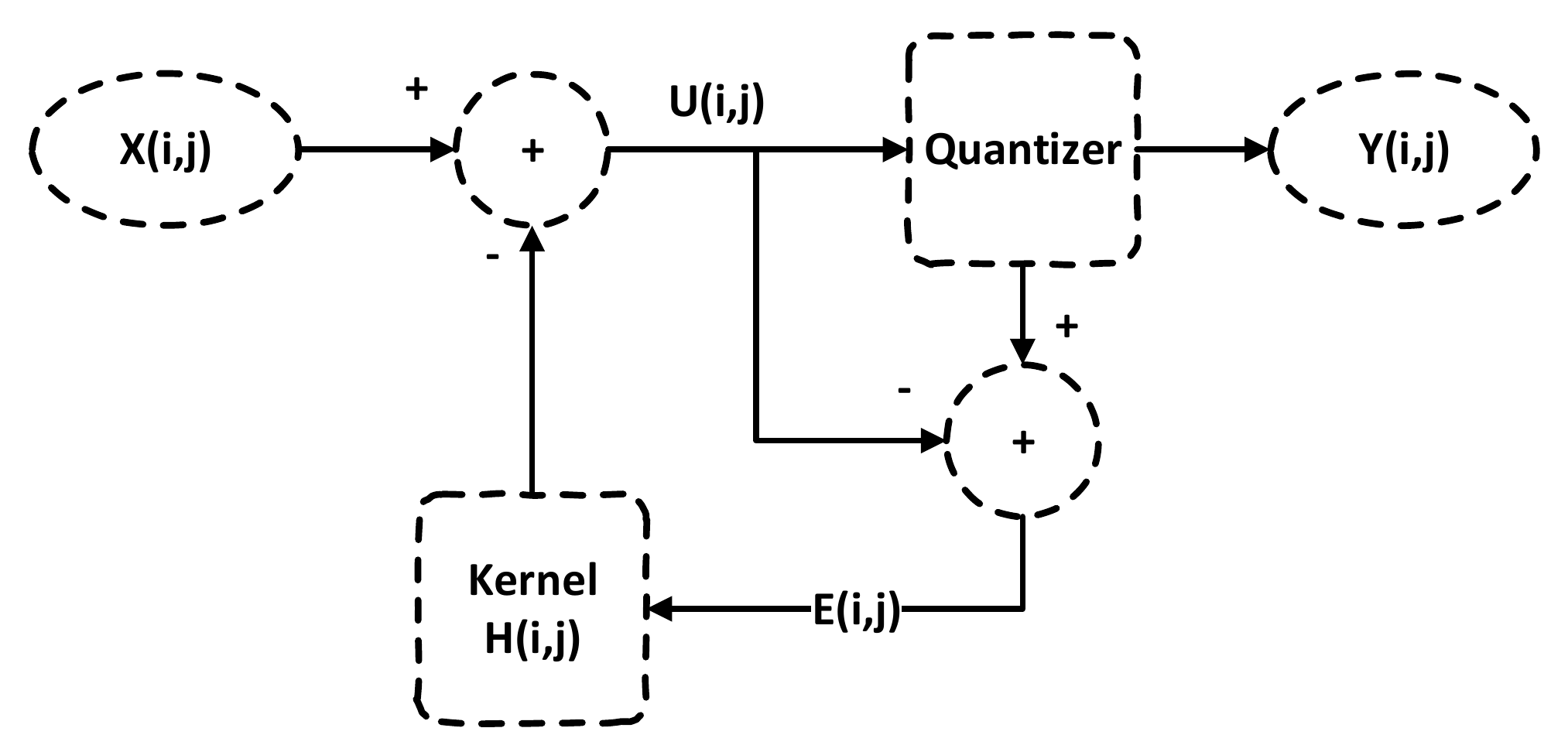}&
\includegraphics[width=0.25\columnwidth,trim=2.5cm 20cm 11cm 1cm,clip]{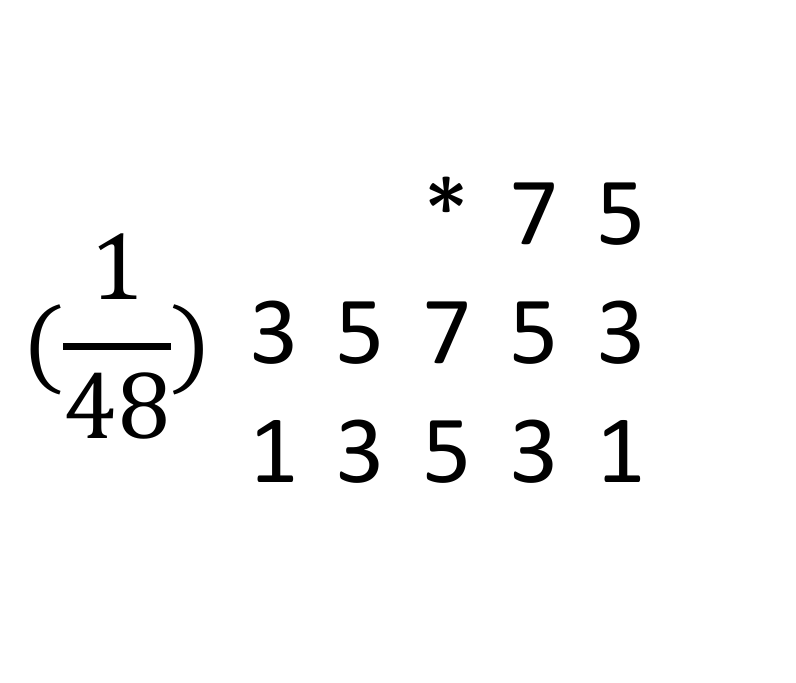}
\\ (a) & (b)
\end{tabular*}
\caption{(a) Error diffusion halftoning procedure, (b) Jarvis's kernel. }
\label{fig:diagramkernel}
\end{figure}

In this paper, we focus on error diffusion halftoning technique that is more famous and popular than others. Also, it is a more complicated technique than ordered dithering, but the main feature is achieving better visual quality. This technique maintains the average intensity level between the input image and the binary version. In this technique, the pixels are quantized in a particular order and the remaining quantization error for the current pixel is diffused forward to local unquantized pixels. The error refers to the difference between the original pixel values and its halftone values. This leads to make the local average intensity of the halftone version close to the original image. The design of error filter plays an important role in error diffusion halftoning technique. Jarvis kernel \cite{ref37} is one of the common error diffusion kernels. In 1976, Jarvis et al published a survey of halftoning technique that represents an error diffusion method with a different error filter. This filter is given in Fig. \ref{fig:diagramkernel}(b), where * represent the current pixel.

Fig. \ref{fig:diagramkernel}(a) is shown the block diagram of error diffusion. In this diagram, quantize units converts input pixel value into a binary one $Y(i, j)$, using a threshold value $T=128$, as shown in Fig. 2(b). Then the error sequence $E(i, j)$ which is the difference between $U(i, j)$ and $Y(i, j)$ is presented to the 2D-filter $H(i,j)$. Fig. \ref{fig:jarvis}(a, c) shows the result of apply error diffusion technique with Jarvis's kernels for the gray and color Lena image.

One of the major application of halftone technique is inverse halftone. The inverse halftoning algorithm is processed to reconstruct the original image with 255 levels from its halftoned version. Today, several inverse halftoning algorithms have been proposed in the literature and most of them are based on low pass filtering. Although their computational complexity is very low, unfortunately, most of these methods perform inverse halftoning with very low quality which may not be acceptable for various applications, such as compression and watermarking. In TRLF, we utilized WInHD technique that is proposed in \cite{ref38} to performs inverse halftoning and reconstructs high-quality gray or color scale images from its halftone one.
Fig. \ref{fig:jarvis}(b, d). illustrate that the WInHD method can create the inverse halftoned image with high image quality. PSNR and SSIM between the original image and the inverse version for gray and color Lena images are 32.16, 0.84 dB and 31.18, 0.98, respectively. For more information about the WInHD, please refer to \cite{ref38}.
\begin{figure*}[t]
\includegraphics[width=1\textwidth]{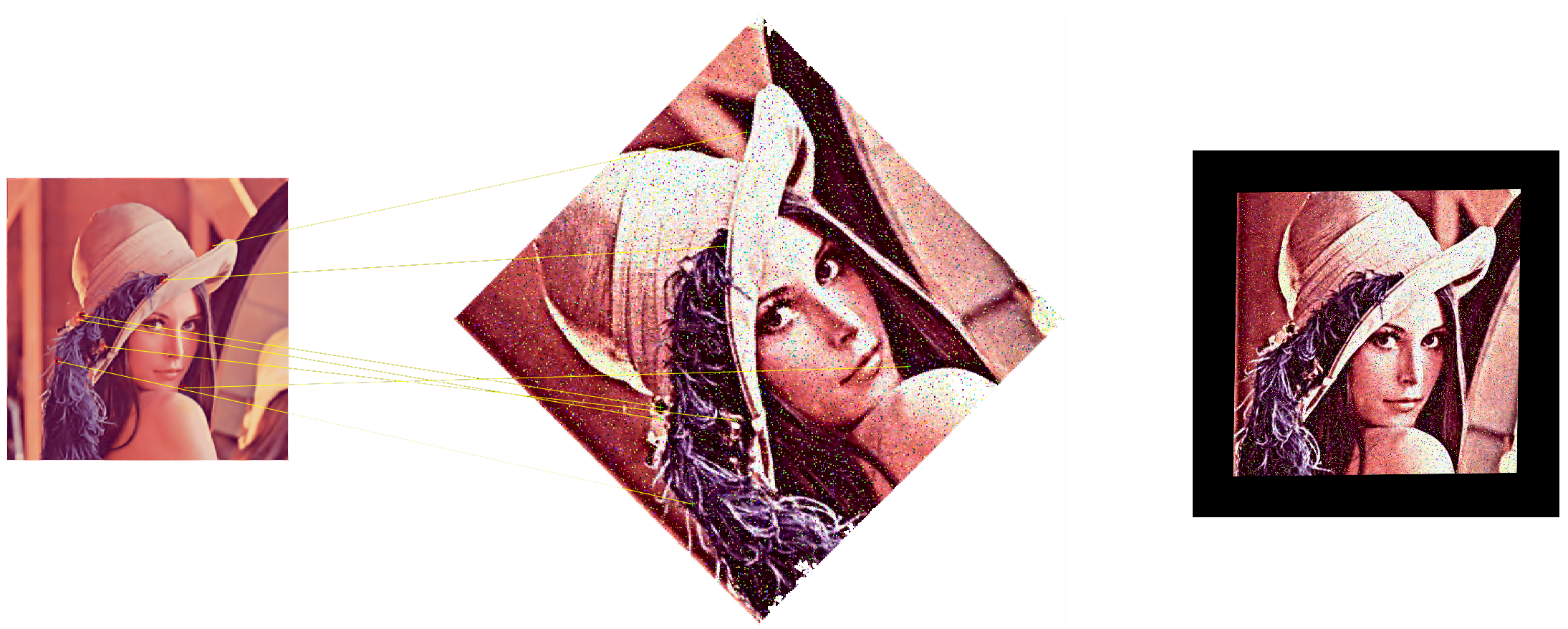}
\center
\begin{tabular}{@{\extracolsep{\fill}}ccc}
(a)&(b)&(b)\\
\end{tabular}
\caption{Result of image registration using SURF algorithm under hybrid attacks such as histogram equalization, Gaussian filter (5,0.5), Sharpening (7,0.8), Salt and pepper noise (0.05), Speckle noise (0.01), JPEG (60), Crop (60 pixels from around), Scale (2) and Rotate (45deg). a) Inverse halftone (WInHD), b) Original image under attacks, c) Reconstructed geometry (Test image Lena).}
\label{fig:jarvis}
\end{figure*}
\begin{figure*}[t]
\includegraphics[width=1\textwidth,trim= 18cm 27cm 27cm 26cm,clip]{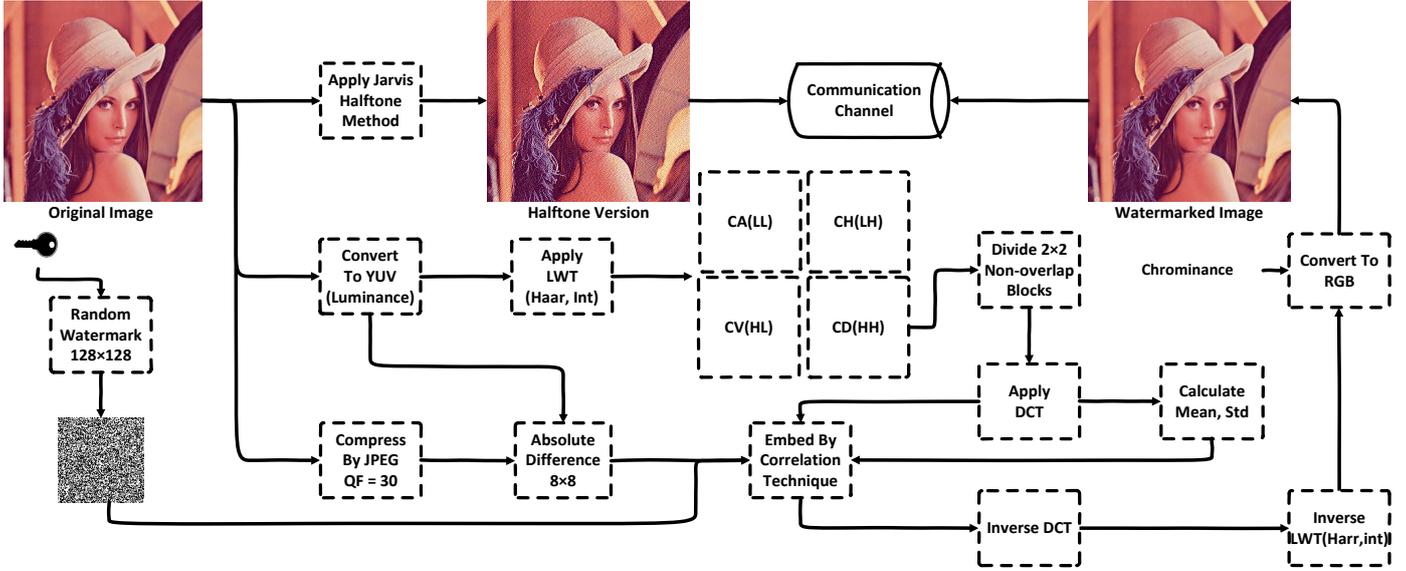}
\caption{Overall generating and embedding watermark procedure. }
\label{fig:diagramencode}
\end{figure*}
\begin{figure*}[t]
\includegraphics[width=1\textwidth,trim= 65cm 16cm 66cm 19cm,clip]{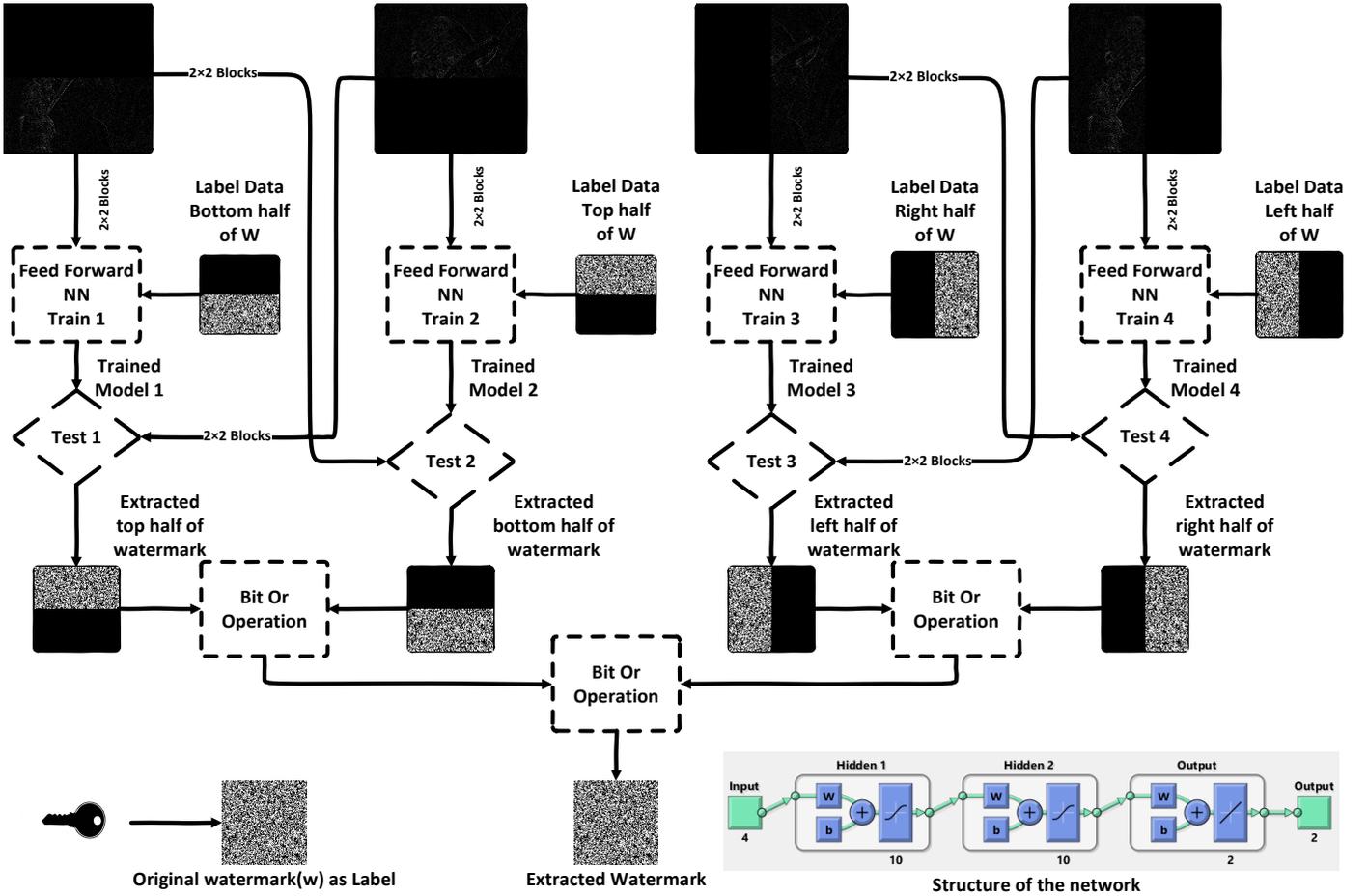}
\caption{Overall watermark extraction procedure. }
\label{fig:nn}
\end{figure*}
\begin{figure*}[t]
\includegraphics[width=1\textwidth,trim= 30cm 25cm 27cm 17cm,clip]{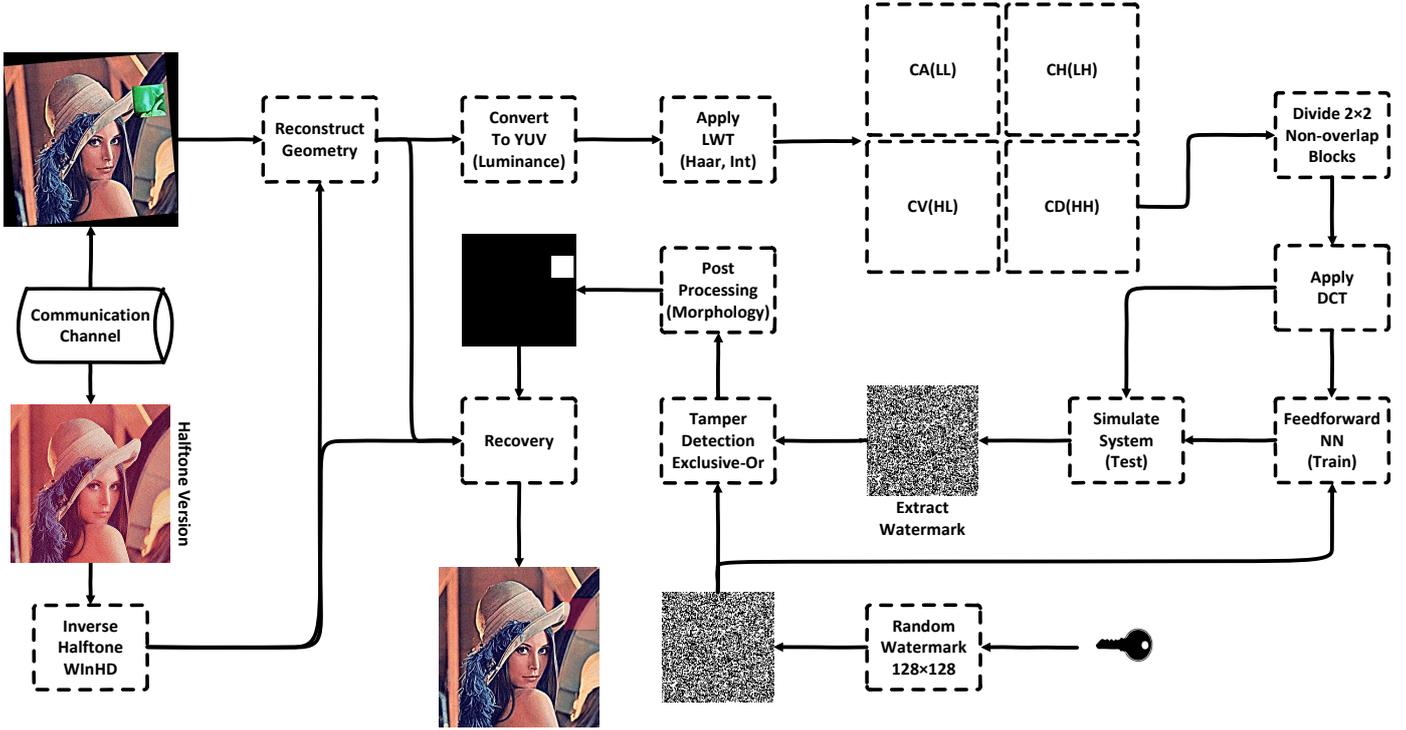}
\caption{Overall tamper detection and recovery procedure. }
\label{fig:diagramdecode}
\end{figure*}
\subsection{Speeded up robust features}\label{sec:surf}

\noindent Recently, local features detector and descriptor such as SIFT and SURF, have been employed for various application in computer vision. These methods, detect and describe distinctive invariant features from digital images which can be used to perform trusty image registration (matching) between various views of objects or scenes. 

Scale Invariant Feature Transform (SIFT) \cite{ref41,ref42} proposed by David Lowe in 2004 (as an extension of his previous work in 1999), for image registration and recognition of the digital images. Lowe's method is popularly and widely used due to its efficiency among different feature detector and descriptor algorithms. Although the SIFT can provide satisfying performances under rotation, scaling, translation transformation, the main disadvantages of this technique are low speed and fragility against some operations such as additive noises and image filtering. Based on this, the Speeded Up Robust Features (SURF) detector was presented by Bay et al in 2006 \cite{ref43}. SURF technique speeds up the SIFT’s algorithm without scarifying the quality of the detected points. In other words, SURF computes distinctive invariant local features more quickly and provide better performance rather than SIFT. Basically, SURF algorithm consists of three major phases:

\noindent\textbf{Interest point detection:} In this phase, in order to speed up feature-point detection, used by Hessian matrix based on the integral image. The integral image is used for calculating the sum of pixels values in optional rectangle region of the image. Then, the interest points are determined by using a $2\times2$ Hessian matrix $H(X, \sigma)$ that is defined by equation (\ref{eq:hessian}):
\begin{equation}
H(X,\sigma) = \begin{bmatrix}L_{xx}(X,\sigma)&L_{xy}(X,\sigma)\\L_{xy}(X,\sigma)&L_{yy}(X,\sigma)\end{bmatrix}
\label{eq:hessian}
\end{equation}
Where $L{xx}$, $L{xy}$, $L{yy}$ are the convolution operations between Gaussian second order derivatives (Laplacian), and the image $I$ in point $X=(x,y)$ at scale $\sigma$.
Furthermore, to decrease the computational complexity, SURF uses equation (\ref{eq:hessianapproximate}) as an approximation for $H(X, \sigma)$:
\begin{equation}
H_{approximate} = \begin{bmatrix}D_{xx}&D_{xy}\\D_{xy}&D_{yy}\end{bmatrix}
\label{eq:hessianapproximate}
\end{equation}
where $D_{xx}$, $D_{xy}$ and $D_{yy}$ denote the convolution results of box filter (approximate of LOG) and image. 
Finally, the determinant of Hessian matrix is calculated by equation (\ref{eq:dethessian}):
\begin{equation}
det(H_{approximate}) = D_{xx}D_{yy} - (0.9D_{xy})^2
\label{eq:dethessian}
\end{equation}
These responses are stored as preselected feature points in a blob response map over different scales space. Subsequently, using non-maximum suppression (NMS) in a $3\times3\times3$ neighborhood around each sample point (over three neighbor scales) to detect candidate interest points.

\noindent\textbf{Interest point description:} In the descriptor phase, in order to perform the invariability, each interest points described by the main orientation of feature point and the sum of Haar wavelet responses. First, the main direction is determined by detecting the vector of the summation of Gaussian weighted Haar wavelet responses based on information from a circular region around detected interest point. In the following, a rectangular region over the interest points split into smaller $4\times4$ square sub-regions. Then, for each sub-region, the horizontal and vertical Haar wavelet responses are collected from $5\times5$ samples. Finally, $dx$ and $dy$ from each sub-regions are utilized to form a four-dimensional description vector in regards with equation (\ref{eq:descriptor}) which generates the interest point descriptor.
\begin{equation}
V = (\Sigma dx, \Sigma d|x|,\Sigma dy, \Sigma d|y|)
\label{eq:descriptor}
\end{equation}

So, the dimensionality of SURF's descriptor is $4\times4\times4$. These descriptors are robust to rotation, scaling, translation transformation in the image and also robust to illumination variations, image filtering, and noise. 

\noindent\textbf{Interest-point matching and geometric transformation estimation:} In the last phase, after describing interest points, compare features descriptors to find the best matching pairs in the whole of feature points. Eventually, estimate the geometric distortions of two consecutive images after the feature-point matching. To do so, at least three matched-point pairs of the produced matching-points set should be selected, and  $3\times3$ transformation matrix should be calculated. Finally, the transformation matrix to reconstruct the attacked image need to be applied. For details, refers to \cite{ref43}.
\subsection{Feed-forward neural network}
\noindent The most common artificial neural network (ANN) is the feed-forward neural network (FNN) \cite{ref46}. This network is the first and simplest type of network architecture. Therefore, it is widely used in classifier problems. Particularly, TRLF focuses on multilayer FNN. This kind of networks has input, output, and hidden units. In this topology, the information moves forwarded from the input neurons, through the hidden neurons, reaching the output neurons. In other words, each layer except output layer connects to the next layer (not previous layers). So, there are no cycles or loops in the network. The hidden neurons extract important features contained in the data. Furthermore, it is very important to adjust the number of neurons in hidden layers by cross-validation. In general, each connection between neurons has a weighted value that describes the relational degree between two neurons. 

Assume, a network with $l$ layers which $l_{th}$ layer contains $N^l$ neurons. The forward propagation of input information in the network is defined by calculating the outputs of all neurons in each layer according to equation (\ref{eq:ff1}) and (\ref{eq:ff2}):
\begin{equation}
\psi_i^l =\displaystyle\sum_{j=1}^{N^{l-1}}W_{ij}^{l-1}x_j^{l-1} - \beta_i^l
\label{eq:ff1}
\end{equation}
\begin{equation}
x_i^l = f^l(\psi_i^l)
\label{eq:ff2}
\end{equation}
where $x_i^l$ and $W_{ij}^l$ represent the output of $i_{th}$ neurons in the $l_{th}$ layers, and the weight connection between the $i_{th}$ ($l_{th}$ layer) and $j_{th}$ ($l-1_{th}$ layer) neurons, respectively. The $\psi_i^l$ is the output and $\beta_i^l$ is bias term (threshold coefficient) of the $i_{th}$ neuron in the $l_{th}$ layer. Subsequently, the $f$ is referred to non-linear transfer function (activation function) of hidden layer as the sigmoid function that defined as  equation (\ref{eq:ff3}).
\begin{equation}
f(x) = \frac{1}{1+e^{-gx}}
\label{eq:ff3}
\end{equation}
In TRLF for hidden and output layers, the sigmoid and linear function is used as transfer function, respectively. In equation \ref{eq:ff3}, when $g$ become large, the sigmoid function become a signum function. In summary, the weighted sum of the inputs are calculated by neurons (the activation function), and then passed through a transfer function to generate the output for the neuron. Finally, the network is trained with a back-propagation learning algorithm to adjust the weights of the neural to train the network. To do so, the algorithm starts from the output layer and the error in the output of neurons are propagated backward through hidden layers to the input layer. The weights optimization problem is used to find the best weights that can minimize the mean-squared error between network output and the desired output. The error value is then propagated backward through the network, and weights are updated to decrease the error in each layer. The trained process is repeated continuously until a convergence of weight coefficients are achieved. To train the network the Levenberg–Marquardt back-propagation algorithm is used in TRLF \cite{ref47}. 
\section{Proposed method}
\noindent This section presents an effective semi-fragile watermarking method for tamper detection and recovery based on LWT and FNN, which provide high visual quality and robustness. Most image authentication methods consider fixed threshold step for hiding data, regardless of the characteristics of blocks. However, some blocks require the strong threshold to obtain adequate robustness against various attacks, and in contrast, some blocks require lower threshold based on their contents (texture or flat regions). In TRLF, the threshold step of embedding watermark is determined using JPEG compression; in other words, it is used to distinguish the threshold step of each block. In addition, LWT is chosen over the other transforms as the embedding domain due to its energy compaction property that increases robustness, and low computational time and memory requirements. On the receiving side, FNN is used in the extraction phase, because this network has powerful and good generalization ability.

TRLF consists of two major phases: watermark generation and embedding phase, and tamper detection and recovery phase. The watermark generation and embedding phase is described in Section \ref{sec:embedding}, and Section \ref{sec:detection} focuses on how to detect and reconstruct tampered regions. In the following subsections, we explain each phase in details.

\subsection{Watermark generation and embedding}
\label{sec:embedding}
\noindent An original image as $Cover$ is a $H\times W$ lossless gray or color image, where $H$ is the image height and $W$ is the image width, and both are multiple of $4$. Let $N$ represent the total number of blocks ($N = \rfrac{H}{4} \times \rfrac{W}{4}$). In TRLF, we can detect tampered blocks with the size of $4\times4$, instead of the commonly used $8\times8$ \cite{ref26}. The smaller blocks lead to increased capability in localizing the possible tampered areas. The block diagram of the proposed embedding process is illustrated in Fig. \ref{fig:diagramencode}. This phase essentially includes the following steps.

\textbf{\textit{Step} 1:} First of all, a random binary watermark which is denoted as $W$ with pixel dimensions of  $\rfrac{H}{4} \times \rfrac{W}{4}$ is generated by the secret $Key$ and serves as the authentication code. In the following, the halftone version of the original image (denote by $Halftone_{digest}$), which is used as image digest of the original image, is generated by Jarvis algorithm \cite{ref37} according to section \ref{sec:winhd}. Further, the compression version of the original image is obtained by JPEG (QF=30), and just to be denoted as $Cover_{jpeg}$.

\textbf{\textit{Step} 2:} The YUV color space is employed in $Cover$, and denote luminance channel as $Lum$. If the cover image is in gray-scale mode, just define it as $Lum$. Similarly, this process is done for $Cover_{jpeg}$, and denote result as $Lum_{jpeg}$. Based on the experimental results, we concluded that the YUV space has better performance for visual quality of watermarked image rather than YCbCr.

\begin{figure}[H]
\center
\includegraphics[width=0.5\columnwidth]{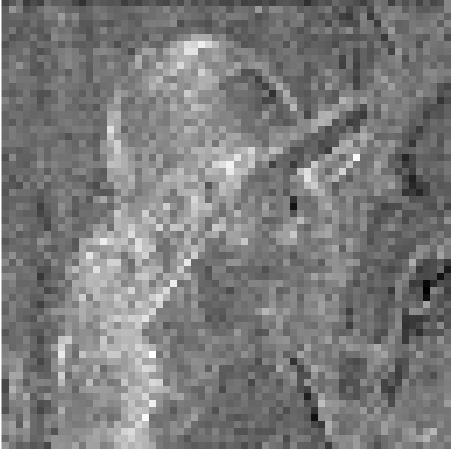}
\caption{The mean absolute difference of Lena image (scaled up).}
\label{fig:dif}
\end{figure}

\textbf{\textit{Step} 3:} In this step, the absolute difference between $Lum$ and $Lum_{jpeg}$ is calculated. To be implemented, the mean absolute difference for each $8\times8$ blocks is computed, and denote result as $Dif$ with pixel dimensions of 64$\times$64. This analysis helps us to consider difference threshold step for various texture and flat regions. Accordingly, the quality of the watermarked image and the robustness of method against various attacks (especially compression) is improved. Finally, $Dif$ is resized to 128$\times$128. The mean absolute difference of Lena image is illustrated in Fig. \ref{fig:dif}.

\textbf{\textit{Step} 4:} A level of LWT is performed for the $Lum$ with the use of the $Haar$ filter. Then the $CD$ sub-band coefficients with pixel dimensions of $256\times256$ is divided into non-overlapping blocks of size $2\times2$. Meanwhile, DCT is employed in each block. The result of decomposition should be a matrix formed by $2\times2$ blocks with the size of $128\times128$. 

Embedding the watermark on $CA$ sub-band would reduce the perceptual quality of the watermarked images. In other words, hiding the watermark in the high-frequency coefficient, render the high imperceptibility for the human eye. On the other hand, we concluded the $CD$ sub-band provides higher quality and robustness for watermarked image rather than other high-frequency sub-bands.

\textbf{\textit{Step} 5:} Mean and standard deviation of $DC$ coefficient for all blocks is computed by equation (\ref{eq:mean}) and (\ref{eq:std}).
\begin{equation}
\mu = \frac{1}{N}{\displaystyle\sum_{i=1}^{\rfrac{H}{4}} \displaystyle\sum_{j=1}^{\rfrac{W}{4}} DC(i, j)}
\label{eq:mean}
\end{equation}
\begin{equation}
\sigma = \sqrt{\frac{1}{N}{\displaystyle\sum_{i=1}^{\rfrac{H}{4}} \displaystyle\sum_{j=1}^{\rfrac{W}{4}} (DC(i, j) - \mu)^2}}
\label{eq:std}
\end{equation}

\textbf{\textit{Step} 6:} Threshold matrix with dimensions of 128$\times$128 is computed by equation (\ref{eq:threshold}).
\begin{equation}
T = T_1 + (T_2 \times Dif(i, j))
\label{eq:threshold}
\end{equation}
This threshold matrix is very important and controls the imperceptibility and robustness of watermark for various block, based on their position. It should be noted that the high $(T_1, T_2)$ decreases the quality of the watermarked image, and increase the robustness. In TRLF, we set $T_1$ and $T_2$ as $5$ and $3$, respectively, which is empirically determined based on the trade-off among invisibility and robustness. Now, for each $W$ bit, correlate the $DC$ coefficient of the corresponding block in $CD$ using equation \ref{eq:w1} and \ref{eq:w0}.
\noindent If the embedded $W$ bit is 1, modifies $DC(i, j)$ using equation (\ref{eq:w1}):
\begin{equation}
DC(i, j) =
  \begin{cases}
    DC(i, j) + T(i, j) & \quad \text{if } DC(i, j) <  \sigma + \mu\\
    DC(i, j)  & \quad \text{otherwise}\\
  \end{cases}
\label{eq:w1}
\end{equation}

\noindent and, if the embedded $W$ bit is 0, modifies $DC(i, j)$ using equation (\ref{eq:w0}):
\begin{equation}
 DC(i, j) =
  \begin{cases}
    DC(i, j) - T(i, j) & \quad \text{if } DC(i, j) >  -\sigma + \mu \\
    DC(i, j)  & \quad \text{otherwise}\\
  \end{cases}
\label{eq:w0}
\end{equation}

\textbf{\textit{Step} 7:} Inverse DCT is performed in each block to reconstruct the $CD$. Then, inverse LWT is employed to obtain the Luminance.  Finally, the $Cover_{watermarked}$ is generated by converting the image to $RGB$ space and send it along with $Halftone_{digest}$ through the communication channel. In order to assurance the security of $Halftone_{digest}$, we can use Arnold Cat Map \cite{ref31}, or other chaotic maps to confuse it.
\begin{figure*}[t]
\center
\begin{tabular*}{1\textwidth}{cc}
\includegraphics[width=0.16\textwidth]{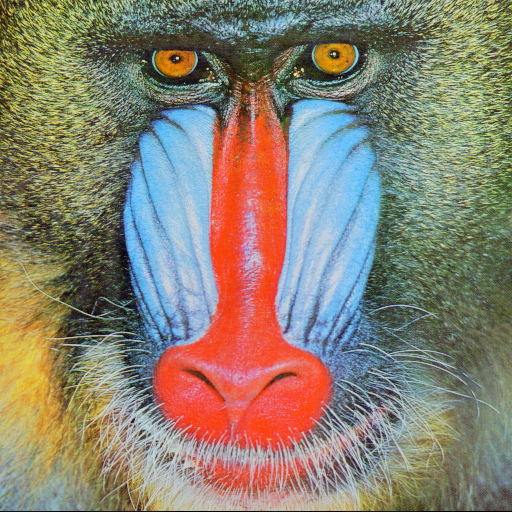}\includegraphics[width=0.16\textwidth]{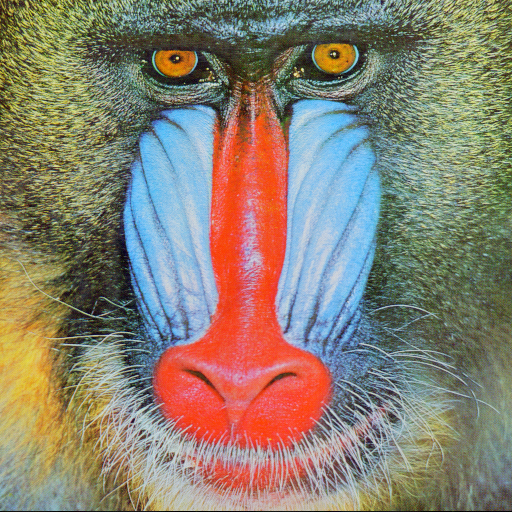}\includegraphics[width=0.16\textwidth]{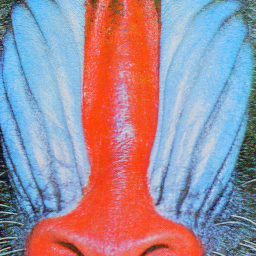}&
\includegraphics[width=0.16\textwidth]{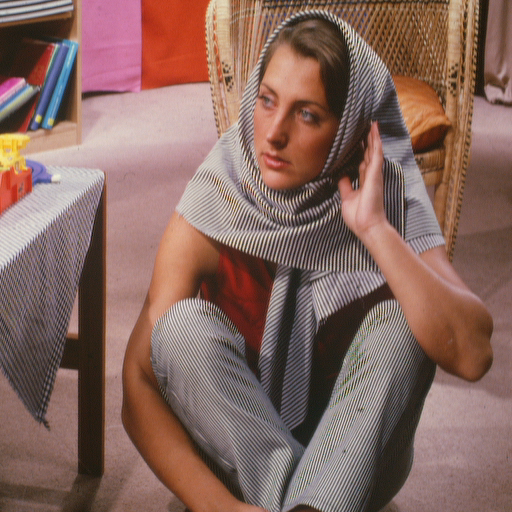}\includegraphics[width=0.16\textwidth]{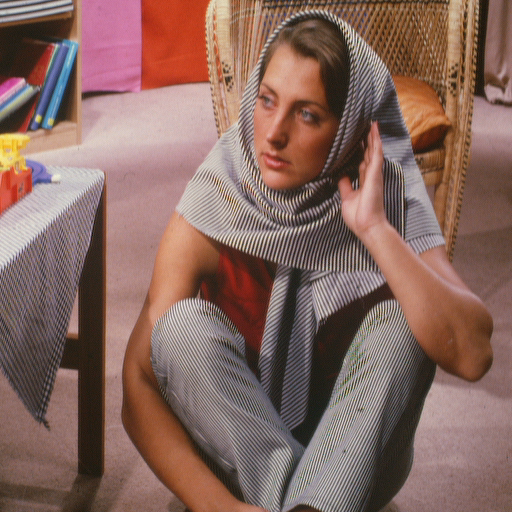}\includegraphics[width=0.16\textwidth]{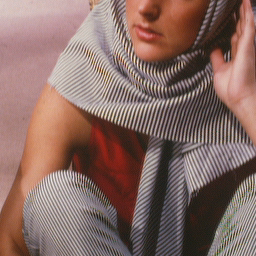}\\
(a) & (b)\\
\includegraphics[width=0.16\textwidth]{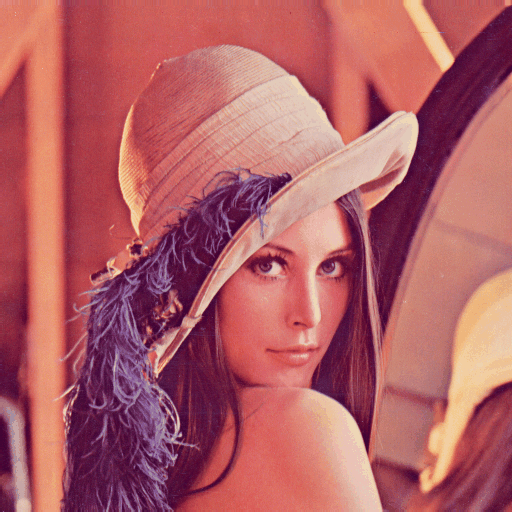}\includegraphics[width=0.16\textwidth]{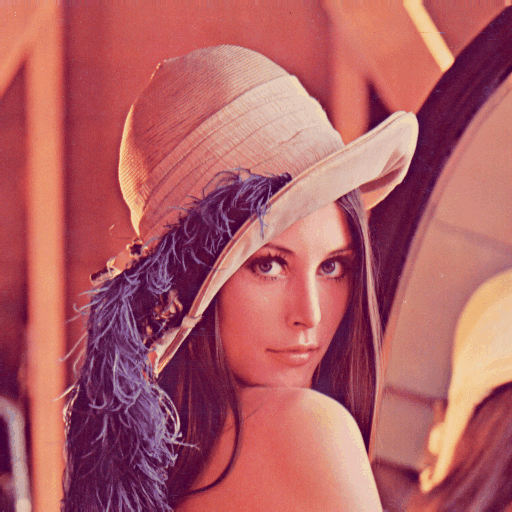}\includegraphics[width=0.16\textwidth]{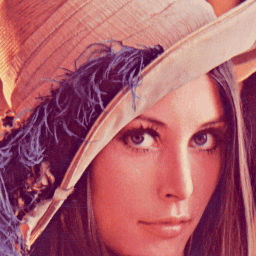}&
\includegraphics[width=0.16\textwidth]{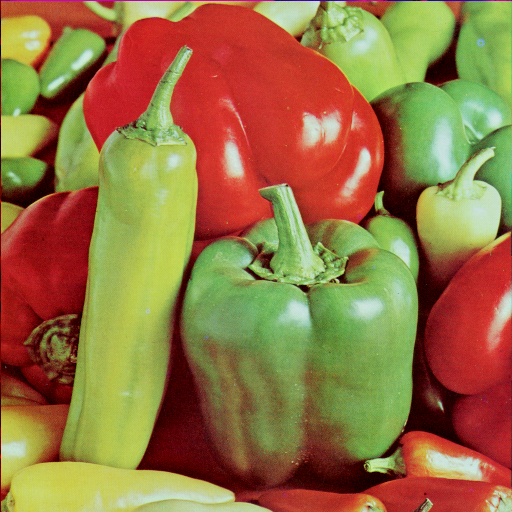}\includegraphics[width=0.16\textwidth]{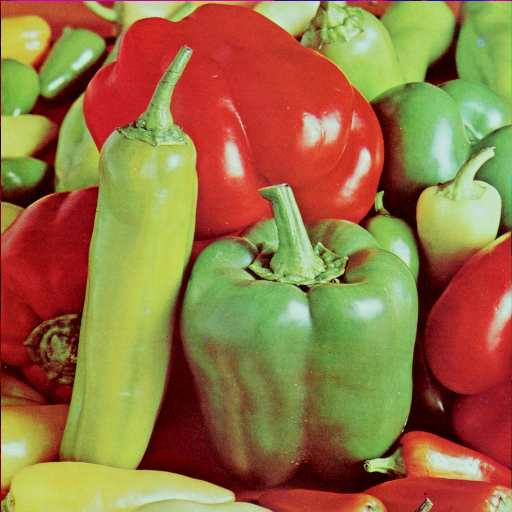}\includegraphics[width=0.16\textwidth]{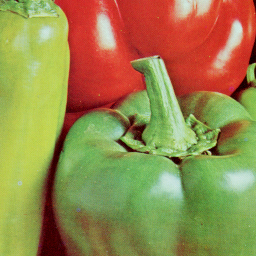}\\
(c) & (d)\\
\includegraphics[width=0.16\textwidth]{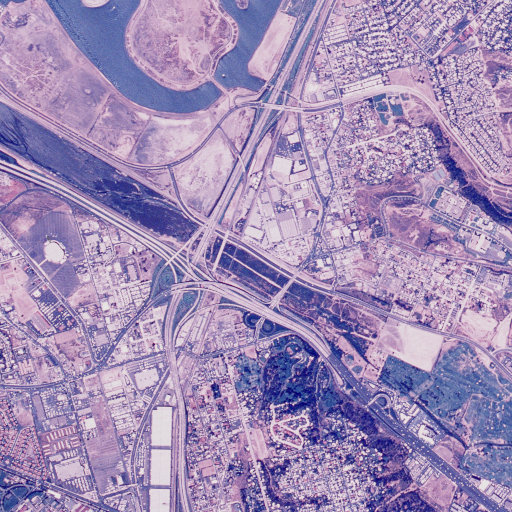}\includegraphics[width=0.16\textwidth]{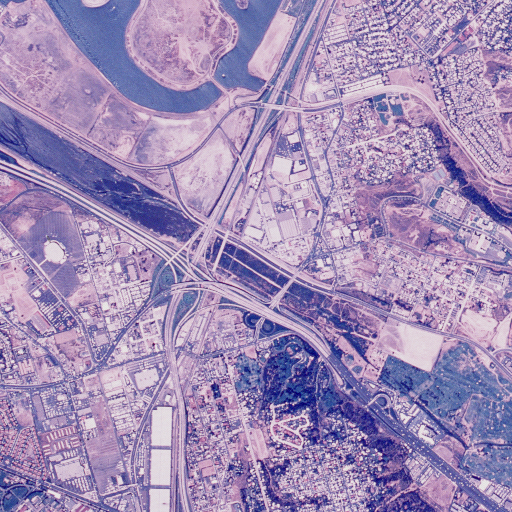}\includegraphics[width=0.16\textwidth]{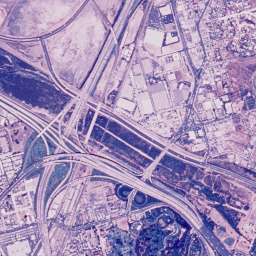}&
\includegraphics[width=0.16\textwidth]{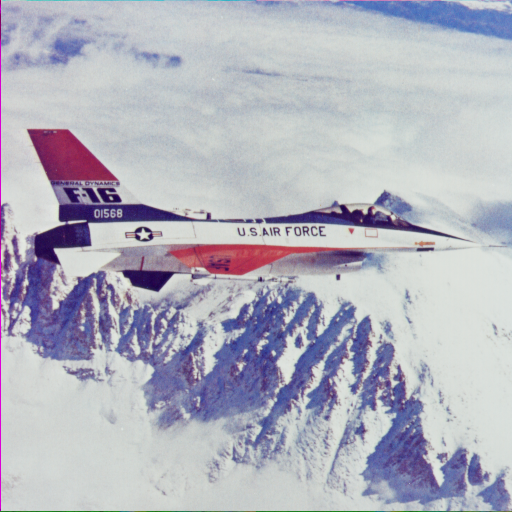}\includegraphics[width=0.16\textwidth]{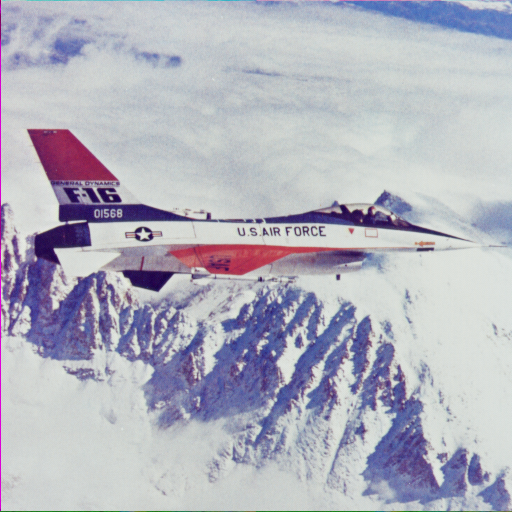}\includegraphics[width=0.16\textwidth]{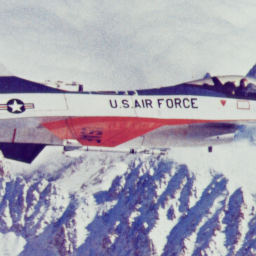}\\
(e) & (f)\\
\includegraphics[width=0.16\textwidth]{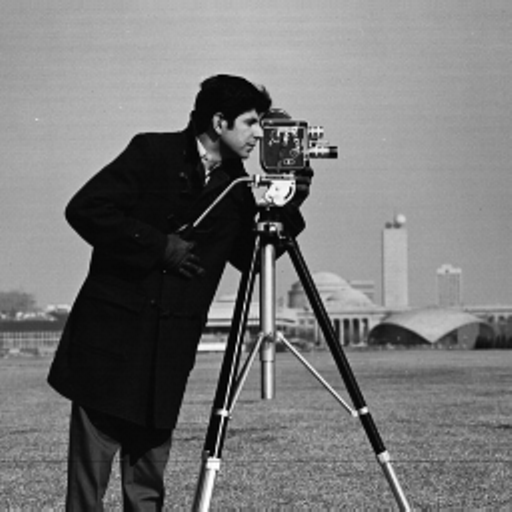}\includegraphics[width=0.16\textwidth]{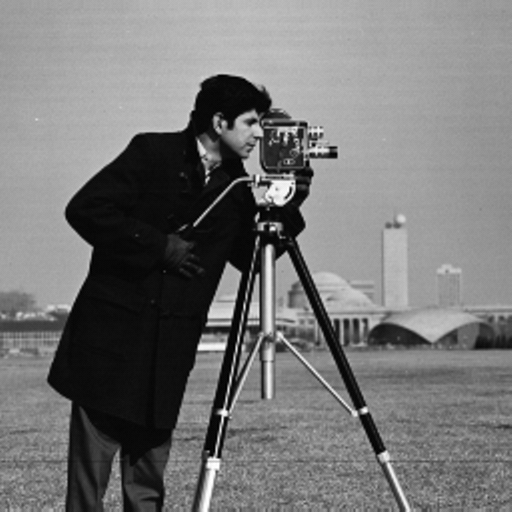}\includegraphics[width=0.16\textwidth]{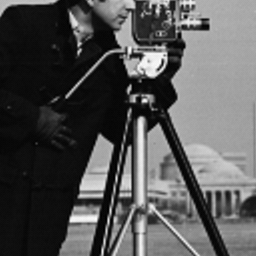}&
\includegraphics[width=0.16\textwidth]{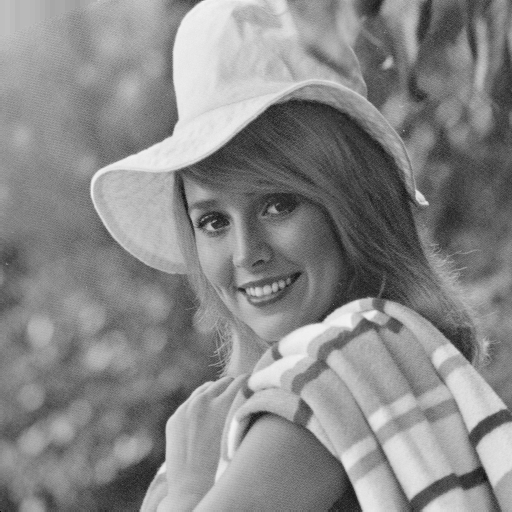}\includegraphics[width=0.16\textwidth]{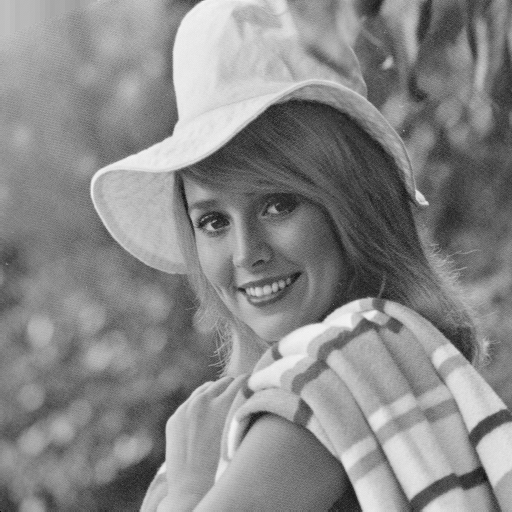}\includegraphics[width=0.16\textwidth]{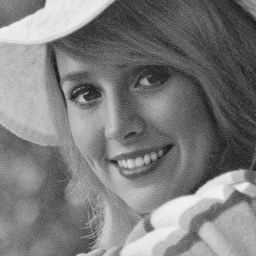}\\
(g) & (h)\\
\end{tabular*}
\caption{Standard test image (512 $\times$ 512) used in our experiments, Watermarked image, Zoom of the watermarked image (center) (a) Baboon (PSNR=41.0983, SSIM=0.9968, MSE=5.0495), (b) Barbara (PSNR=43.8356, SSIM=0.9962, MSE=2.6885), (c) Lena (PSNR=43.4694, SSIM=0.9988, MSE=2.9251), (d) Pepper (PSNR=43.2810, SSIM=0.9987, MSE=3.0548), (e) Santiago (PSNR=40.9769, SSIM=0.9961, MSE=5.1927), (f) F16 (PSNR=44.5105, SSIM=0.9863, MSE=2.3016), (f) Cameraman (PSNR=46.7033, SSIM=0.9869, MSE=1.3892), (h) Elaine (PSNR=42.8292, SSIM=0.9804, MSE=3.3897). }
\label{fig:db_w_zoom}
\end{figure*}
\begin{figure*}[t]
\begin{tabular}{@{}m{0mm}c@{}c@{}c@{}c@{}c@{}c@{}}
&\textbf{$T_1$ = 1}&\textbf{$T_1$ = 3}&\textbf{$T_1$ = 5}&\textbf{$T_1$ = 7}&\textbf{$T_1$ = 9}&\textbf{$T_1$ = 11}\\

&(47.7505, 0.9996)&(46.5219, 0.9994)&(45.0638, 0.9992)&(43.8520, 0.9989)&(42.8457, 0.9986)&(41.9128, 0.9983)\\
\rotatebox{90}{\textbf{$T_2$ = 2}}&
\includegraphics[align=c,width=0.16\textwidth]{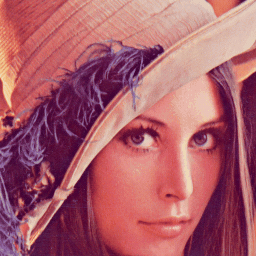}&
\includegraphics[align=c,width=0.16\textwidth]{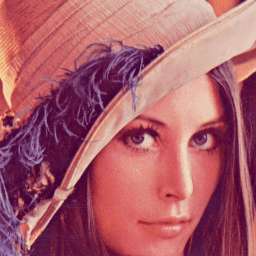}&
\includegraphics[align=c,width=0.16\textwidth]{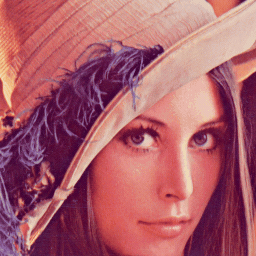}&
\includegraphics[align=c,width=0.16\textwidth]{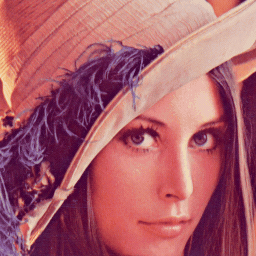}&
\includegraphics[align=c,width=0.16\textwidth]{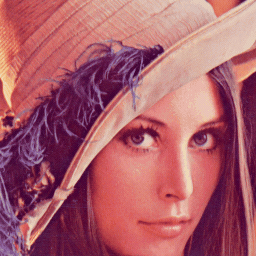}&
\includegraphics[align=c,width=0.16\textwidth]{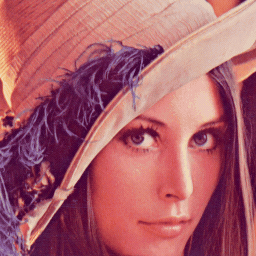}\\

&(45.9278, 0.9993)&(44.6047, 0.9991)&(43.4694, 0.9988)&(42.4836, 0.9985)&(41.5640, 0.9981)&(40.6860, 0.9977)\\
\rotatebox{90}{\textbf{$T_2$ = 3}}&
\includegraphics[align=c,width=0.16\textwidth]{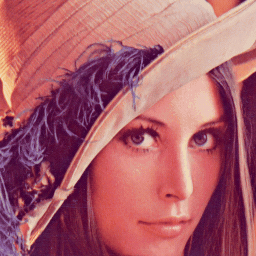}&
\includegraphics[align=c,width=0.16\textwidth]{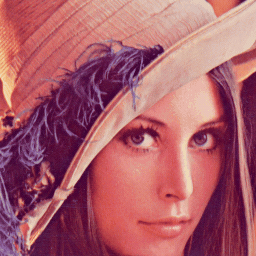}&
\includegraphics[align=c,width=0.16\textwidth]{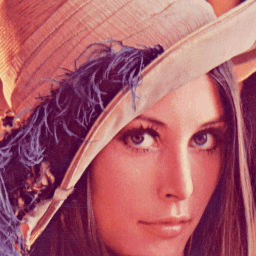}&
\includegraphics[align=c,width=0.16\textwidth]{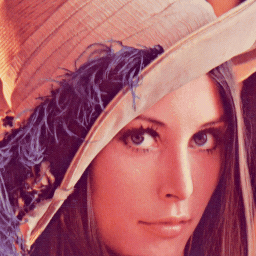}&
\includegraphics[align=c,width=0.16\textwidth]{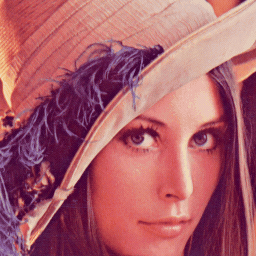}&
\includegraphics[align=c,width=0.16\textwidth]{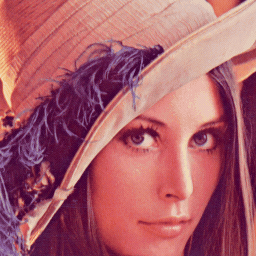}\\

&(44.1382, 0.9990)&(43.0647, 0.9987)&(42.1097, 0.9983)&(41.2168, 0.9980)&(40.3812, 0.9975)&(39.6222, 0.9971)\\
\rotatebox{90}{\textbf{$T_2$ = 4}}&
\includegraphics[align=c,width=0.16\textwidth]{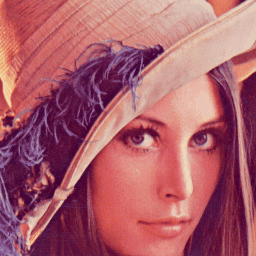}&
\includegraphics[align=c,width=0.16\textwidth]{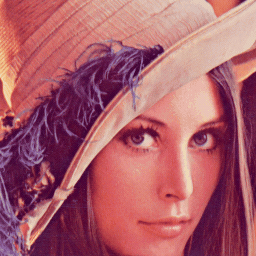}&
\includegraphics[align=c,width=0.16\textwidth]{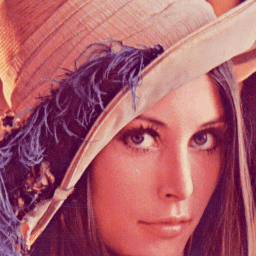}&
\includegraphics[align=c,width=0.16\textwidth]{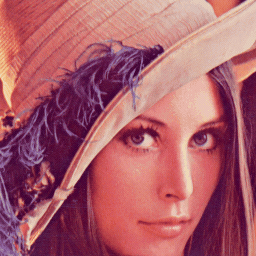}&
\includegraphics[align=c,width=0.16\textwidth]{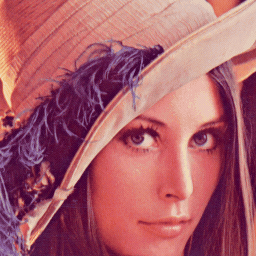}&
\includegraphics[align=c,width=0.16\textwidth]{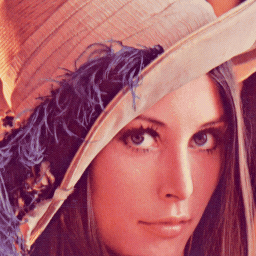}

\end{tabular}
\caption{Zoom of watermarked image with various thresholds ($T_1$, $T_2$), test image Lena.}
\label{fig:lena_zoom}
\end{figure*}
\begin{figure*}[t]
\center
\begin{tabular*}{1\textwidth}{@{}c@{}c@{}}
\includegraphics[width=1\textwidth,trim= 5cm 0cm 5cm 0cm,clip]{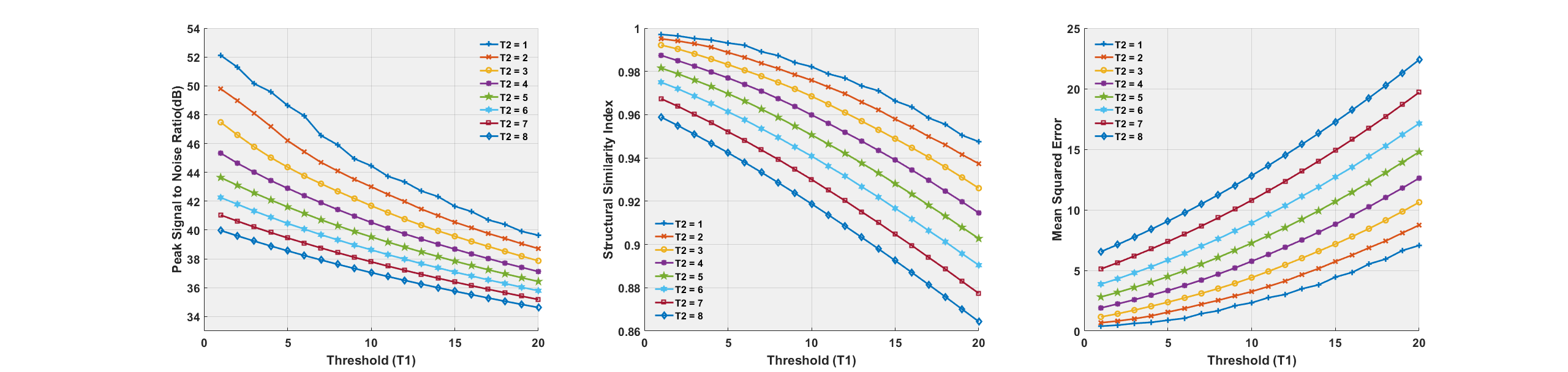}\\ (a) \\
\includegraphics[width=1\textwidth,trim= 5cm 0cm 5cm 0cm,clip]{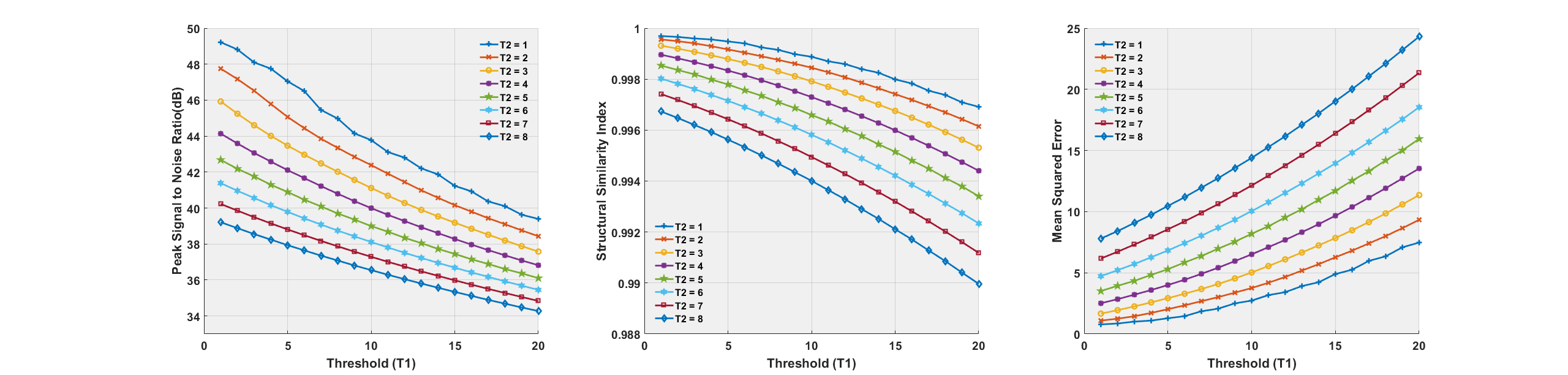} \\ (b)
\end{tabular*}
\caption{The relationship between PSNR, SSIM, MSE and thresholds ($T_1$, $T_2$), test image Lena. (a) Gray, (b) Color. }
\label{fig:relationship_between_PSNR,_SSIM_MSE_T1_T2}
\end{figure*}
\begin{table*}[t]
\center
\footnotesize
\caption{PSNR, SSIM, MSE values of watermarked images for the proposed technique and related works. \\
Note: N/A means image is not available using the previous methods.}
\label{TABLE:compare_psnr_ssim_mse_watermark}
\renewcommand{\arraystretch}{1.5}
\scalebox{1} {
\begin{tabular*}{\textwidth}{@{\extracolsep{\fill} }@{}l@{}c@{}c@{}c@{}c@{}c@{}c@{}c@{}@{}c@{}c@{}c@{}c@{}c@{}c@{}c@{}c@{} }
\cline{1-15}
\multicolumn{1}{c}{\multirow{3}{*}{Image}} & \multicolumn{6}{c}{Proposed Method} &  \multicolumn{2}{c}{\cite{ref31}} & \cite{ref30} & \cite{ref22} & \cite{ref24} & \multicolumn{2}{c}{\cite{ref25}}& \cite{ref28}\\
\cline{2-7} \cline{8-15} & \multicolumn{3}{c}{Color} & \multicolumn{3}{c}{Gray} 
& \multicolumn{2}{c}{Gray}  & Gray & Gray & Gray & \multicolumn{2}{c}{Gray}& Gray \\
\cline{2-4}\cline{5-7}\cline{8-15} &PSNR&SSIM&MSE&PSNR&SSIM&MSE&PSNR&SSIM&PSNR&PSNR&PSNR&PSNR&SSIM&PSNR \\ 
\cline{1-15}
Baboon&41.0983&0.9968&5.0495&41.7502&0.9908&4.3457&27.01&0.90&41.09&N/A&41.30&35.13&0.95&43.1\\
Barbara&43.8356&0.9962&2.6885&44.4620&0.9877&2.3274&N/A&N/A&N/A&N/A&N/A&N/A&N/A&43.3\\
Lena&43.4694&0.9988&2.9251&44.3641&0.9832&2.3805&34.67&0.95&41.76&36.73&41.04&35.02&0.87&43.2\\
Pepper&43.2810&0.9987&3.0548&44.0413&0.9819&2.5642&34.51&0.95&41.24&N/A&40.51&N/A&N/A&N/A\\
Santiago&40.9769&0.9961&5.1927&41.5456&0.9934&4.5553&N/A&N/A&N/A&35.91&N/A&N/A&N/A&N/A\\
F16&44.5105&0.9863&2.3016&45.4974&0.9858&1.8337&32.81&0.91&41.04&N/A&40.35&35.16&0.86&N/A\\
Cameraman&N/A&N/A&N/A&46.7033&0.9869&1.3892&32.98&0.93&41.80&N/A&40.18&35.57&0.86&N/A\\
Elaine&N/A&N/A&N/A&42.8292&0.9804&3.3897&34.79&0.93&N/A&N/A&N/A&N/A&N/A&N/A\\
\cline{1-15}
\end{tabular*}}
\end{table*}
\begin{figure*}[t!]
\center
\begin{tabular*}{\textwidth}{@{}c@{}}
\includegraphics[width=1\textwidth,trim= 3cm 0cm 3cm 0cm,clip]{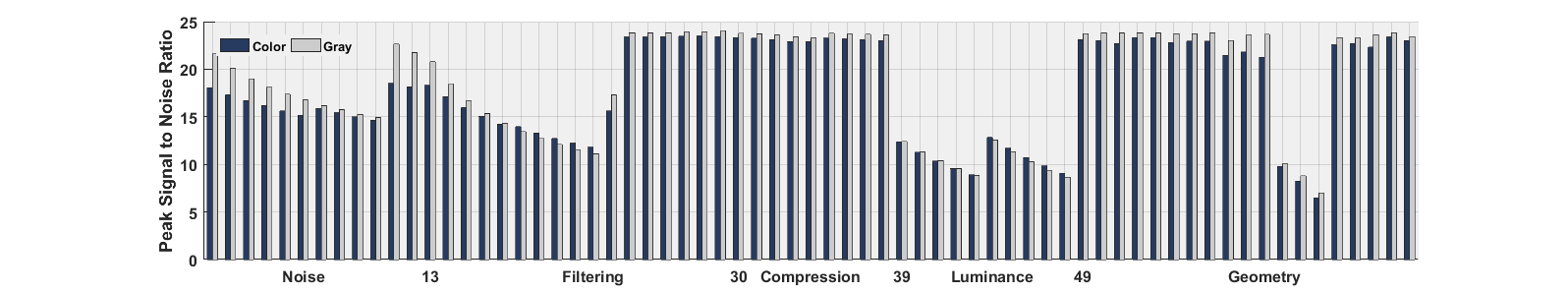}\\
\includegraphics[width=1\textwidth,trim= 3cm 0cm 3cm 0cm,clip]{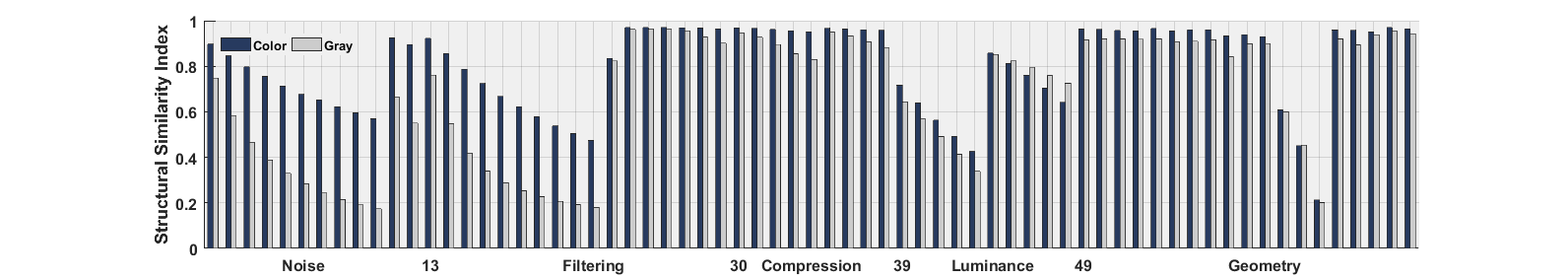}\\
\includegraphics[width=1\textwidth,trim= 3cm 0cm 3cm 0cm,clip]{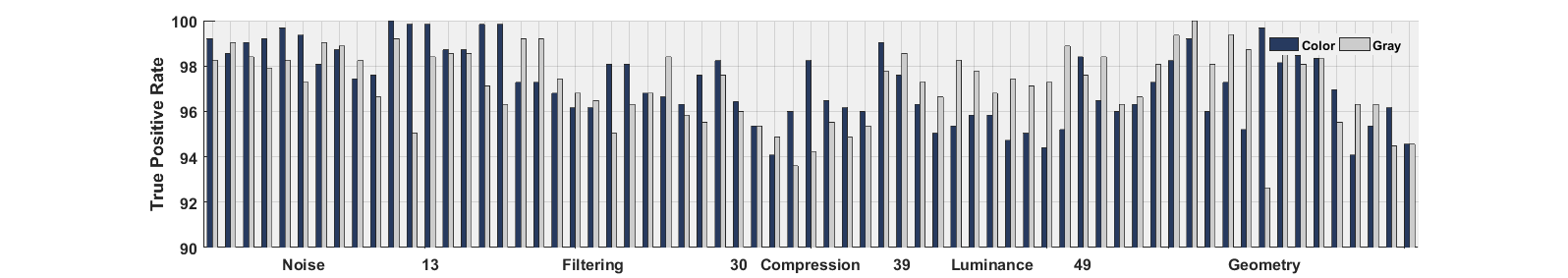}\\
\includegraphics[width=1\textwidth,trim= 3cm 0cm 3cm 0cm,clip]{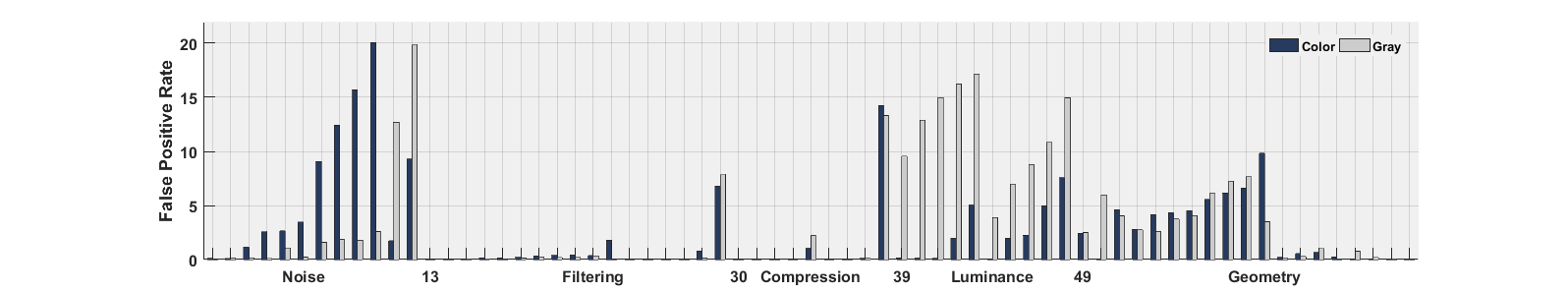}
\end{tabular*}
\caption{Results of test color and gray watermarked image Fig. \ref{fig:testimage}. under various attacks. 1-10 salt and pepper noise ($\rho$ = 0.01 to 0.1 by step = 0.01), 11-12 speckle noise ($\sigma$ = 0.005 to 0.01 by step = 0.005), 13-22 sharpening (size = 5, $\sigma$ = 1, strength  = 2 to 20 by step = 2), 23 histogram equalization, 24-29 gaussian smoothing (size = 3, $\sigma$ = 0.1 to 0.6 by step 0.1), 30-34 JPEG2000 (QF = 90 to 50 by step 10), 35-38 JPEG (QF = 90 to 60 by step 10), 39-43 darken (60 to 100 by step 10), 44-48 lighten (60 to 100 by step 10), 49-59 rotation (1, 2, 3, 4, 5, 10, 15, 20, 25, 30, 35 degree), 60-61 translation ('100,100' and '150,150'), 62 cropping (25\%), 63-67 scale (1.1 to 1.5 by step 0.1).}
\label{fig:diagramrobustness}
\end{figure*}
\begin{table*}[t]
\center
\footnotesize
\caption{Time complexity (seconds) of watermark embedding, A) Generate halftone, B) Color convert (RGB to YUV), C) Embedding, D) Color convert (YUV to RGB) E) Total, F) Inverse halftone, G) Estimate geometrics, H) Color convert (RGB to YUV), I) Extract confinements, J) Train$\times$4, K) Test$\times$4, L) Post processing, M) Recovery, N) Total, O) Sum, Time order report under 200$\times$200 tampering, and various attacks such as Histogram equalization, sharpening (size = 29, $\sigma$ = 7, strength = 0.8), salt and pepper noise ($\rho$ = 0.01), cropping (42\% around), scale (2), rotate (90 deg).\\}
\label{TABLE:time_complexity}
\renewcommand{\arraystretch}{1.5}
\begin{tabular*}{\textwidth}{@{\extracolsep{\fill}}@{}l@{}c@{}c@{}c@{}c@{}c@{}c@{}c@{}c@{}c@{}c@{}c@{}c@{}c@{}c@{}c@{}}
\cline{1-16}
\multicolumn{1}{l}{\multirow{2}{*}{Image}}&\multicolumn{5}{c}{Embedding phase}&\multicolumn{9}{c}{Authentication phase}&Sum\\
\cline{2-6}\cline{7-15}\cline{16-16}
&A&B&C&D&E&F&G&H&I&J&K&L&M&N&O\\
\cline{1-16}
Baboon&0.47&0.01&2.62&0.02&3.12&2.58&0.60&0.01&0.79&0.58&0.01&1.06&0.03&5.66&8.78\\
Barbara&0.46&0.01&2.60&0.03&3.10&2.64&0.57&0.01&0.82&0.59&0.01&1.03&0.03&5.70&8.80\\
Lena&0.48&0.01&2.61&0.02&3.12&2.62&0.52&0.01&0.79&0.51&0.01&0.94&0.03&5.43&8.55\\
Pepper&0.48&0.01&2.64&0.02&3.15&2.62&0.51&0.01&0.83&0.53&0.01&0.95&0.04&5.50&8.65\\
Santiago&0.48&0.01&2.61&0.03&3.13&2.62&0.68&0.01&0.81&0.52&0.01&1.11&0.03&5.79&8.92\\
F16&0.47&0.01&2.58&0.02&3.08&2.62&0.57&0.01&0.82&0.52&0.01&1.02&0.04&5.61&8.69\\
Camera man&0.37&0&2.63&0&3&1.12&0.49&0&0.84&0.56&0.01&1.07&0.02&4.11&7.11\\
Elaine&0.37&0&2.63&0&3&1.13&0.48&0&0.85&0.54&0.01&1.18&0.03&4.22&7.22\\
\cline{1-16}
\end{tabular*}
\end{table*}
\begin{table*}[t]
\center
\footnotesize
\caption{Prominent features and performance comparison with other methods.
\\A) Domain, B) Robustness, C) Detection Process, 
D) Payload (bit per block), E) Localization (Block Size), F) Time Complexity, G) Security, H) Quality of watermarked image, I) Quality of Recovered image ($\times$ no recover)
J) Copy-move, K) Collage, L) Vector-quantization.\\
Note: N/R and F/W means not reported and future work, respectively.}
\label{TABLE:compare_performance1}
\renewcommand{\arraystretch}{1.5}
\begin{tabular*}{\textwidth}{@{\extracolsep{\fill}}@{}l@{}c@{}c@{}c@{}c@{}c@{}c@{}c@{}c@{}c@{}c@{}c@{}c@{}}
\cline{1-13}
\multicolumn{1}{l}{\multirow{2}{*}{Method}}&
\multicolumn{3}{c}{Classification}&
\multicolumn{6}{c}{Features}&
\multicolumn{3}{c}{Tampering Type}\\
\cline{2-4}\cline{5-10}\cline{11-13}
&A&B&C&D&E&F&G&H&I&J&K&L\\
\cline{1-13}
TRLF&LWT&Semi-fragile&Semi-blind&1&$4\times4$&Low&Random binary&High&High&Detect&Detect&Detect\\
Phadikar et al. \cite{ref25}&IWT&Semi-fragile&Blind&2&$8\times8$&High&Random binary&Low&Low&Detect&F/W&F/W\\
Roldan et al. \cite{ref26}&IWT&Semi-fragile&Blind&1&$8\times8$&N/R&Push aside\cite{ref10}&Low&Medium&N/R&N/R&N/R\\
Preda. \cite{ref27}&DWT&Semi-fragile&Blind&1&$4\times4$&N/R&Random binary&High&$\times$&N/R&N/R&Detect\\
Yaoran et al. \cite{ref28}&DCT&Semi-fragile&Blind&6&$8\times8$&Low&Medium&High&$\times$&N/R&Detect&N/R\\
Al-Otum et al. \cite{ref29}&DWT&Semi-fragile&Blind&1&$8\times8$&N/R&Random binary&Medium&$\times$&N/R&N/R&N/R\\
Qi  et al. \cite{ref30}&DWT+SVD&Semi-fragile&Semi-blind&1&$4\times4$&N/R&Hash function&Medium&$\times$&N/R&N/R&N/R\\
Rhouma et al. \cite{ref31}&DWT&Semi-fragile&Blind&Dynamic&$4\times4$&N/R&Cat map&Low&Low&N/R&N/R&N/R\\
\cline{1-13}
\end{tabular*}
\end{table*}
\begin{table*}[t!]
\center
\footnotesize
\caption{Compare Robustness against various attacks with other methods.
\\A) Histogram equalization, B) Gaussian filter, C) Average, D) Median, E) Sharpening, F) Salt and pepper, G) Gaussian, H) Speckle, I) JPEG, J) JPEG2000, K) Translation ,L) Cropping , M) Scale, N) Rotation. \\
Note: N/R and F/W means not reported and future work, respectively.}
\label{TABLE:compare_performance2}
\renewcommand{\arraystretch}{1.5}
\begin{tabular*}{\textwidth}{@{\extracolsep{\fill}}l@{}c@{}c@{}c@{}c@{}cc@{}c@{}cc@{}cc@{}c@{}c@{}c@{}}
\cline{1-15}
\multicolumn{1}{l}{\multirow{2}{*}{Method}}&
\multicolumn{5}{c}{Image filters}&
\multicolumn{3}{c}{Noise}&
\multicolumn{2}{c}{Compression}&
\multicolumn{4}{c}{Geometric attacks}\\
\cline{2-6}\cline{7-9}\cline{10-11}\cline{12-15}
&A&B&C&D&E&F&G&H&I&J&K&L&M&N\\
\cline{1-15}
TRLF&High&High&Low&Low&High&High&Medium&High&Medium&Medium&High&High&High&High\\
Phadikar et al. \cite{ref25}&Low&N/R&N/R&High&N/R&Medium&High&High&Medium&N/R&N/R&N/R&N/R&Low\\
Roldan et al. \cite{ref26}&High&N/R&N/R&Medium&N/R&Medium&High&High&Medium&N/R&N/R&N/R&N/R&Low\\
Preda. \cite{ref27}&N/R&N/R&N/R&N/R&N/R&N/R&N/R&N/R&Medium&N/R&N/R&N/R&N/R&N/R\\
Yaoran et al. \cite{ref28}&N/R&N/R&N/R&N/R&N/R&N/R&N/R&N/R&Medium&N/R&N/R&N/R&N/R&N/R\\
Al-Otum et al. \cite{ref29}&N/R&N/R&N/R&High&N/R&High&High&N/R&High&N/R&N/R&Low&N/R&N/R\\
Qi  et al. \cite{ref30}&N/R&High&N/R&Medium&N/R&Medium&N/R&N/R&High&High&F/W&F/W&F/W&F/W\\
Rhouma et al. \cite{ref31}&N/R&N/R&Low&N/R&N/R&Low&N/R&N/R&Medium&N/R&N/R&N/R&N/R&Low\\
\cline{1-15}
\end{tabular*}
\end{table*}
\begin{figure*}[t]
\center
\begin{tabular*}{1\textwidth}{c}
\includegraphics[width=1\textwidth,trim= 5cm 0cm 4cm 0cm,clip]{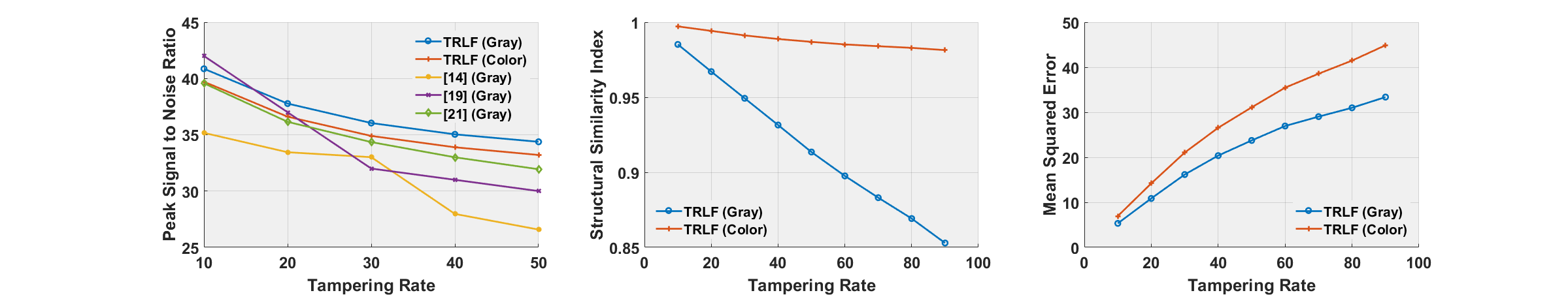}
\end{tabular*}
\caption{PNSR, SSIM, and MSE of recovered image with different tampering rates, test image Lena \cite{ref10, ref15, ref17}.}
\label{fig:diagramrecover}
\end{figure*}
\subsection{Tamper detection and recovery}
\label{sec:detection}
\noindent The proposed tamper detection and recovery method is semi-blind. This means that the tamper detection and recovery phase does not require the cover image, but in TRLF, image digest is transfered to reconstruct the geometry of the suspicious image and also recover tampered regions.

The intentional and non-intentional alterations may be subjected to the watermarked image which has been received through an open communication channel. The watermark is extracted from this suspicious image, which is possibly tampered to analyze it. If the received image is considered as the tamper, the image digest is used to reconstruct the tampered regions. Extracting the watermark from the watermarked image is the inverse of the watermark embedding procedure which the block diagrams is shown in Fig.\ref{fig:diagramdecode}. The tamper detection and recovery phase of TRLF are described as follows:

\textbf{\textit{Step} 1:} The random binary watermark ($W$) is generated in the same way as in the previous phase. In the following, the inverse halftone using WInHD \cite{ref38} is applied on $Halftone_{digest}$, according \label{sec:winhd} to reconstructed image digest, which is denote as $Cover_{digest}$.

\textbf{\textit{Step} 2:} The geometry of $Cover_{watermarked}$ (suspicious image) is reconstructed accordingly in section \ref{sec:surf}. To do so, transform can be estimated by using $Cover_{watermarked}$ (maybe attacked) and $Cover_{digest}$ based on SURF algorithm. Next, the transformed matrix is applied on $Cover_{watermarked}$.

\textbf{\textit{Step} 3:} Similarly, as in step 2 of the last phase, the YUV color space is employed in $Cover_{watermarked}$, to obtained luminance channel ($Lum$). If the received image is the gray-scale mode, just to be defined it as $Lum$.

\textbf{\textit{Step} 4:} The $CD$ sub-band coefficients of the watermarked image are obtained in the same way as in the previous phase. Then, the $CD$ sub-band coefficients are divided into non-overlapping $2\times2$ blocks. 

\textbf{\textit{Step} 5:} DCT is applied in each block to generate feature set $F_{(i, j)}(n)$ with pixel dimensions of $128\times128$, and $n$ is a total number of coefficient in each block, which is equal to $4$.

\textbf{\textit{Step} 6:} The total number of blocks are $128\times128$, in which, 8192 blocks are utilized for training phase, and the remained blocks are used to generate the test data.
The network can be trained by utilizing the set of training data with structure 4-10-10-2 (4 input data, two hidden layers with 10 neurons, and 2 output). 
For each $2\times2$ block (4 coefficients) are introduced to the network, and then feed-froward update the connection weights using the desired output. This process is repeated until the feed-froward algorithm converges. Afterward, the trained model is employed to classify the test set. This process is done four times for top, bottom, right, and left half of $CD$ sub-band, according to Fig. \ref{fig:nn}.
The desired output is a binary value of corresponding block.
\noindent For example for top half:
\begin{align}
f &=  \{Train_{(i, j)}, Label_{(i, j)}\}\nonumber \\
     &= \{([F_{(i, j)}(1), F_{(i, j)}(2), F_{(i, j)}(3), F_{(i, j)}(4)], W_{(i, j)}) | \nonumber \\
     &\quad\quad\quad\quad i \in (1, 2, ..., 64), j \in (1, 2, ..., 128) \}
\end{align}

\noindent where the trained model has been constructed by FNN as $f$. $F_{(i, j)}(1), ..., F_{(i, j)}(4)$ are DCT coefficients of the corresponding $(i, j)$th block, and $W_{(i, j)}$ is its label. By using trained model $f$, corresponding output for the front half is predicted. Thus, the bottom half  of watermark can be obtained by
\begin{align}
W'_{(i, j)} &= f(Test_{(i, j)}) \nonumber \\
            &= \{(f[F_{(i, j)}(1), F_{(i, j)}(2), F_{(i, j)}(3), F_{(i, j)}(4)]) | \nonumber \\
              &\quad\quad\quad\quad i \in (65, 66, ..., 128), j \in (1, 2, ..., 128) 
\end{align}

\textbf{\textit{Step} 7:} The exclusive-or is applied between $W$ and $W'$ to detect tampered regions, and the result has been defined as $Tampered_{region}$. If $Tampered_{region}(i, j) = 1$, it means that the pixel at $(i, j)$ location is tampered. On contrary, if $Tampered_{region}(i, j) = 0$, it represents an accurate pixel. 
\begin{equation}
Tampered_{region}  = W \oplus  W'
\end{equation}

Next, the isolated pixels which their length of connections is less than three and do not have 8 neighbors are eliminated and updates $Tampered_{region}$. Having updated $Tampered_{region}$, the closing operation (morphological operations) by using a disk structuring element whose width is three pixels, is employed to fill the gaps. At the end, $Tampered_{region}$ is extend to the watermarked image.

\textbf{\textit{Step} 8:} After tampered detection stage, all image blocks are marked as valid or invalid. The recovery step triggered for reconstructing all the invalid blocks using the $Cover_{digest}$. To do so, each pixel in the invalid block is replaced by the corresponding pixel in $Cover_{digest}$. Experimental results demonstrate that the recovered image has better visual quality, and also removes blocky artifact rather than other previous methods in the literature.
\section{Experimental result}
\noindent In this section, 1) data set and its implementation, 2) evaluation metric, and 3) some experiment to the evaluated performance of TRLF in the context of tamper detection and recovery are presented.
\subsection{Data set and implementation}
\begin{figure}[H]
\center
\includegraphics[width=0.4\columnwidth]{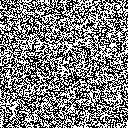}
\caption{The original binary watermark of size 128$\times$128.}
\label{fig:watermark}
\end{figure}

\noindent TRLF is tested on eight standard images (selected from USC-SIPI image database) such as Baboon, Barbara, Lena, Pepper, Santiago, F16, Cameraman, and Elaine as shown in Fig. \ref{fig:db_w_zoom}. The size of the images are 512$\times$512 pixels and all of them are color image except Cameraman and Elaine and are stored in bitmap format. A large number of objects and high variety of categories which contain various texture, edge, smooth and flat regions make these images very suitable to demonstrate the efficacy of TRLF. In our experiment, a random binary sequence of size $128\times128$ shown in Fig.\ref{fig:watermark}. is used as a watermark (with a payload of 0:0625 bpp).
\noindent TRLF is implemented on a computer with a 3.3 GHz Intel Core i5 processor, 4.00 GB memory.
\subsection{Evaluation metrics}
\noindent Before discussing the experimental results, we describe the performance evaluation metrics in advance. In order to evaluate the visual quality of the watermarked image, image digest, and recovered image, an objective metrics such as peak signal to noise ratio (PSNR) \cite{ref44}, mean squared error (MSE), and structure similarity index (SSIM) \cite{ref45} are used as standard metrics. Higher values of PSNR and SSIM and lower value of MSE metrics indicate better perceptual quality and higher transparency (imperceptibility) of images. Transparency means the embedded watermark in the cover image should not reduce the quality. This feature is evaluated by PSNR that shows the degree of invisibility of watermark. This metric compares pixel by pixel similarity between the cover image and watermarked image and is computed by equation (\ref{eq:psnr}) and (\ref{eq:mse}).

\begin{equation}
PSNR = 10.log_{10}[\frac{Max(x(i,j)^2)}{MSE}]
\label{eq:psnr}
\end{equation}
\begin{equation}
MSE = \frac{1}{H \times W} \displaystyle\sum_{i=1}^{H} \displaystyle\sum_{j=1}^{W} [x(i,j)-y(i,j)] ^ 2
\label{eq:mse}
\end{equation}

\noindent where $x$ and $y$ are the cover and watermarked image, respectively. Subsequently, $(i,j)$, $H$ and $W$ denote the coordinate of the pixel, the height, and the width of images, respectively. Obviously, if the cover and watermarked image are completely different, then MSE increases while PSNR decreases. Therefore, by increasing PSNR, the degradation is decreased, and a higher quality value of the watermarked image is achieved. 

PSNR metric is not enough consistent with human visual system, and do not consider the local content or structure of the image. Therefore, in this paper SSIM is been used as the evaluation metric for image quality assessment. The human visual characteristics is sensitive to measures such as brightness, contrast and structure.  It could be more suitable for defining the distortion limit. Assuming $i$ and $j$ are local units corresponding block, brightness comparison function as equation (\ref{eq:brightness}), contrast comparison function as equation (\ref{eq:contrast})  and structure comparison function equation (\ref{eq:structure}) are defined: 
\begin{equation}
l(x,y) = \frac{2\mu_x\mu_y+C_1}{\mu_x^2+\mu_y^2+C_1}
\label{eq:brightness}
\end{equation}
\begin{equation}
c(x,y) = \frac{2\sigma_x\sigma_y+C_2}{\sigma_x^2+\sigma_y^2+C_2}
\label{eq:contrast}
\end{equation}
\begin{equation}
s(x,y) = \frac{\sigma_{xy}+C_3}{\sigma_x\sigma_y+C_3}
\label{eq:structure}
\end{equation}

\noindent  $\mu_x$ and $\mu_y$ are the mean values of blocks, $\sigma_x^2$, $\sigma_y^2$ and $\sigma_{xy}$ are variance and covariance between two images. By assuming the combination of three functions and $c_3=\frac{c_2}{2}$, SSIM image is evaluated by equation (\ref{eq:ssim}). $C_1$, $C_2$, and $C_3$ are constants.
\begin{equation}
SSIM(x,y) = [l(x,y).c(x,y).s(x,y)]
\label{eq:ssim}
\end{equation}
\noindent The quality performance of TRLF is evaluated by PSNR, SSIM, and MSE measures which are used to assess imperceptibility of the watermarked image. 

In addition, the watermarking methods are evaluated based on resistance characteristics against attacks. The robustness of TRLF is examined using Normalized Correlation (NC), and Bit Error Rate (BER) between the embedded and extracted watermark, which are defined as equation (\ref{eq:NC}) and (\ref{eq:BER}).
\begin{equation}
NC = \frac{\displaystyle\sum_{i=1}^{h}\displaystyle\sum_{j=1}^{w}W(i,j) . W'(i, j)}
{h\times w}
\label{eq:NC}
\end{equation}
\begin{equation}
BER = \frac{\displaystyle\sum_{i=1}^{h}\displaystyle\sum_{j=1}^{w}W(i,j) \oplus W'(i, j)}
{h\times w}
\label{eq:BER}
\end{equation}
where $W(i,  j)$ and $W'(i, j)$ denote the $(i, j)_{th}$ pixel value of the embedded and extracted watermark, respectively. $h$ and $w$ denote height and width of the watermark, respectively. These metrics are computed to find the similarity between the embedded and extracted watermark. NC and BER values are close to one and zero, respectively, to demonstrate the accurate extraction of the watermark. 

In order to measure the tamper detection efficiency and accuracy of TRLF, two distinct measures are adopted as the true positive rate (TPR) and false positive rate (FPR) which are defined as equation (\ref{eq:TPR}) and (\ref{eq:FPR}).
\begin{equation}
TPR = \frac{TP}{TP+FN}\times 100\%
\label{eq:TPR}
\end{equation}
\begin{equation}
FPR = \frac{FP}{FP+TN}\times 100\%
\label{eq:FPR}
\end{equation}
where TP (number of True Positive blocks) denotes the number of tampered blocks that truly detected as tampered blocks, FN (number of False Negative blocks) represent the number of tampered blocks incorrectly judged as unmodified, FP (number of False Positive blocks), denotes the number of unmodified blocks incorrectly detected as tampered, and TN (number of True Negative blocks) represent the number of unmodified blocks truly identified as unmodified. Evidently, TP+FN and FP+TN are the number of valid and invalid blocks, respectively. High TPR and low FPR illustrates better robustness against various attacks and higher tamper localization. Thus, in this paper, the reliability is evaluated with NC, BER, TPR, and FPR measures.
\subsection{Performance evaluation of TRLF}
\noindent In this subsection, numerous experiments are presented to illustrate the performance of TRLF in terms of imperceptibility and robustness by comparison of previous works. Imperceptibility and robustness are two main properties that a watermarking method should have. TRLF achieved high robustness with good imperceptibility. This is proven by computing PSNR, SSIM, and MSE between the cover and watermarked images, and NC and BER values between the embedded and extracted watermark. In addition, the efficiency and accuracy of tamper detection phase are evaluated by calculating TPR and FPR metrics.

Recently, watermarking methods often suffer from geometric attacks including rotation, translation, cropping, and scaling operations. The methods of performing geometry attacks make changes to coordinates in which the watermark is embedded. Many watermarking methods ignore the status of the locations at which the watermark is embedded. Hence, the watermark is destroyed and incorrectly extracted from the cover image. In TRLF as mentioned in the previous sections, SURF algorithm is used to reconstruct the geometry of the watermarked image. Furthermore, in order to improve tamper detection rate under various attacks, a novel watermark extraction method is performed by using FNN with good generalization ability. We extract the watermark from an attacked watermarked image and comparing the extracted watermark with the embedded one, to localize the modification area. 

The cover, watermarked image, and also zoom of the watermarked image are given in Fig. \ref{fig:db_w_zoom}. From these figures, it can be seen that there is no perceptual difference between the cover and the watermarked images. The obtained values of PSNR, SSIM, and MSE demonstrate that the good imperceptibility is obtained by TRLF.

The zoom, PSNR and SSIM of the watermarked image Lena with various thresholds $(T_1, T_2)$ are given in Fig. \ref{fig:lena_zoom}. As can be seen, no visual degradation can be detected by the naked eye even when selected higher values for thresholds $(T_1, T_2)$. PSNR and SSIM metrics is used to confirm this subjective statement. In the meantime, as described previously, a trade-off between transparency and robustness is made as the watermark is embedded. The relation between PSNR, SSIM, and MSE values and various thresholds $(T_1, T_2)$ are demonstrated in Fig. \ref{fig:relationship_between_PSNR,_SSIM_MSE_T1_T2}. The smaller thresholds decrease the robustness of the watermark and lead to obtaining higher visual quality for the watermarked image. On the contrary, the image distortion is caused by larger thresholds, which will affect the visual quality of the cover image by appearing a distortion in the watermarked image, but it leads to a stronger robustness. In TRLF, a threshold matrix (mean absolute difference as $Dif$) as described in the previous section is used to avoid the low imperceptibility and to improve robustness. Based on various tests, $T_1$ and $T_2$ are assigned to 5 and 3, respectively. At this values, PSNR, SSIM, and MSE values of several images are illustrated in Table \ref{TABLE:compare_psnr_ssim_mse_watermark}.

PSNR, SSIM, and MSE values between the watermarked images and their corresponding cover images for color and gray-scale are listed in Table \ref{TABLE:compare_psnr_ssim_mse_watermark}. In addition, the quality of watermarked image which obtained by the similar semi-fragile method is investigated in this table. Unfortunately, in literature no evaluation were reported for these methods, to show the quality and resistant of the method against various attacks in color mode. So, we illustrate the performance of TRLF based on grayscale images. It is clear that TRLF retain better visual quality for watermarked image. As can be seen, TRLF gives higher considerable PSNR and SSIM (close to 1 in most cases) values compared to others methods. In addition, as shown in Table \ref{TABLE:compare_psnr_ssim_mse_watermark}. TRLF obtains better results for flat images compared to texture image.

Tables. \ref{TABLE:time_complexity} represent computed times for various step on several images. These relatively short times make TRLF become suitable for real-time image processing applications such as data hiding, image and video tamper detection and recovery in particular. This low embedding and authentication time and high quality of watermarked image indicate that TRLF is fast and efficient.

\begin{figure}[H]
\center
\begin{tabular}{cc}
\includegraphics[width=0.42\columnwidth]{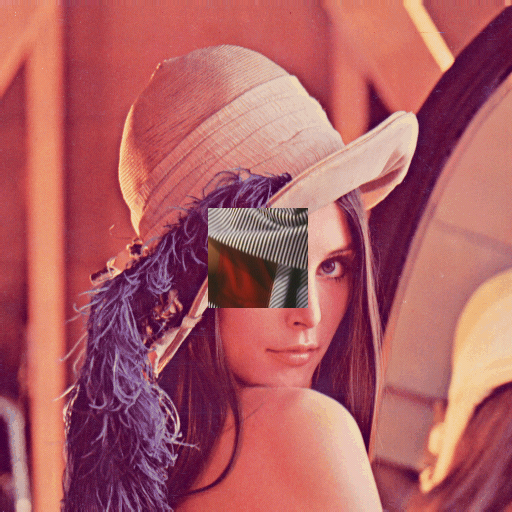} & \includegraphics[width=0.42\columnwidth]{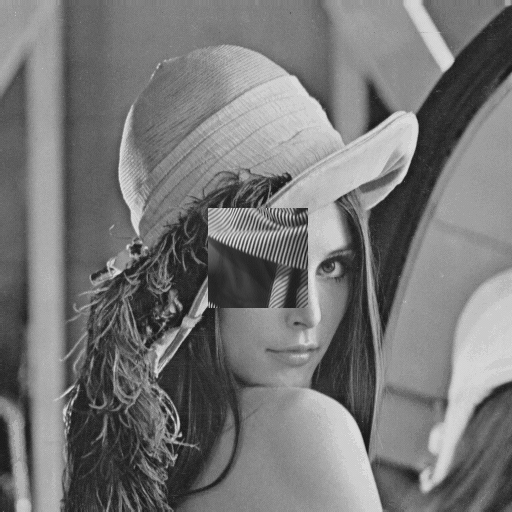} \\
(a) & (b)
\end{tabular}
\caption{Tampered image (100$\times$100 with Barbara in center) for testing performance against various attacks. a) Lena Color, b) Lena Gray.}
\label{fig:testimage}
\end{figure}

The authentication performance of TRLF is investigated by applying several attacks. It included common image processing operations such as additive noise, filtering, compression, luminance, and geometric operations. For this purpose, the watermarked Lena image is tampered with covering 100$\times$100 by Barbara in center, according to Fig. \ref{fig:testimage}.

The summarization of the results is shown in Fig. \ref{fig:diagramrobustness}. It illustrates the performance evaluation of TRLF in terms of TPR and FPR. 
Furthermore, the degraded quality of the attacked image in terms of PSNR and SSIM are given in Fig. \ref{fig:diagramrobustness}. As can be seen, TPR of the proposed tamper detection is actually high for various attacks. Also, FPR is fairly low for the most attacks. Thus, TRLF can handle the problem of the false positive rate in most cases. The analysis of TRLF in terms of TPR and FPR clearly verified the superiority and efficiency of the tamper detection and localization method compared to other the-state-of-the-art methods. 

Not only TRLF can detect the tampered regions, but also performs exact recovery for the destroyed part of the watermarked image. As mentioned in the previous section, in TRLF, the invalid regions reconstructed by image digest, which were generated based on inverse halftoning technique. The use of halftoning technique for generating image digest leads to effective tamper correction. The result of Jarvis and WInHD algorithm to generate halftone and inverse halftone for Lena image is illustrated in Fig. \ref{fig:jarvis}. 
\begin{figure}[h!]
\center
\begin{tabular}{cc}
\includegraphics[width=0.42\columnwidth]{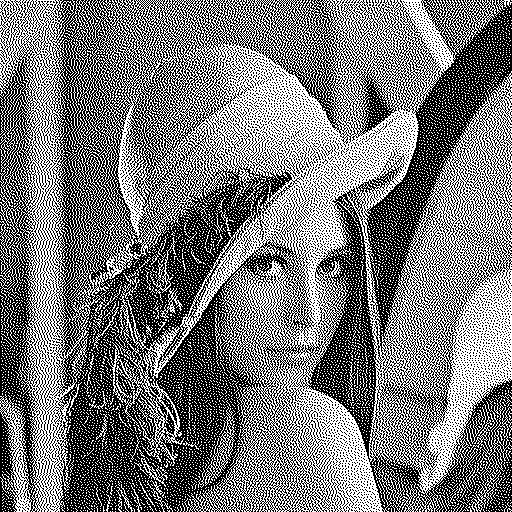} & \includegraphics[width=0.42\columnwidth]{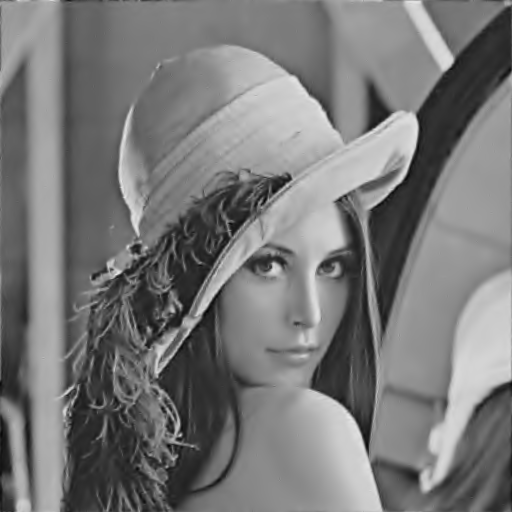} \\
(a) & (b)\\
\includegraphics[width=0.42\columnwidth]{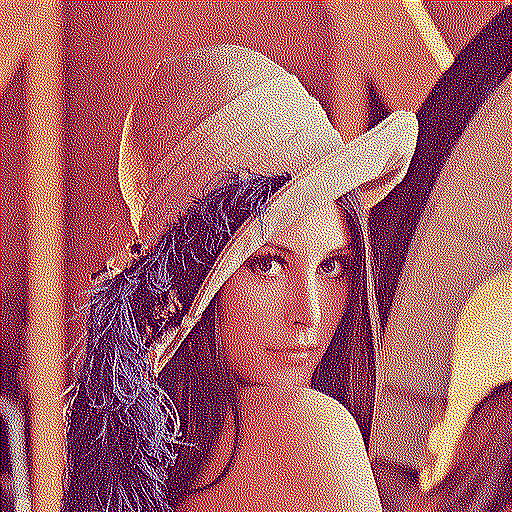} & \includegraphics[width=0.42\columnwidth]{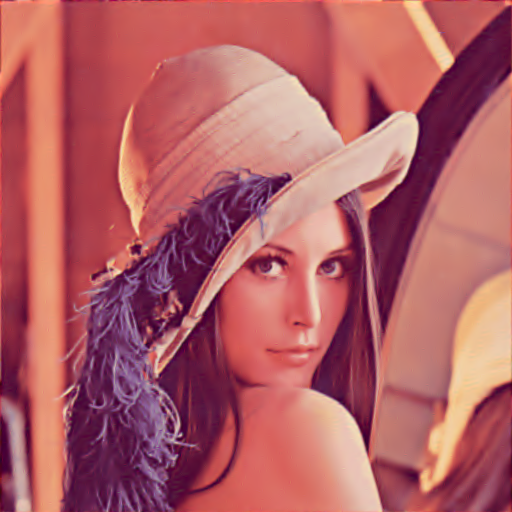}\\
(c) & (d)
\end{tabular}
\caption{Result of Jarivs and WInHD algorithm for Gray and Color Lena Image. a) and c) halftone version, b) and d) inverse halftone.}
\label{fig:jarvis}
\end{figure}
PSNR, SSIM and MSE values of image digests (inverse halftone) with respect to the cover images are listed in Table \ref{TABLE:compare_psnr_ssim_mse_halftone}.
\begin{table*}[t!]
\center
\footnotesize
\caption{PSNR, SSIM, and MSE of Inverse Halftone (WInHD technique) compared to fragile methods for gray and color images.\\
*: Calculate the average intensity value of each $2\times2$ block, and obtain 5 MSBs. Then, assign them to all pixel's 5 MSBs of current block.\\
**: Calculate the average intensity value of each $4\times4$ block. Then, assign them to all pixels of current block.}
\label{TABLE:compare_psnr_ssim_mse_halftone}
\renewcommand{\arraystretch}{1.5}
\begin{tabular*}{\textwidth}{@{\extracolsep{\fill} }@{}l@{}c@{}c@{}c@{}c@{}c@{}c@{}c@{}c@{}c@{}c@{}c@{}c@{}c@{}c@{}c@{}c@{}c@{}c@{} }
\cline{1-19}
\multicolumn{1}{c}{\multirow{3}{*}{Image}} &\multicolumn{6}{c}{Halftoning technique}&\multicolumn{6}{c}{Block based $2\times2$ and 5MSBs *}&\multicolumn{6}{c}{Block based $4\times4$ and 8Bits **}\\
\cline{2-7}\cline{8-13}\cline{14-19}
&\multicolumn{3}{c}{Gray} &  \multicolumn{3}{c}{Color}&\multicolumn{3}{c}{Gray} &  \multicolumn{3}{c}{Color}&\multicolumn{3}{c}{Gray} &  \multicolumn{3}{c}{Color}\\
\cline{2-4}\cline{5-7}\cline{8-10}\cline{11-13}\cline{14-16}\cline{17-19}
&PNSR&SSIM&MSE&PNSR&SSIM&MSE&PNSR&SSIM&MSE&PNSR&SSIM&MSE&PNSR&SSIM&MSE&PNSR&SSIM&MSE\\
\cline{1-19}
Baboon&24.22&0.73&246.41&23.66&0.85&279.79&23.15&0.72&314.68&22.65&0.82&353.35&20.85&0.45&534.96&20.33&0.66&603.23\\
Barbara&25.37&0.76&188.84&25.26&0.83&193.48&25.36&0.78&189.47&25.26&0.86&193.71&23.34&0.62&301.22&23.14&0.74&315.92\\
Lena&32.17&0.85&39.49&31.18&0.98&49.54&30.02&0.85&64.78&29.59&0.98&71.49&26.75&0.75&137.56&26.75&0.95&137.35\\
Pepper&31.18&0.84&49.61&29.71&0.98&69.52&29.03&0.84&81.23&28.32&0.97&95.82&26.01&0.74&163.08&25.51&0.95&183.04\\
Santiago&24.16&0.76&249.56&24.48&0.86&231.73&22.95&0.73&329.71&23.43&0.84&294.93&20.10&0.42&635.12&20.61&0.67&565.22\\
F16&31.55&0.91&45.50&30.37&0.93&59.74&28.20&0.89&98.45&28.10&0.90&100.80&24.70&0.78&220.36&24.89&0.84&210.90\\
Cameraman&33.95&0.91&26.16&-&-&-&30.03&0.91&64.54&-&-&-&25.67&0.82&176.18&-&-&-\\
Elaine&31.41&0.75&47.04&-&-&-&30.53&0.78&57.61&-&-&-&28.21&0.69&98.11&-&-&-\\
\cline{1-19}
\end{tabular*}
\end{table*}
PSNR, SSIM and MSE of the recovered image relative to various tampered rates are shown in Fig. \ref{fig:diagramrecover}. We compare the quality of recovered image with fragile (spatial domain) methods in term of PSNR measure. It is clear that by using halftoning technique, we can obtain a better perceptual quality, compared to other methods which were used 5 or 6 Most Significant Bits to generate the image digest, and reconstructed tampered region. So, we can see clearly the real content of the tampered area.

The performance of TRLF under different types of attacks is evaluated by comparing FPR values of tamper detection phase with similar semi-fragile methods \cite{ref25, ref26} with tamper detection and recovery capability in Table. \ref{tbl:FPR}. It should be note, the block size of TRLF and \cite{ref25, ref26} are $4\times4$ and $8\times8$ pixels, respectively. The evaluation results show that TRLF can effectively thwart different attacks.  
\begin{table}[h!]
\centering
\footnotesize
\caption{False positive rate of TRLF compared with \cite{ref25, ref26} under various attacks.  A) Dynamic range change [50, 200], B) Histogram equalization, C) Median filtering (3$\times$3), D) Mean filtering (3$\times$3), E) Salt and pepper noise ($\sigma$ = 0.05), F) Rotation (3 deg), G) JPEG (QF = 70), H) Gaussian noise ($\sigma^2$ = 0.009). Fr and N/R means fragile and not reported, respectively.}
\label{TABLE:compare_attacks}
\renewcommand{\arraystretch}{1.5}
\begin{tabular*}{\columnwidth}{@{\extracolsep{\fill} }@{}l@{}c@{}c@{}c@{}c@{}c@{}c@{}c@{}c@{}}
\cline{1-9}
Method&A&B&C&D&E&F&G&H\\ 
\cline{1-9}
TRLF (Color)&0&0&Fr&Fr&0&0&0&8.12\\
TRLF (Gray)&0&0&Fr&Fr&0&0&0&3.40\\
Phadikar et al.\cite{ref25} (Gray)&0&7.72&0.70&0.84&0&0&0&N/R\\
Rosales et al.\cite{ref26} (Gray)&0.4&0.09&7.2&N/R&1.4&2.3&N/R&1.2\\
\cline{1-9}
\end{tabular*}
\label{tbl:FPR}
\end{table}

Performance comparison of TRLF with other fragile and semi-fragile methods is demonstrated in Table. \ref{TABLE:compare_performance1} and \ref{TABLE:compare_performance2}. The comparison is made with respect to main features of tamper detection and recovery methods. As can be seen, in most of the cases, TRLF is superior compared to other schemes. 
\begin{figure*}[t]
\center
\begin{tabular}{ccccc}
\includegraphics[width=0.18\textwidth]{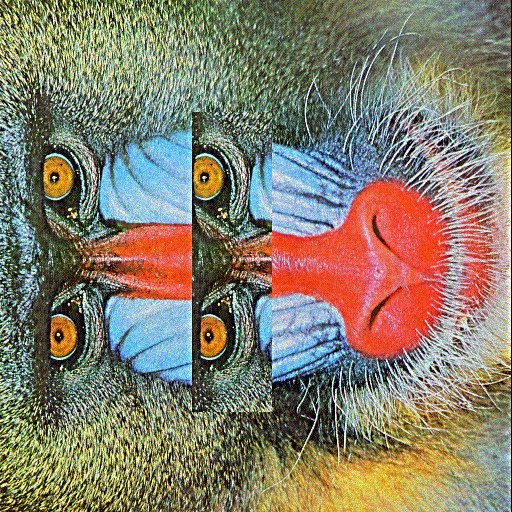}&\includegraphics[width=0.18\textwidth]{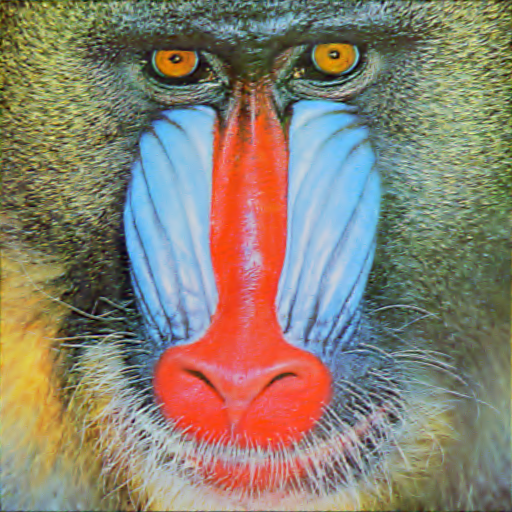}&\includegraphics[width=0.18\textwidth]{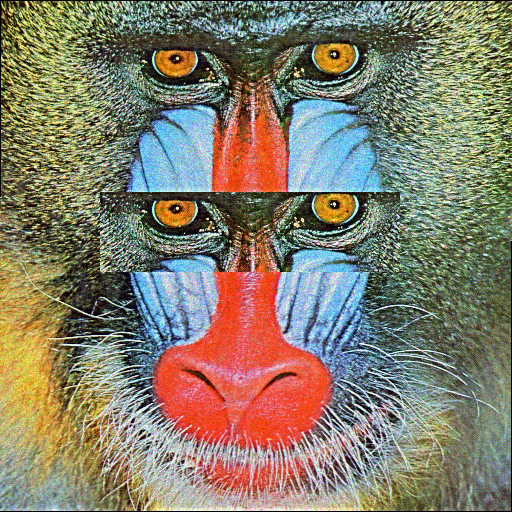}&\includegraphics[width=0.18\textwidth]{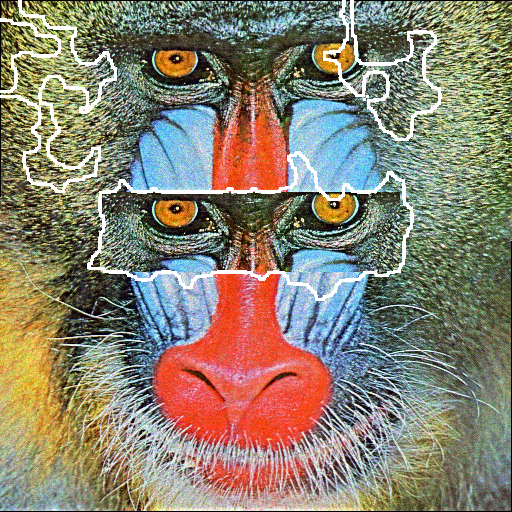}&\includegraphics[width=0.18\textwidth]{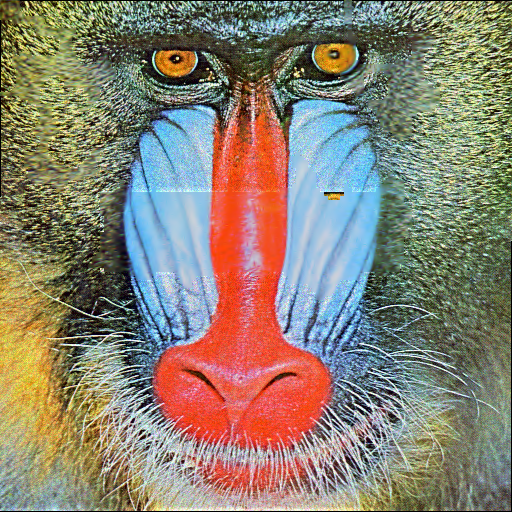}\\
(a) & (b) & (c) & (d) & (e) \\
\includegraphics[width=0.18\textwidth]{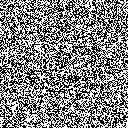}&\includegraphics[width=0.18\textwidth]{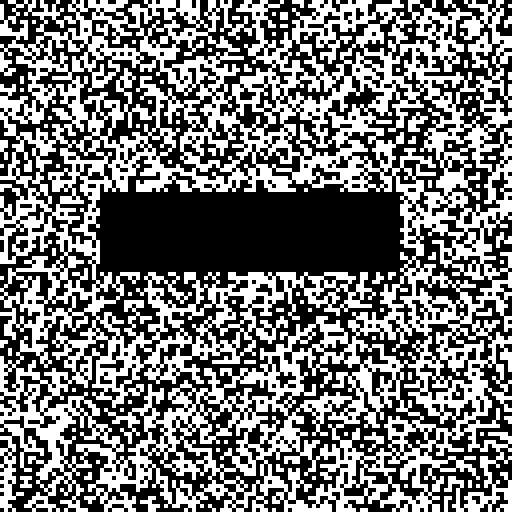}&\includegraphics[width=0.18\textwidth]{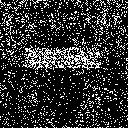}&\includegraphics[width=0.18\textwidth]{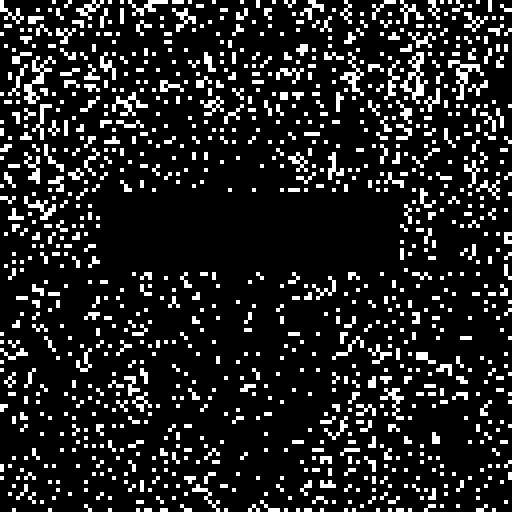}&\includegraphics[width=0.18\textwidth]{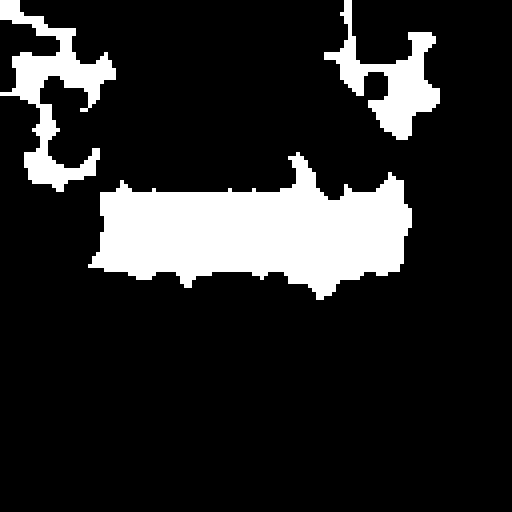}\\
(f) & (g) & (h) & (i) & (j)
\end{tabular}
\caption{Tamper detection and recovery performance under copy-move tampering (10\%), and hybrid attacks such as sharpening (size = 13, $\sigma$ = 3, strength = 0.8), speckle noise (0.01), rotation (90 deg).
(a) Watermarked image under tampering and attacking, (b) Image digest (Inverse halftone), (c) Reconstrued geometry (PSNR = 16.33, SSIM = 0.79, MSE = 1512), (d) Tamper Detection, (e) Recovered tampered regions, (f) Extracted watermark, (g) Compare valid region (NC = 0.72, BER = 0.14), (h) Authenticate by watermark (exclusive-or), (i) Error, (j) Post processing (TPR = 99.13, FPR = 6.81). }
\label{fig:resbaboon}
\end{figure*}
\begin{figure*}[t!]
\center
\begin{tabular}{ccccc}
\includegraphics[width=0.18\textwidth]{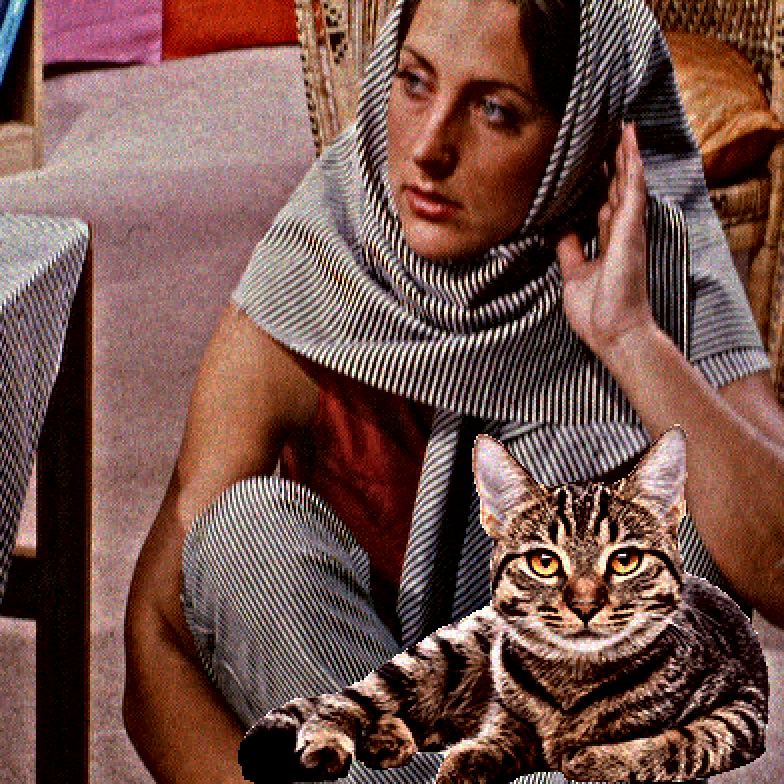}&\includegraphics[width=0.18\textwidth]{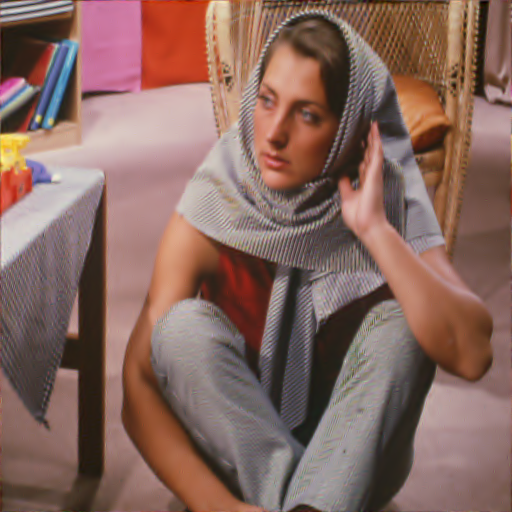}&\includegraphics[width=0.18\textwidth]{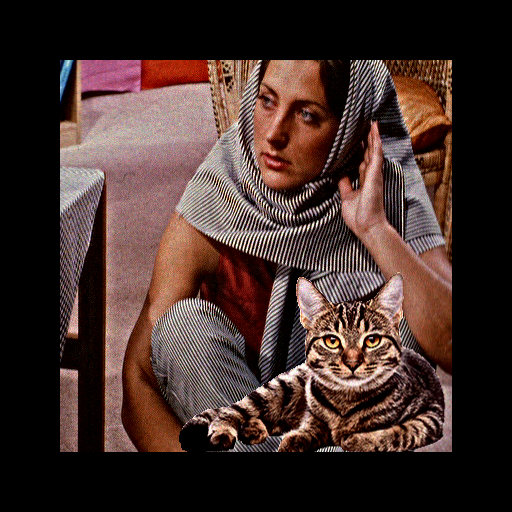}&\includegraphics[width=0.18\textwidth]{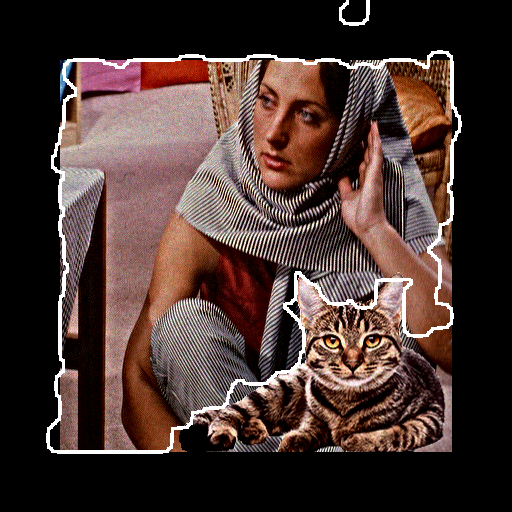}&\includegraphics[width=0.18\textwidth]{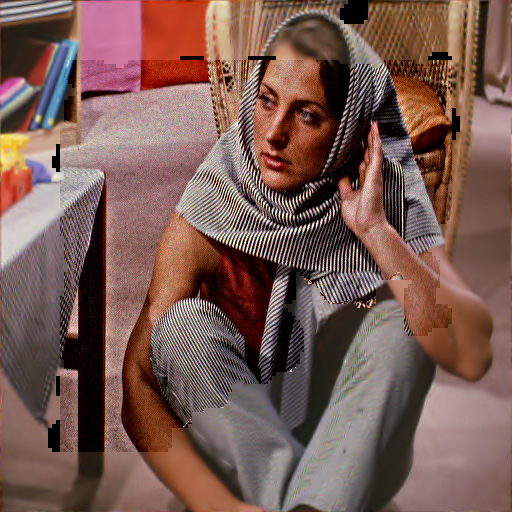}\\
(a) & (b) & (c) & (d) & (e) \\
\includegraphics[width=0.18\textwidth]{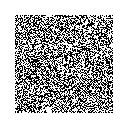}&\includegraphics[width=0.18\textwidth]{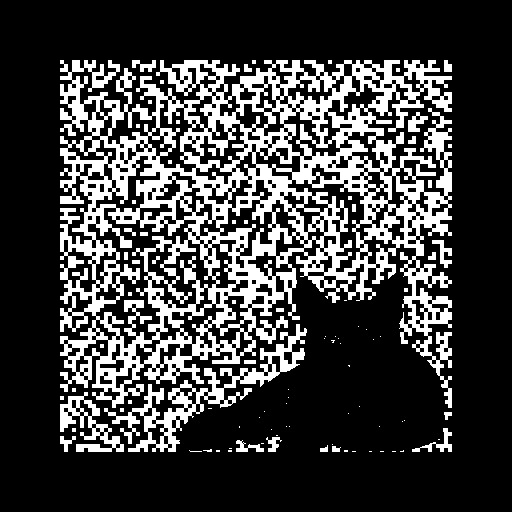}&\includegraphics[width=0.18\textwidth]{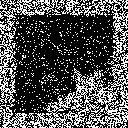}&\includegraphics[width=0.18\textwidth]{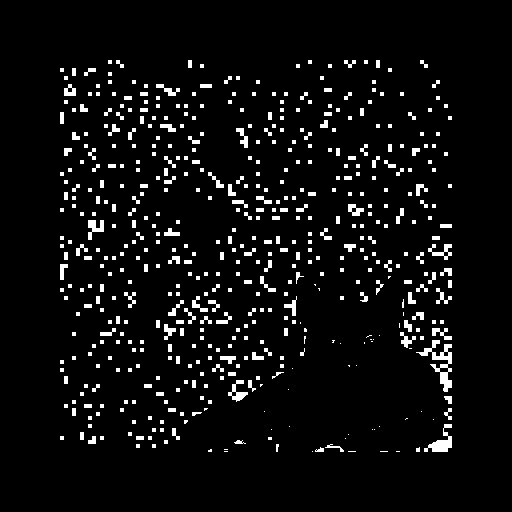}&\includegraphics[width=0.18\textwidth]{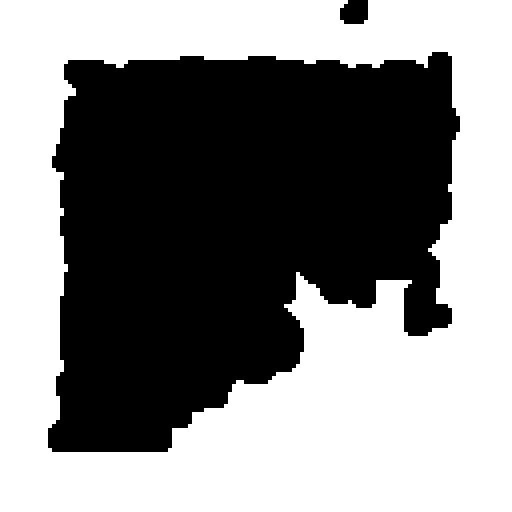}\\
(f) & (g) & (h) & (i) & (j)
\end{tabular}
\caption{Tamper detection and recovery performance under add object and removing tampering (50\%), and hybrid attacks such as darken (30), sharpening (size = 29, $\sigma$ = 7, strength = 0.8), speckle noise ($\sigma$ = 0.005), scale (2), cropping (42\% around) and JPEG2000 (QF  = 60). (a) Watermarked image under tampering and attacking, (b) Image digest (Inverse halftone), (c) Reconstrued geometry (PSNR = 9.63, SSIM = 0.36, MSE = 7080), (d) Tamper Detection, (e) Recovered tampered regions, (f) Extracted watermark, (g) Compare valid region (NC = 0.89, BER = 0.05), (h) Authenticate by watermark (exclusive-or), (i) Error, (j) Post processing (TPR = 98.44, FPR = 5.59).}
\label{fig:resbarbara}
\end{figure*}
\begin{figure*}[t]
\center
\begin{tabular}{ccccc}
\includegraphics[width=0.18\textwidth]{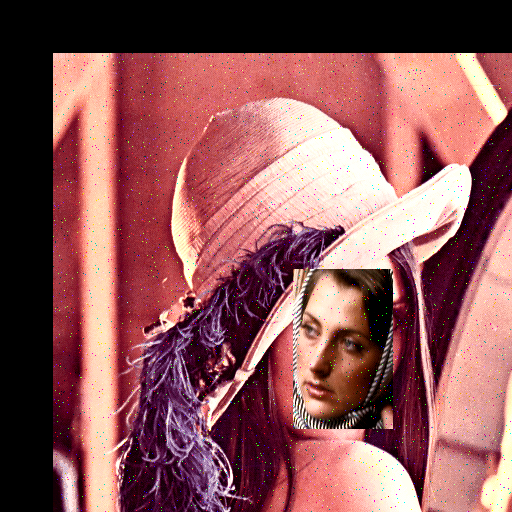}&\includegraphics[width=0.18\textwidth]{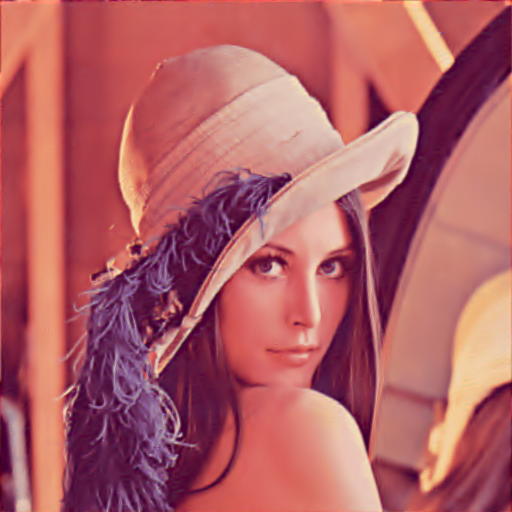}&\includegraphics[width=0.18\textwidth]{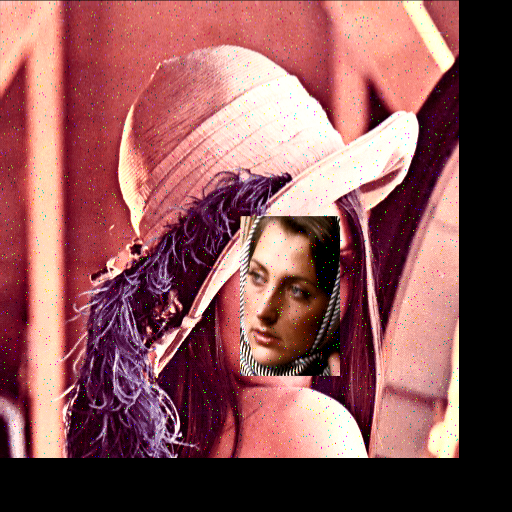}&\includegraphics[width=0.18\textwidth]{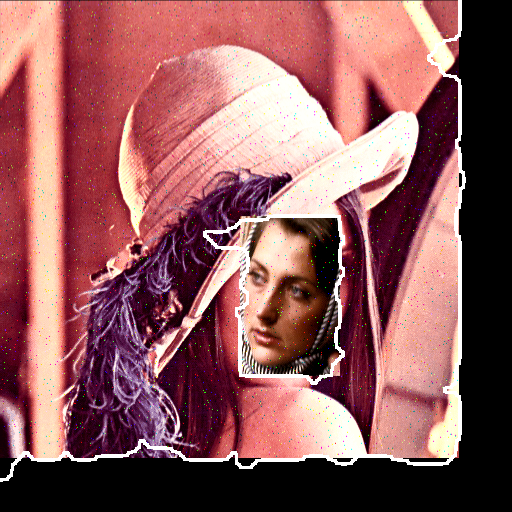}&\includegraphics[width=0.18\textwidth]{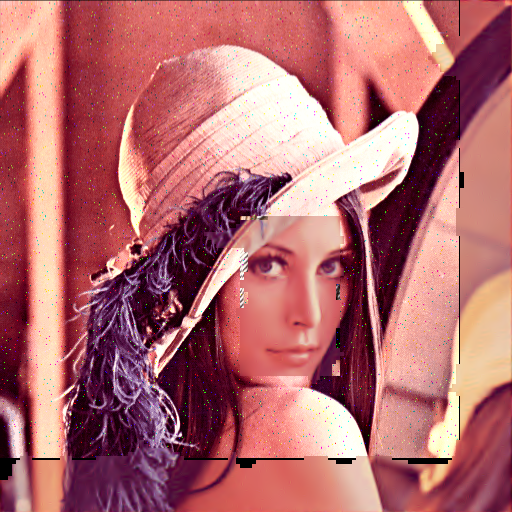}\\
(a) & (b) & (c) & (d) & (e) \\
\includegraphics[width=0.18\textwidth]{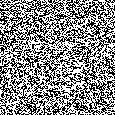}&\includegraphics[width=0.18\textwidth]{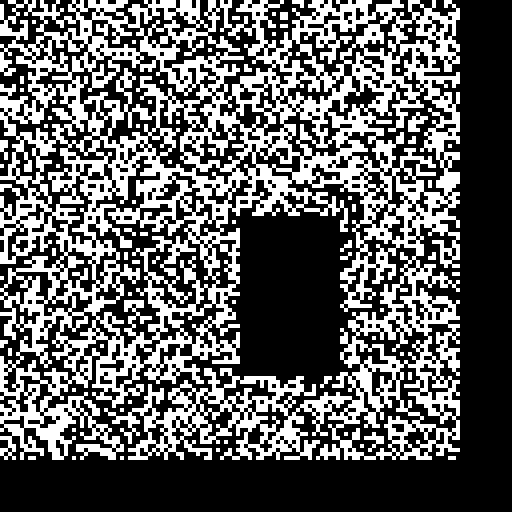}&\includegraphics[width=0.18\textwidth]{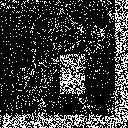}&\includegraphics[width=0.18\textwidth]{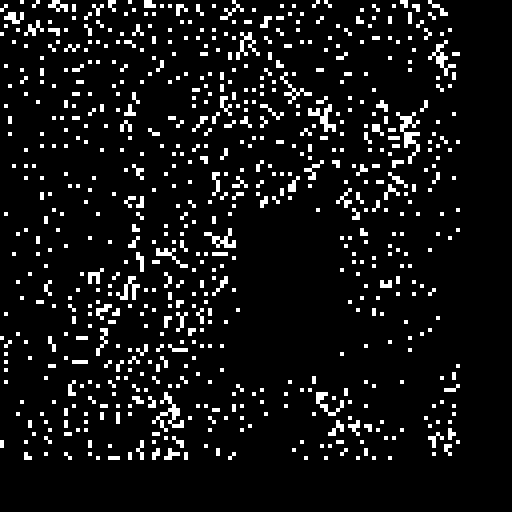}&\includegraphics[width=0.18\textwidth]{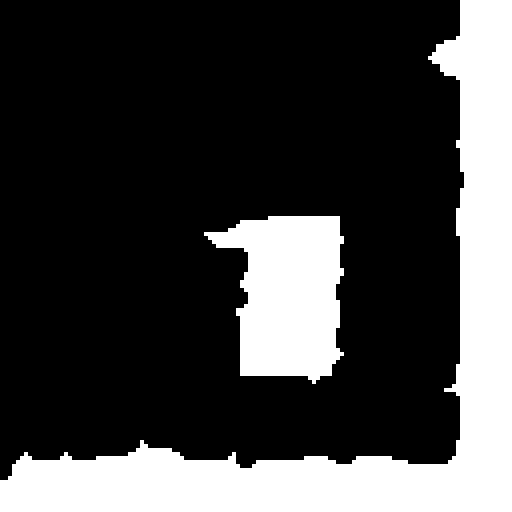}\\
(f) & (g) & (h) & (i) & (j)
\end{tabular}
\caption{Tamper detection and recovery performance under covering (vector-quantization) and removing tampering (26\%), and hybrid attacks such as histogram equalization, salt and pepper noise ($\rho$ = 0.01), translation (50, 50). (a) Watermarked image under tampering and attacking, (b) Image digest (Inverse halftone), (c) Reconstrued geometry (PSNR = 10.69, SSIM = 0.63, MSE = 5547), (d) Tamper Detection, (e) Recovered tampered regions, (f) Extracted watermark, (g) Compare valid region (NC = 0.83, BER = 0.08), (h) Authenticate by watermark (exclusive-or), (i) Error, (j) Post processing (TPR = 97.95, FPR = 1.62). }
\label{fig:reslena}
\end{figure*}
\begin{figure*}[t!]
\center
\begin{tabular}{ccccc}
\includegraphics[width=0.18\textwidth]{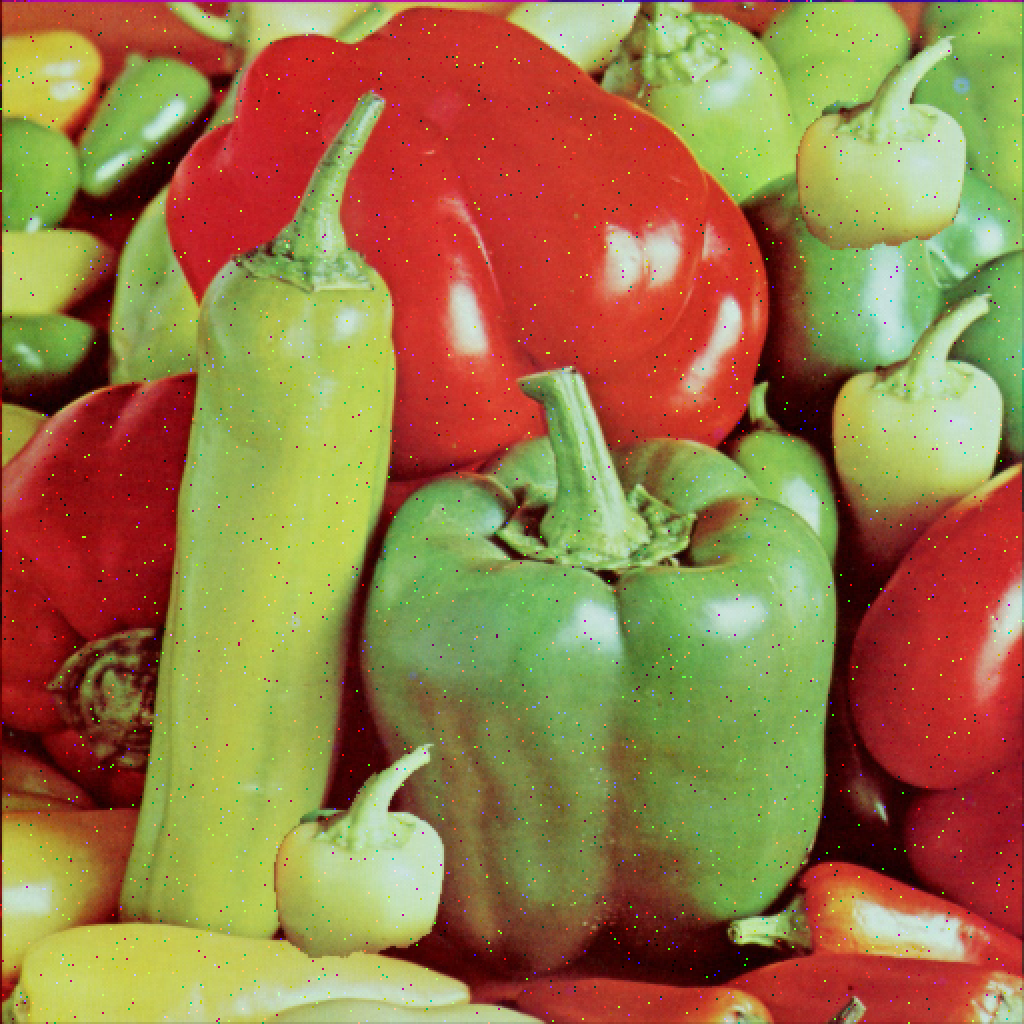}&\includegraphics[width=0.18\textwidth]{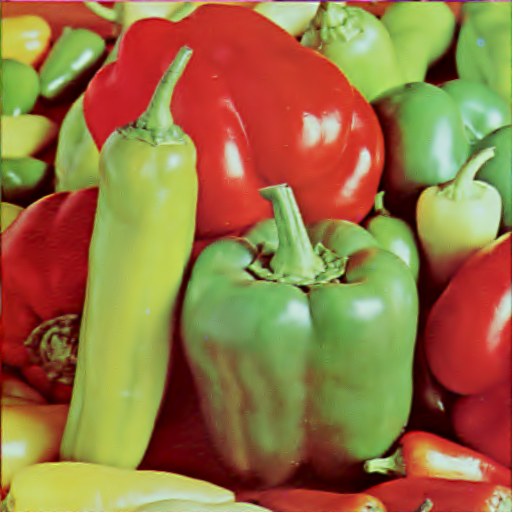}&\includegraphics[width=0.18\textwidth]{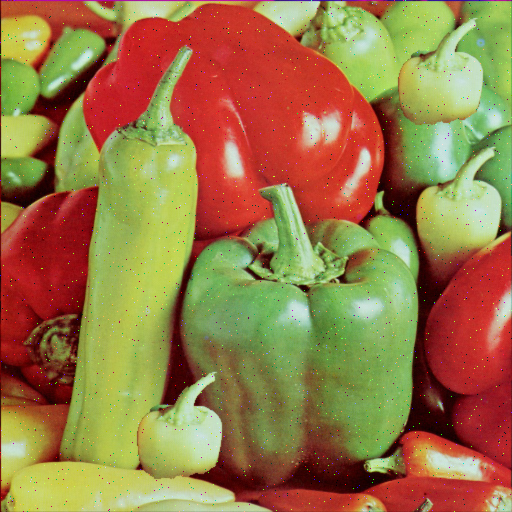}&\includegraphics[width=0.18\textwidth]{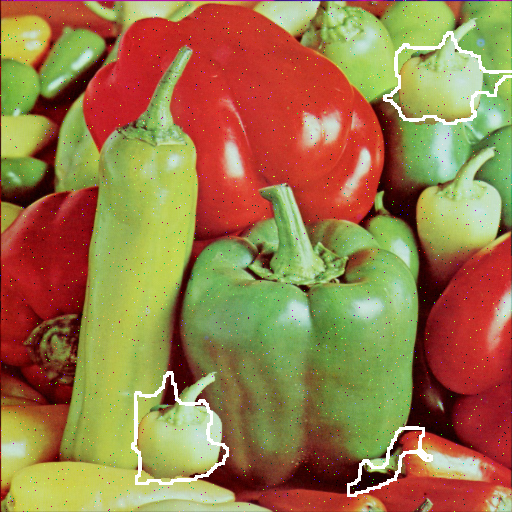}&\includegraphics[width=0.18\textwidth]{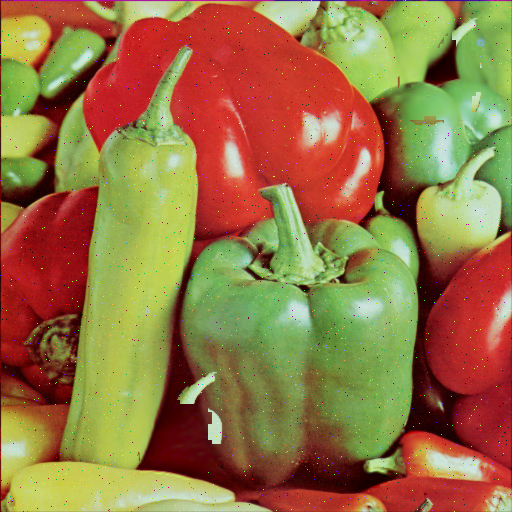}\\
(a) & (b) & (c) & (d) & (e) \\
\includegraphics[width=0.18\textwidth]{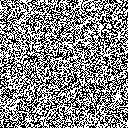}&\includegraphics[width=0.18\textwidth]{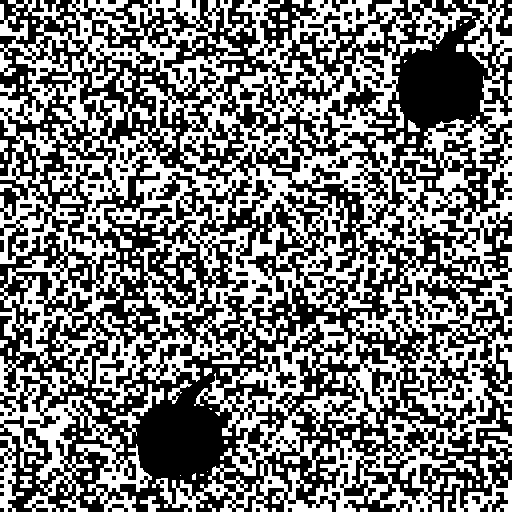}&\includegraphics[width=0.18\textwidth]{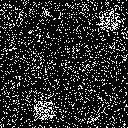}&\includegraphics[width=0.18\textwidth]{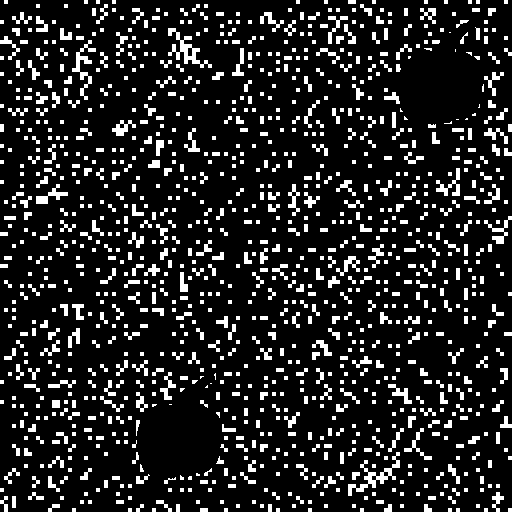}&\includegraphics[width=0.18\textwidth]{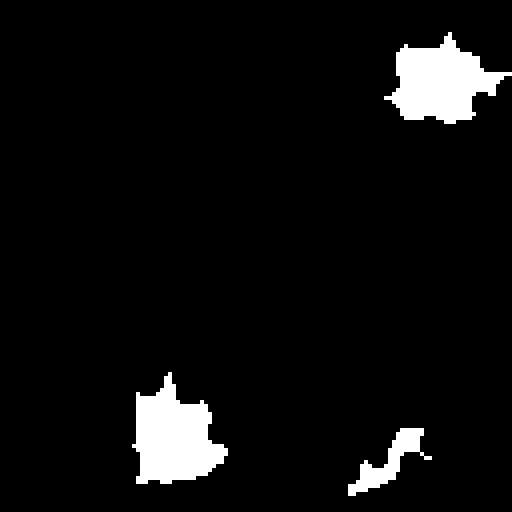}\\
(f) & (g) & (h) & (i) & (j)
\end{tabular}
\caption{Tamper detection and recovery performance under copy-move tampering (5\%), and hybrid attacks such as gaussian smoothing (size = 7, $\sigma$ = 0.5), salt and pepper noise ($\rho$ = 0.05),  JPEG (QF = 70) and scale (2). (a) Watermarked image under tampering and attacking, (b) Image digest (Inverse halftone), (c) Reconstrued geometry (PSNR = 19.56, SSIM =0.92, MSE = 720), (d) Tamper Detection, (e) Recovered tampered regions, (f) Extracted watermark, (g) Compare valid region (NC = 0.71, BER = 0.14), (h) Authenticate by watermark (exclusive-or), (i) Error, (j) Post processing (TPR = 87.28, FPR = 1.72).}
\label{fig:respepper}
\end{figure*}
\begin{figure*}[t]
\center
\begin{tabular}{ccccc}
\includegraphics[width=0.146\textwidth]{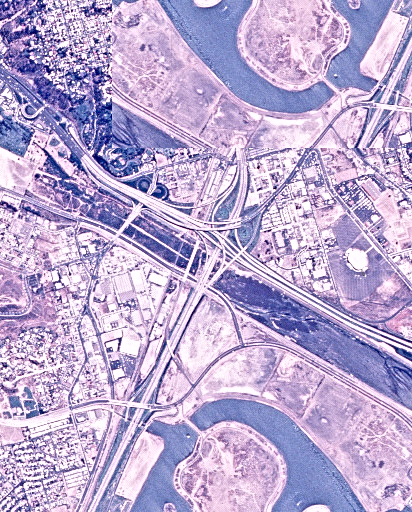}&\includegraphics[width=0.18\textwidth]{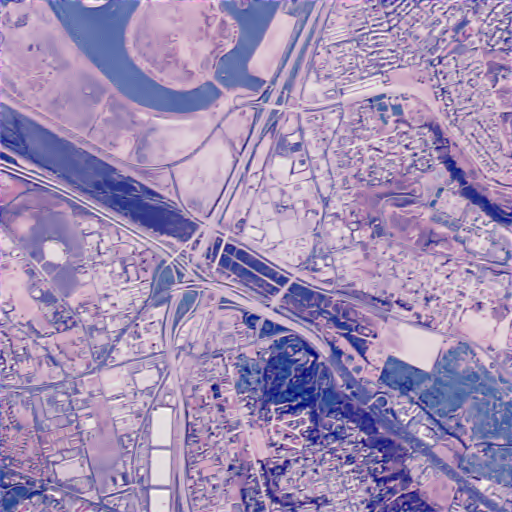}&\includegraphics[width=0.18\textwidth]{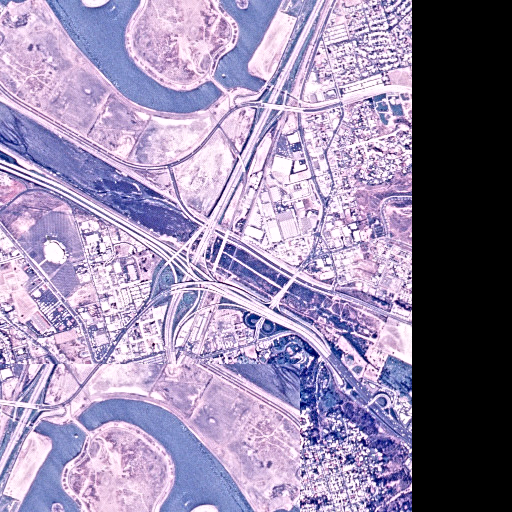}&\includegraphics[width=0.18\textwidth]{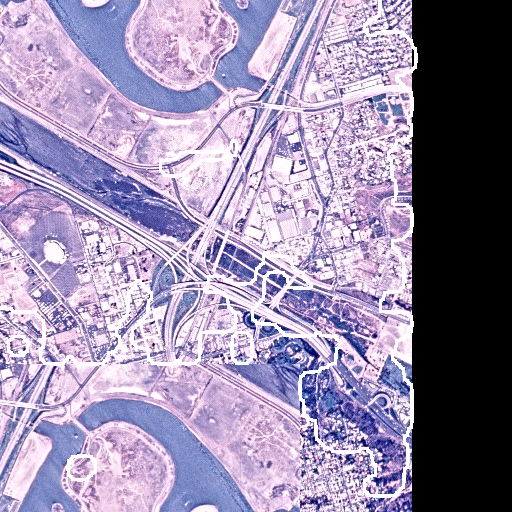}&\includegraphics[width=0.18\textwidth]{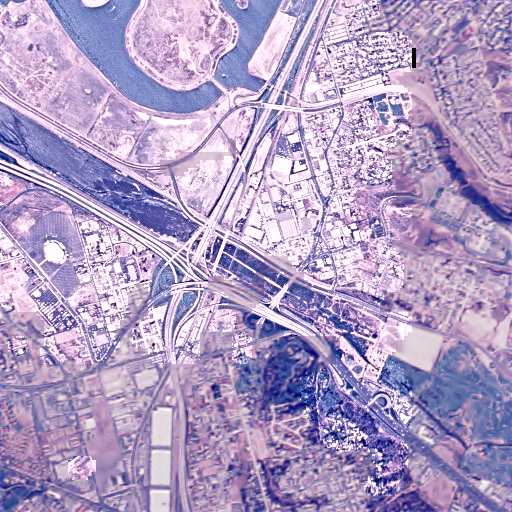}\\
(a) & (b) & (c) & (d) & (e) \\
\includegraphics[width=0.18\textwidth]{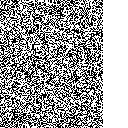}&\includegraphics[width=0.18\textwidth]{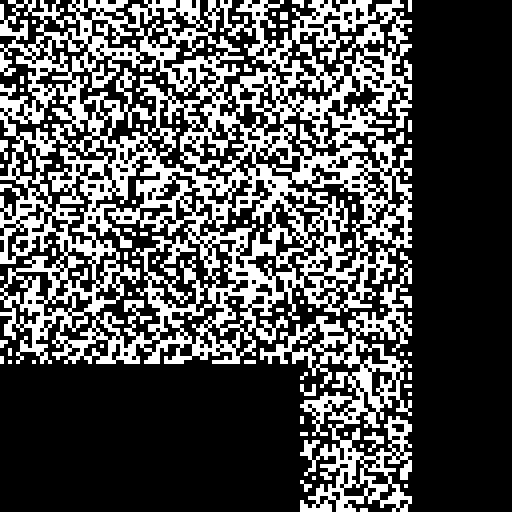}&\includegraphics[width=0.18\textwidth]{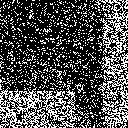}&\includegraphics[width=0.18\textwidth]{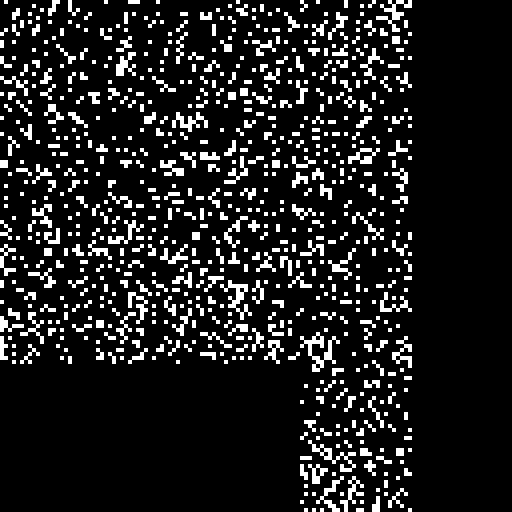}&\includegraphics[width=0.18\textwidth]{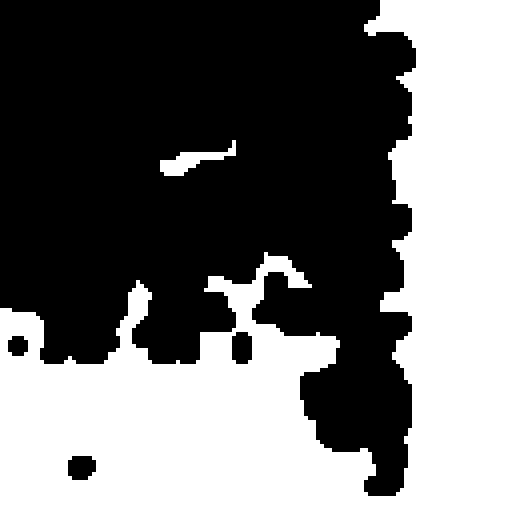}\\
(f) & (g) & (h) & (i) & (j)
\end{tabular}
\caption{Tamper detection and recovery performance under copy-move tampering (36\%), and hybrid attacks such as lighten (30), sharpening (size = 21, $\sigma$ = 5, strength = 0.8), JPEG (QF = 60), rotation (180 deg), translation (-100, 0), and cropping (20\% right). (a) Watermarked image under tampering and attacking, (b) Image digest (Inverse halftone), (c) Reconstrued geometry (PSNR = 9.74, SSIM =0.60, MSE = 6898), (d) Tamper Detection, (e) Recovered tampered regions, (f) Extracted watermark, (g) Compare valid region (NC = 0.75, BER = 0.12), (h) Authenticate by watermark (exclusive-or), (i) Error, (j) Post processing (TPR = 99.31, FPR = 13.25). }
\label{fig:ressantigo}
\end{figure*}
\begin{figure*}[t!]
\center
\begin{tabular}{ccccc}
\includegraphics[width=0.18\textwidth]{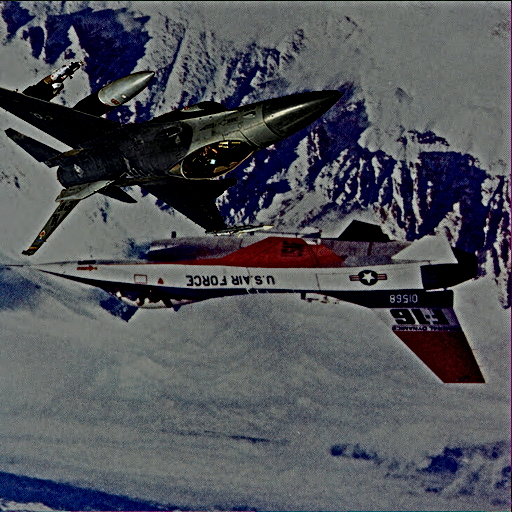}&\includegraphics[width=0.18\textwidth]{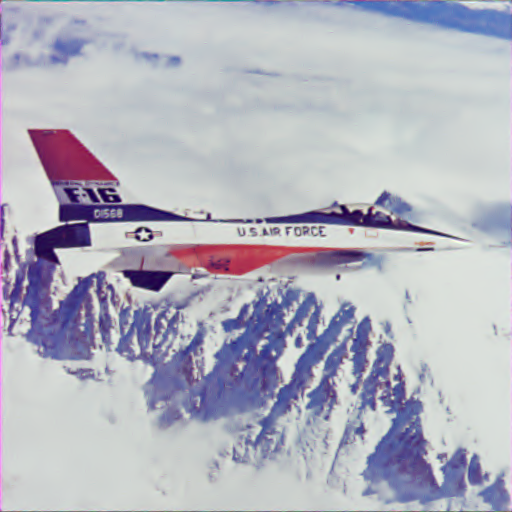}&\includegraphics[width=0.18\textwidth]{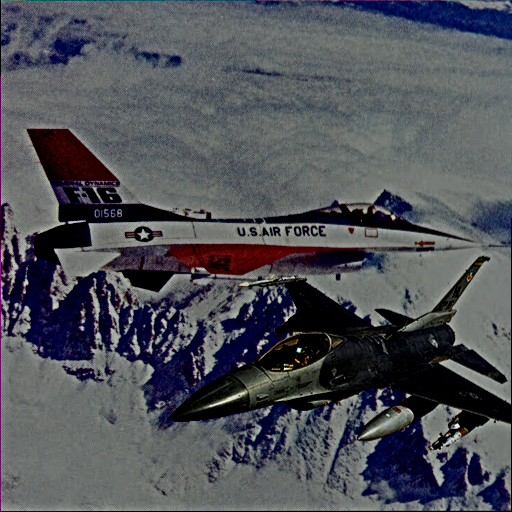}&\includegraphics[width=0.18\textwidth]{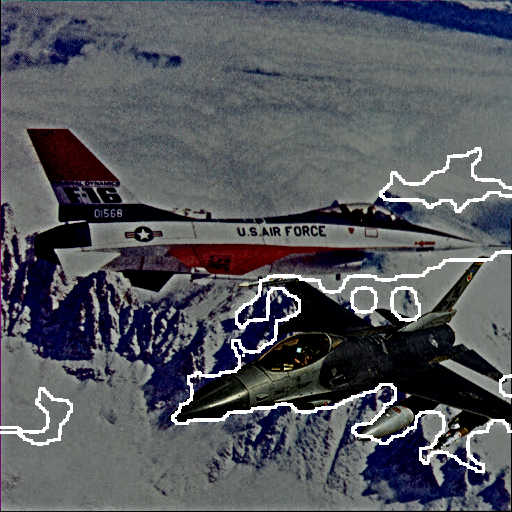}&\includegraphics[width=0.18\textwidth]{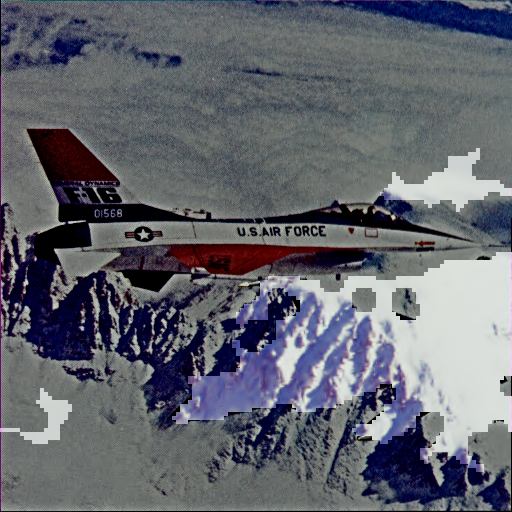}\\
(a) & (b) & (c) & (d) & (e) \\
\includegraphics[width=0.18\textwidth]{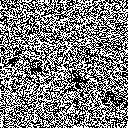}&\includegraphics[width=0.18\textwidth]{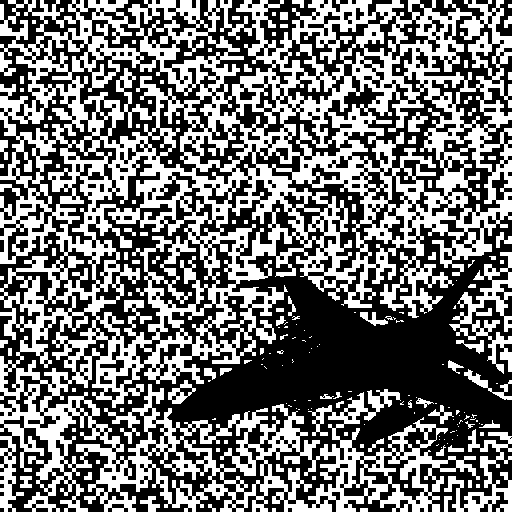}&\includegraphics[width=0.18\textwidth]{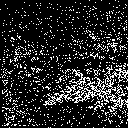}&\includegraphics[width=0.18\textwidth]{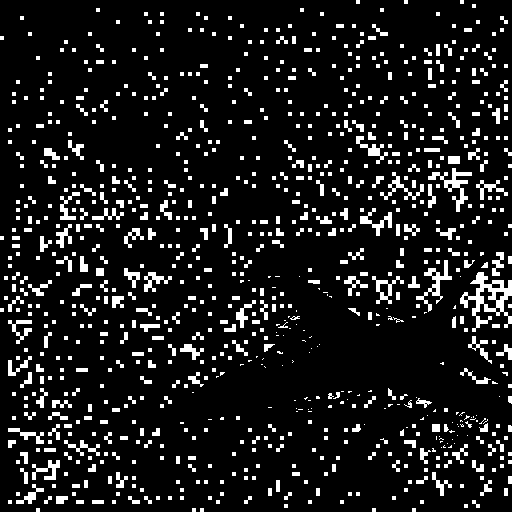}&\includegraphics[width=0.18\textwidth]{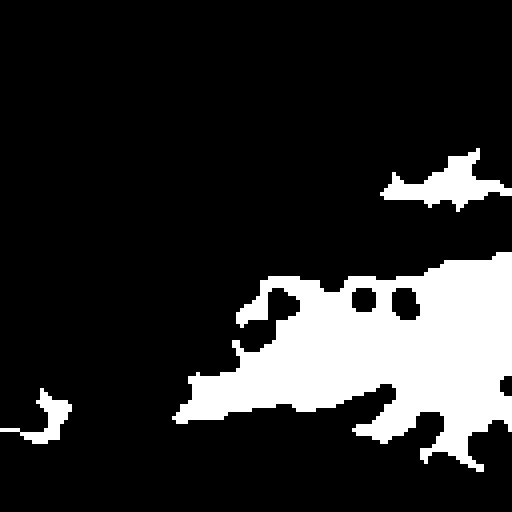}\\
(f) & (g) & (h) & (i) & (j)
\end{tabular}
\caption{Tamper detection and recovery performance under add object tampering (10\%), and hybrid attacks such as darken (100), sharpening (size = 21, $\sigma$ = 5, strength = 0.8), JPEG2000 (QF = 60), rotation (180 deg). (a) Watermarked image under tampering and attacking, (b) Image digest (Inverse halftone), (c) Reconstrued geometry (PSNR = 7.76, SSIM = 0.53, MSE = 1088), (d) Tamper Detection, (e) Recovered tampered regions, (f) Extracted watermark, (g) Compare valid region (NC = 0.78, BER = 0.10), (h) Authenticate by watermark (exclusive-or), (i) Error, (j) Post processing (TPR = 95.85, FPR = 7.52).}
\label{fig:resf16}
\end{figure*}
\begin{figure*}[t]
\center
\begin{tabular}{ccccc}
\includegraphics[width=0.18\textwidth]{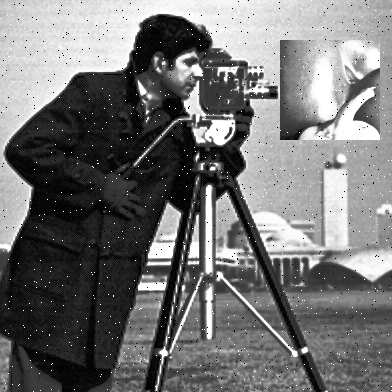}&\includegraphics[width=0.18\textwidth]{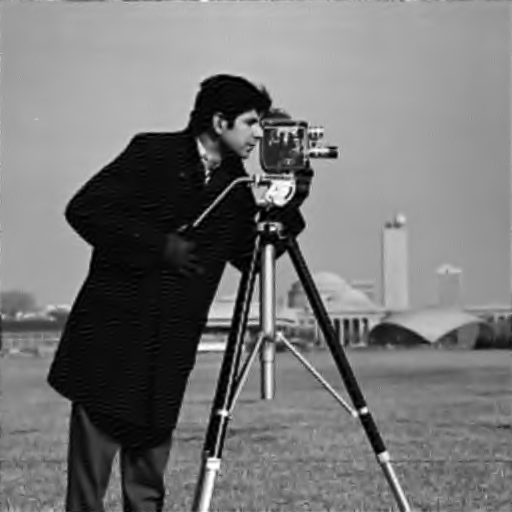}&\includegraphics[width=0.18\textwidth]{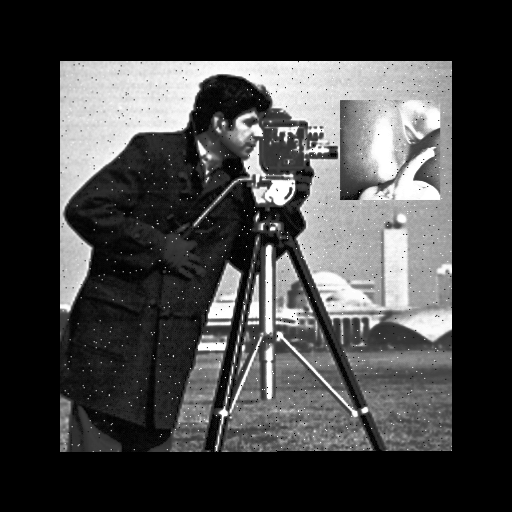}&\includegraphics[width=0.18\textwidth]{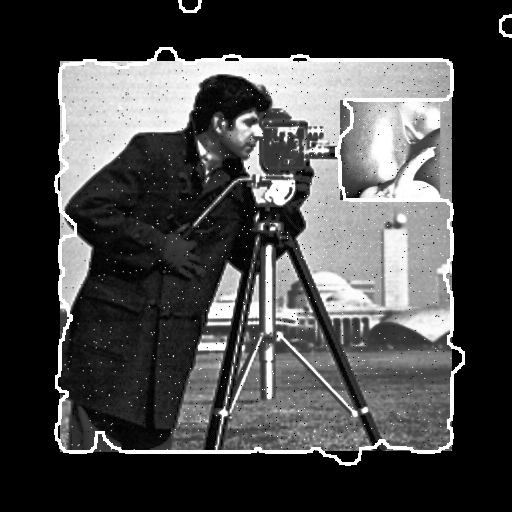}&\includegraphics[width=0.18\textwidth]{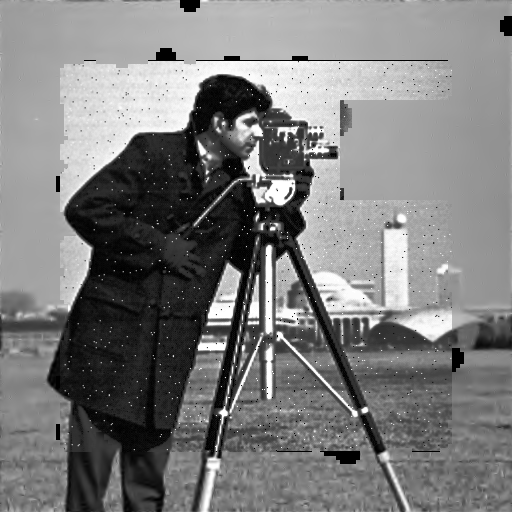}\\
(a) & (b) & (c) & (d) & (e) \\
\includegraphics[width=0.18\textwidth]{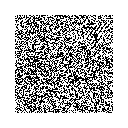}&\includegraphics[width=0.18\textwidth]{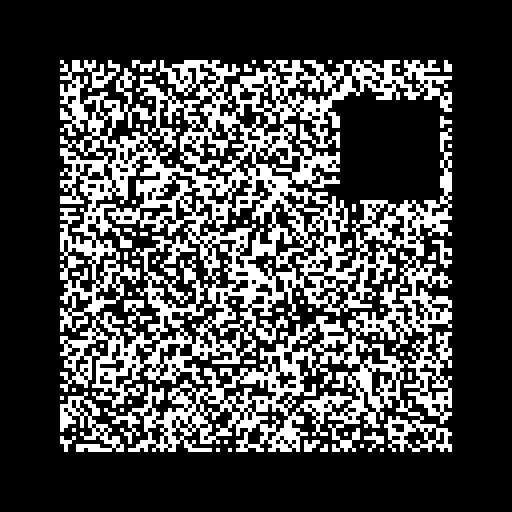}&\includegraphics[width=0.18\textwidth]{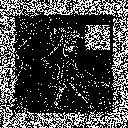}&\includegraphics[width=0.18\textwidth]{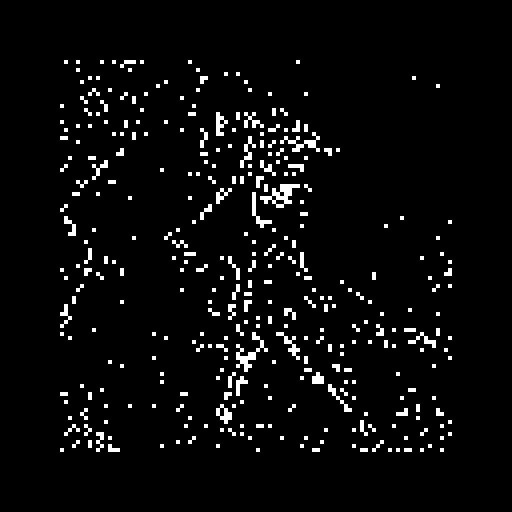}&\includegraphics[width=0.18\textwidth]{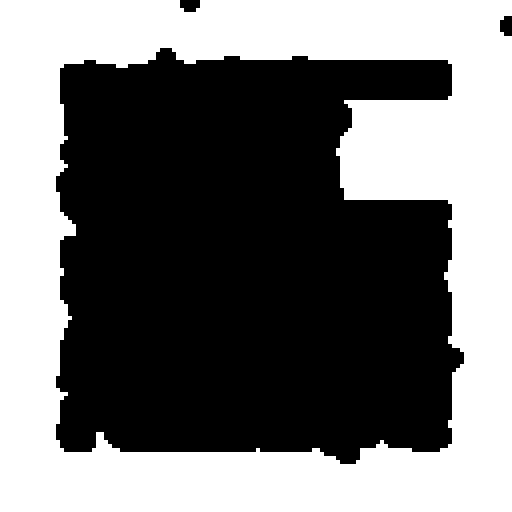}\\
(f) & (g) & (h) & (i) & (j)
\end{tabular}
\caption{Tamper detection and recovery performance under add object and removing tampering (45\%), and hybrid attacks such as histogram equalization, salt and pepper noise ($\rho$ = 0.01), cropping (42\% around). (a) Watermarked image under tampering and attacking, (b) Image digest (Inverse halftone), (c) Reconstrued geometry (PSNR = 8.37, SSIM = 0.36, MSE = 9444), (d) Tamper Detection, (e) Recovered tampered regions, (f) Extracted watermark, (g) Compare valid region (NC = 0.91, BER = 0.04), (h) Authenticate by watermark (exclusive-or), (i) Error, (j) Post processing (TPR = 98.44, FPR = 2.46).}
\label{fig:rescameraman}
\end{figure*}
\begin{figure*}[t!]
\center
\begin{tabular}{ccccc}
\includegraphics[width=0.18\textwidth]{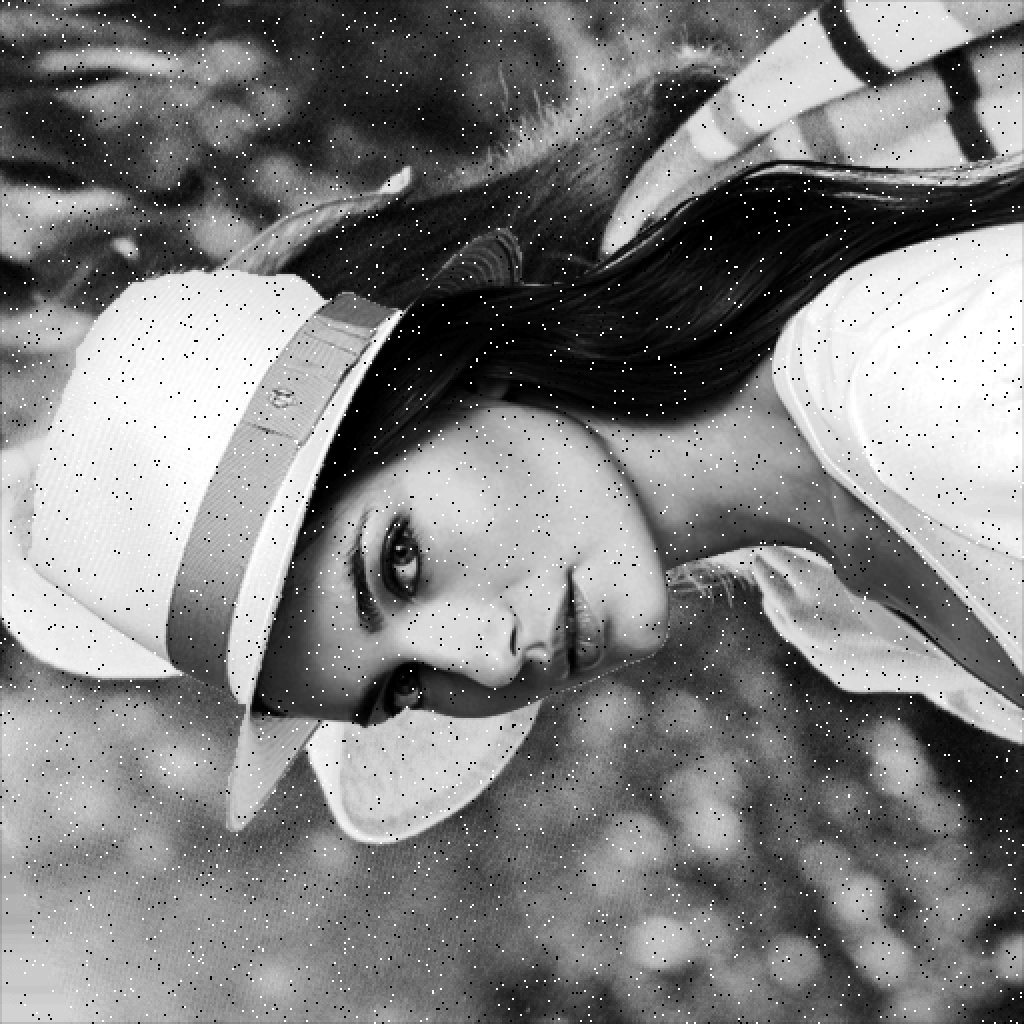}&\includegraphics[width=0.18\textwidth]{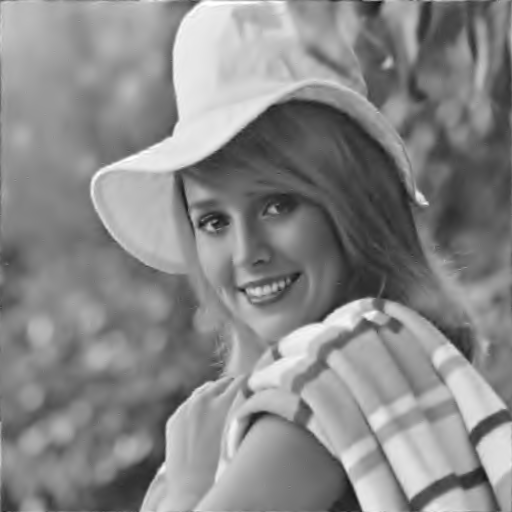}&\includegraphics[width=0.18\textwidth]{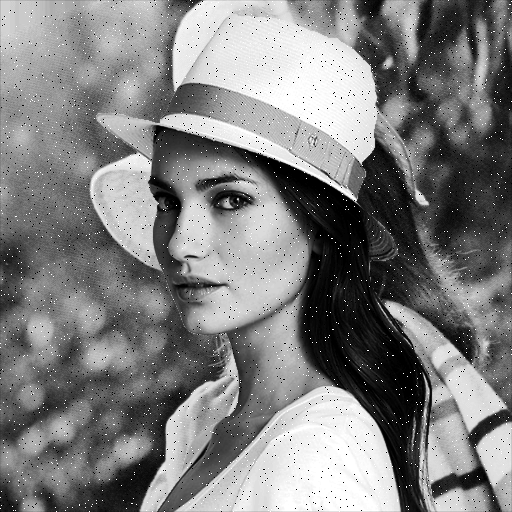}&\includegraphics[width=0.18\textwidth]{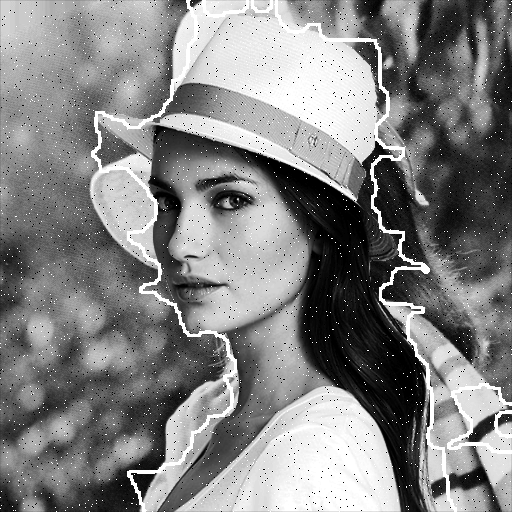}&\includegraphics[width=0.18\textwidth]{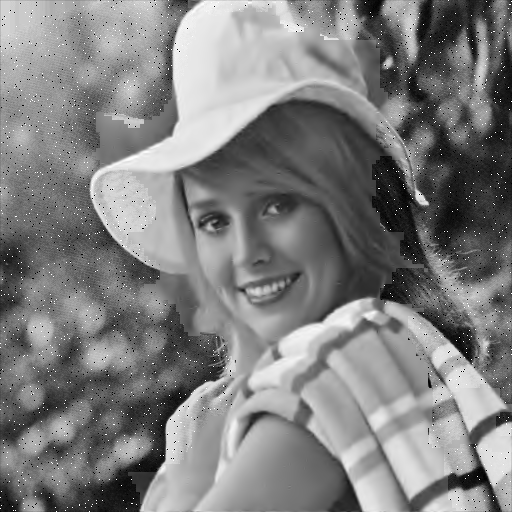}\\
(a) & (b) & (c) & (d) & (e) \\
\includegraphics[width=0.18\textwidth]{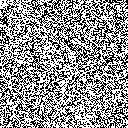}&\includegraphics[width=0.18\textwidth]{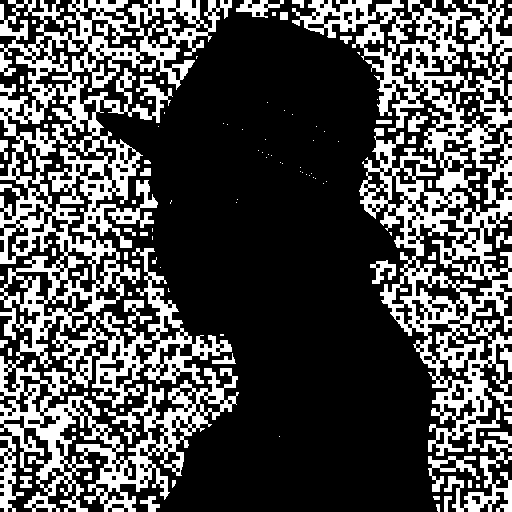}&\includegraphics[width=0.18\textwidth]{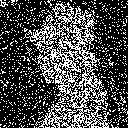}&\includegraphics[width=0.18\textwidth]{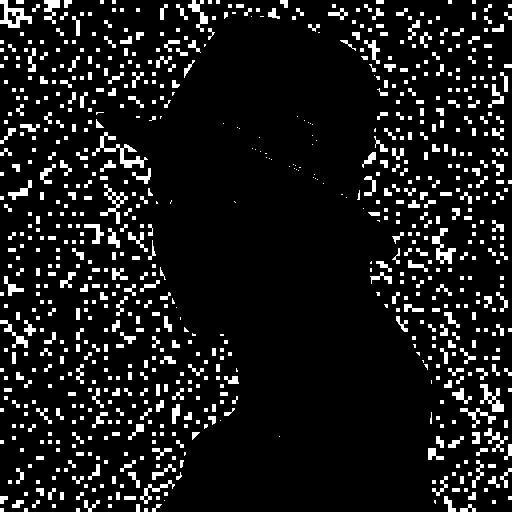}&\includegraphics[width=0.18\textwidth]{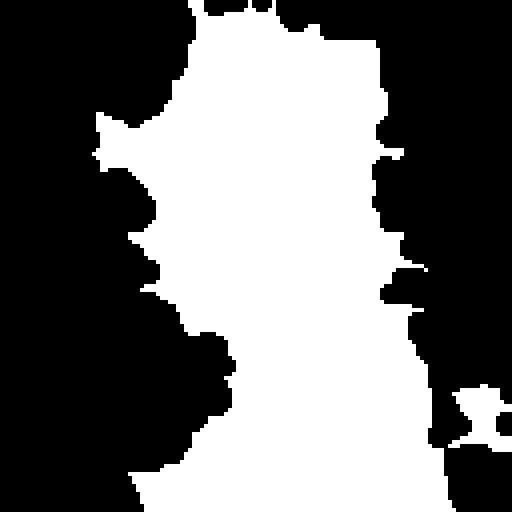}\\
(f) & (g) & (h) & (i) & (j)
\end{tabular}
\caption{Tamper detection and recovery performance under add object tampering (40\%), and hybrid attacks such as histogram equalization, gaussian smoothing (size = 5, $\sigma$ = 0.5), salt and pepper noise ($\rho$ = 0.02), JPEG (QF = 70), scale (2), and rotation (90 deg). (a) Watermarked image under tampering and attacking, (b) Image digest (Inverse halftone), (c) Reconstrued geometry (PSNR = 11.24, SSIM = 0.29, MSE = 4878), (d) Tamper Detection, (e) Recovered tampered regions, (f) Extracted watermark, (g) Compare valid region (NC = 0.80, BER = 0.10), (h) Authenticate by watermark (exclusive-or), (i) Error, (j) Post processing (TPR = 99.37, FPR = 8.91).}
\label{fig:reselaine}
\end{figure*}

To further demonstrate the superiority and efficacy of TRLF for tamper detection and recovery, we test its performance under different conditions. For this purpose, the watermarked image is destroyed and manipulated with varying attacks and tampering ratios as shown in Figs. \ref{fig:resbaboon}, \ref{fig:resbarbara}, \ref{fig:reslena}, \ref{fig:respepper}, \ref{fig:ressantigo}, \ref{fig:resf16}, \ref{fig:rescameraman}, and \ref{fig:reselaine}. 

In these figures, the tampered images are generated based on different types of tampering such as copy-move, add an object, covering, removing, vector-quantization, collage, etc. Thereinafter, the different hybrid attacks are applied in the tampered image. In the copy-move tampering, a part of the watermarked image was copied and pasted somewhere else within the image. The vector-quantization is kind of tampering, which is a part of another watermarked image was copied and pasted in the same location in the current watermarked image. In collage tampering, the copied region pasted in somewhere else to generate the fake image. In these figures, sub-images representation is as follows: (a) watermarked image under tampering and attack, (b) image digest which was generated by using inverse halftone technique, (c) reconstructed geometry of tampered image (we report also PSNR and SSIM between the watermarked and manipulated images), (d) tampered regions, (e) recovered tampered regions, (f) extracted watermark, (g) compared valid regions, (h) authenticate by watermark (exclusive-or), (i) error, and (j) post processing.

To prove the copy-move detection, we copy Baboon eyes and paste it in a different location, in Fig. \ref{fig:resbaboon}. Afterward, the various attacks include sharpening (size = 13, $\sigma$ = 3, strength = 0.8), speckle noise (0.01), rotation (90 deg) are applied. PSNR, SSIM, and MSE between the watermarked image and attacked image are 16.33, 0.79 and 1512, respectively. NC and BER between embedded and extracted watermark are 0.72 and 0.14, respectively. The TRP and FPR of the tamper detection phase are 99.13 and 6.81, respectively. The experimental result demonstrates that TRLF has the desirable performance for texture image. It clearly shows that TRLF successfully localizes tampered regions.

We add a nary object to the watermarked image, in Fig. \ref{fig:resbarbara}. Furthermore, several attacks such as darken (30), sharpening (size = 29, $\sigma$ = 7, strength = 0.8), speckle noise ($\sigma$ = 0.005), scale (2), cropping (42\% around) and JPEG2000 (QF  = 60) are applied. PSNR, SSIM, and MSE between the watermarked image and attacked image are 9.63, 0.36 and 7080, respectively. NC and BER between embedded and extracted watermark are 0.89 and 0.05, respectively. The TRP and FPR of the tamper detection phase are 98.44 and 5.59, respectively. The experiment result indicates that TRLF can locate exactly the added object.

Next, we show the capacity of TRLF to detect the vector-quantization tampering. For this purpose, the watermarked image is 26\% manipulated in Fig. \ref{fig:reslena}. Here, to prove vector-quantization tampering, we copy the head of Barbara which watermarked with the same key and pastes it in the same location in Lena image. Moreover, the various attack, such as histogram equalization, salt and pepper noise ($\rho$ = 0.01), translation (50,50) are applied. PSNR, SSIM, and MSE between the watermarked image and attacked image are 10.69, 0.63, 5547, respectively. NC and BER between embedded and extracted watermark are 0.83 and 0.08, respectively. The TRP and FPR of the tamper detection phase are 97.95 and 1.62, respectively. As we have seen, TRLF has afforded high TPR values.

The tampered image obtained by copying a part of the watermarked image and pasted in two different locations, in Fig. \ref{fig:respepper}. Subsequently, the different attacks including Gaussian smoothing (size = 7, $\sigma$ = 0.5), salt and pepper noise ($\rho$ = 0.05), JPEG (QF = 70) and scale (2) are applied to tampered image to destroy the authentication watermark. PSNR, SSIM, and MSE between the watermarked image and attacked image are 19.56, 0.92 and 720, respectively. NC and BER between embedded and extracted watermark are 0.71 and 0.14, respectively. The TRP and FPR of the tamper detection phase are 87.28 and 1.72, respectively. The experiments show that TRLF can locate precisely the invalid regions.

The tampered image generated by copying part of the map and pasted in corner of the watermarked image, in Fig. \ref{fig:ressantigo}. Thereinafter, the various attacks including lighten (30), sharpening (size = 21, $\sigma$ = 5, strength = 0.8), JPEG (QF = 60), rotation (180 deg), translation (-100, 0), and cropping (20\% right) are applied. PSNR, SSIM, and MSE between the watermarked image and attacked image are 9.74, 0.60 and 6898, respectively. NC and BER between embedded and extracted watermark are 0.75 and 0.12, respectively. The TRP and FPR of the tamper detection phase are 99.31 and 13.25, respectively. Again, the experimental results demonstrate that TRLF has a relatively satisfactory performance for the texture image.

We added extra warplane to the watermarked image, in Fig. \ref{fig:resf16}. Then, in order to damage the watermark, several attacks such as darken (100), sharpening (size = 21, $\sigma$ = 5, strength = 0.8), JPEG2000 (QF = 60), and rotation (180 deg) are applied. PSNR, SSIM, and MSE between the watermarked image and attacked image are 7.76, 0.53 and 1088, respectively. NC and BER between embedded and extracted watermark are 0.78 and 0.10, respectively. The TRP and FPR of the tamper detection phase are 95.85 and 7.52, respectively. The results indicate the outstanding performance of TRLF.

Next, in Fig. \ref{fig:rescameraman} the tampered image obtained by copying part of watermarked Pepper and pasted in another position in Cameraman. Then, in order to destroy the watermark, several attacks such as histogram equalization, salt and pepper noise ($\rho$ = 0.01), cropping (42\% around) are applied. PSNR, SSIM, and MSE between the watermarked image and attacked image are 8.37, 0.36 and 9444, respectively. NC and BER between embedded and extracted watermark are 0.91 and 0.04, respectively. The TRP and FPR of the tamper detection phase are 98.44 and 2.46, respectively. It is observed that the essential information like NC and TRP values are effectively high, and BER and FPR are very low. In addition, we are able to detect vector-quantization and collage tampering attack because the embedded watermark is dependent on the cover image content. 

In the last test, the tampered image generated by covering Elaine's body with another woman in Fig. \ref{fig:reselaine}. Furthermore, the hybrid attacks such as histogram equalization, Gaussian smoothing (size = 5, $\sigma$ = 0.5), salt and pepper noise ($\rho$ = 0.02), JPEG (QF = 70), scale (2), and rotation (90 deg) are applied in tampered image. PSNR, SSIM, and MSE between the watermarked image and attacked image are 11.24, 0.29 and 4878, respectively. NC and BER between embedded and extracted watermark are 0.80 and 0.10, respectively. The TRP and FPR of the tamper detection phase are 99.37 and 8.91, respectively. We can see that TRLF detects the tampered regions correctly. The experimental clearly shows that TRLF works well under these heavy attacks.

These experiments clearly show that TRLF successfully localizes the tampered regions under several tampering and hybrid attacks (causes heavy degradation). TPR and FPR prove that TRLF has an extremely high accuracy of tampering localization, but in some cases, the authentication process may mark erroneously some blocks without tamper as modified. As can be seen, the extracted watermark is very close to the embedded watermark in most cases. The minimum NC values of the extracted watermark in these test is 0.71, which is still high to use in the watermarking method with copyright protection. In other words, we can use the binary logo instead of a random binary sequence as the watermark. In addition, we have proved the superiority of TRLF for the color image with the above experiments.

In summary, the experimental results illustrated that TRLF is perfectly efficient, effective, and robust to most common attacks including additive noise, filtering, compression, luminance, and geometric operations, because of using the powerful authentication based on FNN. Even if the watermarked image is attacked by hybrids attacks, TPR and FPR values are still admissible. Furthermore, it performs well in transparency and imperceptibility of the watermarked image. Totally, compared with other existing techniques, TRLF exhibited superior performance in term of tamper detection rate under hybrid attacks. 

It should be noted, when watermarked images are subjected to attacks such as JPEG compression, median and average filters, the robustness of TRLF is slightly lower and more fragile than other methods. On the other side, TRLF is using $4\times4$ blocks. Hence, the localization is better in compression to other semi-fragile methods which used $8\times8$ blocks. In recovery step, the safe compression version of the original image is used to reconstruct the tampered region. Obviously, TRLF effectively recovered the tampered region with no error. In addition, the blocking artifacts are also removed because each pixel in the tampered region has reliable value at the corresponding position in image digest. In other words, the recovery system is not block-based.
\section{Conclusion and future work}
\noindent An effective semi-fragile watermarking method for tamper detection and recovery based on LWT and FNN has been proposed in this paper. In this work, the random watermark bits are embedded into the high-frequency band of LWT using correlation technique. To do so, instead of using a fixed threshold step, $DC$ coefficient of each block is correlated by a specific threshold according to the content regions. This analysis improves the quality of the watermarked image, especially when the image contains the texture (rough) regions.  In additions, it helps to perform the adaptive control of the embedding capacity through the desired threshold step. The high quality of watermarked images clearly demonstrates that embedding process is well optimized and the visual quality after embedding is also promising. The major contribution of TRLF is the application of FNN algorithm to extract the watermark. To the best of our knowledge, it is the first time that this mechanism is used for image authentication which can further be explored. When the watermarked image is destroyed by various attacks, FNN is used to extract the binary watermark from the watermarked image. This is done, with the maximum possible correlation, which can further reduce the error and improve the rate of the true positive. In the recovery phase, the inverse halftone of host image has used as image digest to recover tampered regions. Extensive experimental results indicate that the proposed algorithm is superior to other the-state-of-the-art methods which reported in the literature in terms of quality of watermarked image and robustness. TRLF was investigated under various conditions of different types of attacks. The results revealed that TRLF is not only robust to singular attacks such as noise adding, compression, image filtering, point processing (luminance) and geometrical ones, but also against various hybrid attacks. Hence, TRLF successfully localizes tampered region and has a high detection rate. Moreover, it can detect copy-move, vector-quantization, and collage tampering. The high quality, efficiency, and simplicity of TRLF make it suitable for real-time applications. The binary operations and memory efficiency make TRLF simpler and suitable to run on low power devices. Based on the advantages exhibited above, TRLF outperforms the related works reported in the literature, in terms of superiority, efficiency, and effectiveness for image tamper detection and recovery. 

Future works are required to further reduce the false positive rates for compression attack and to improve the robustness against median and averaging filters, without affecting the quality of the watermarked image. These constrain will be addressed in coming future works. In addition, extending the blind method is another future work.
\bibliographystyle{elsarticle-num}

\begin{thebibliography}{10}
\expandafter\ifx\csname url\endcsname\relax
  \def\url#1{\texttt{#1}}\fi
\expandafter\ifx\csname urlprefix\endcsname\relax\def\urlprefix{URL }\fi
\expandafter\ifx\csname href\endcsname\relax
  \def\href#1#2{#2} \def\path#1{#1}\fi

\bibitem{ref51}
F.~Y. Shih, Digital Watermarking and Steganography: Fundamentals and
  Techniques, Second Edition, CRC press, 2017.

\bibitem{ref1}
F.~Husain, A survey of digital watermarking techniques for multimedia data, MIT
  International Journal of Electronics and Communication Engineering 2~(1)
  (2012) 37--43.

\bibitem{ref2}
P.~Singh, R.~Chadha, A survey of digital watermarking techniques, applications
  and attacks, International Journal of Engineering and Innovative Technology
  (IJEIT) 2~(9) (2013) 165--175.

\bibitem{ref3}
X.~yang Wang, C.~peng Wang, H.~ying Yang, P.~pan Niu, A robust blind color
  image watermarking in quaternion fourier transform domain, Journal of Systems
  and Software 86~(2) (2013) 255 -- 277.
\newblock \href {http://dx.doi.org/http://dx.doi.org/10.1016/j.jss.2012.08.015}
  {\path{doi:http://dx.doi.org/10.1016/j.jss.2012.08.015}}.

\bibitem{ref4}
M.~Ali, C.~W. Ahn, An optimized watermarking technique based on self-adaptive
  \{DE\} in dwt–svd transform domain, Signal Processing 94 (2014) 545 -- 556.
\newblock \href
  {http://dx.doi.org/http://dx.doi.org/10.1016/j.sigpro.2013.07.024}
  {\path{doi:http://dx.doi.org/10.1016/j.sigpro.2013.07.024}}.

\bibitem{ref5}
M.~Ali, C.~W. Ahn, M.~Pant, A robust image watermarking technique using \{SVD\}
  and differential evolution in \{DCT\} domain, Optik - International Journal
  for Light and Electron Optics 125~(1) (2014) 428 -- 434.
\newblock \href
  {http://dx.doi.org/http://dx.doi.org/10.1016/j.ijleo.2013.06.082}
  {\path{doi:http://dx.doi.org/10.1016/j.ijleo.2013.06.082}}.

\bibitem{ref6}
A.~M. Abdelhakim, H.~I. Saleh, A.~M. Nassar, A quality guaranteed robust image
  watermarking optimization with artificial bee colony, Expert Systems with
  Applications 72 (2017) 317 -- 326.
\newblock \href
  {http://dx.doi.org/http://dx.doi.org/10.1016/j.eswa.2016.10.056}
  {\path{doi:http://dx.doi.org/10.1016/j.eswa.2016.10.056}}.

\bibitem{ref7}
P.-P. Zheng, J.~Feng, Z.~Li, M.~quan Zhou, A novel \{SVD\} and ls-svm
  combination algorithm for blind watermarking, Neurocomputing 142 (2014) 520
  -- 528, \{SI\} Computational Intelligence Techniques for New Product
  Development.
\newblock \href
  {http://dx.doi.org/http://dx.doi.org/10.1016/j.neucom.2014.04.005}
  {\path{doi:http://dx.doi.org/10.1016/j.neucom.2014.04.005}}.

\bibitem{ref8}
R.~P. Singh, N.~Dabas, V.~Chaudhary, Nagendra, Online sequential extreme
  learning machine for watermarking in \{DWT\} domain, Neurocomputing 174, Part
  A (2016) 238 -- 249.
\newblock \href
  {http://dx.doi.org/http://dx.doi.org/10.1016/j.neucom.2015.03.115}
  {\path{doi:http://dx.doi.org/10.1016/j.neucom.2015.03.115}}.

\bibitem{ref9}
H.-H. Tsai, Y.-S. Lai, S.-C. Lo, A zero-watermark scheme with geometrical
  invariants using \{SVM\} and \{PSO\} against geometrical attacks for image
  protection, Journal of Systems and Software 86~(2) (2013) 335 -- 348.
\newblock \href {http://dx.doi.org/http://dx.doi.org/10.1016/j.jss.2012.08.040}
  {\path{doi:http://dx.doi.org/10.1016/j.jss.2012.08.040}}.

\bibitem{ref40}
S.~H. Soleymani, A.~H. Taherinia, Double expanding robust image watermarking
  based on spread spectrum technique and bch coding, Multimedia Tools and
  Applications (2016) 1--19\href {http://dx.doi.org/10.1007/s11042-016-3734-2}
  {\path{doi:10.1007/s11042-016-3734-2}}.

\bibitem{ref49}
K.~Sreenivas, V.~Kamkshi~Prasad,
  \href{https://doi.org/10.1007/s13042-017-0641-4}{Fragile watermarking schemes
  for image authentication: a survey}, International Journal of Machine
  Learning and Cybernetics\href {http://dx.doi.org/10.1007/s13042-017-0641-4}
  {\path{doi:10.1007/s13042-017-0641-4}}.
\newline\urlprefix\url{https://doi.org/10.1007/s13042-017-0641-4}

\bibitem{ref48}
B.~B. Haghighi, A.~H. Taherinia, A.~Harati,
  \href{http://www.sciencedirect.com/science/article/pii/S1047320317301876}{Trlh:
  Fragile and blind dual watermarking for image tamper detection and
  self-recovery based on lifting wavelet transform and halftoning technique},
  Journal of Visual Communication and Image Representation 50 (2018) 49 -- 64.
\newblock \href {http://dx.doi.org/https://doi.org/10.1016/j.jvcir.2017.09.017}
  {\path{doi:https://doi.org/10.1016/j.jvcir.2017.09.017}}.
\newline\urlprefix\url{http://www.sciencedirect.com/science/article/pii/S1047320317301876}

\bibitem{ref10}
T.-Y. Lee, S.~D. Lin, Dual watermark for image tamper detection and recovery,
  Pattern Recognition 41~(11) (2008) 3497 -- 3506.
\newblock \href
  {http://dx.doi.org/http://dx.doi.org/10.1016/j.patcog.2008.05.003}
  {\path{doi:http://dx.doi.org/10.1016/j.patcog.2008.05.003}}.

\bibitem{ref11}
C.-S. Hsu, S.-F. Tu, Probability-based tampering detection scheme for digital
  images, Optics Communications 283~(9) (2010) 1737 -- 1743.
\newblock \href
  {http://dx.doi.org/http://dx.doi.org/10.1016/j.optcom.2009.12.073}
  {\path{doi:http://dx.doi.org/10.1016/j.optcom.2009.12.073}}.

\bibitem{ref12}
Z.~Qian, G.~Feng, X.~Zhang, S.~Wang, Image self-embedding with high-quality
  restoration capability, Digital Signal Processing 21~(2) (2011) 278 -- 286.
\newblock \href {http://dx.doi.org/http://dx.doi.org/10.1016/j.dsp.2010.04.006}
  {\path{doi:http://dx.doi.org/10.1016/j.dsp.2010.04.006}}.

\bibitem{ref13}
J.~Zhang, Q.~Zhang, H.~Lv, A novel image tamper localization and recovery
  algorithm based on watermarking technology, Optik - International Journal for
  Light and Electron Optics 124~(23) (2013) 6367 -- 6371.
\newblock \href
  {http://dx.doi.org/http://dx.doi.org/10.1016/j.ijleo.2013.05.040}
  {\path{doi:http://dx.doi.org/10.1016/j.ijleo.2013.05.040}}.

\bibitem{ref14}
X.~Tong, Y.~Liu, M.~Zhang, Y.~Chen, A novel chaos-based fragile watermarking
  for image tampering detection and self-recovery, Signal Processing: Image
  Communication 28~(3) (2013) 301 -- 308.
\newblock \href
  {http://dx.doi.org/http://dx.doi.org/10.1016/j.image.2012.12.003}
  {\path{doi:http://dx.doi.org/10.1016/j.image.2012.12.003}}.

\bibitem{ref15}
S.~Dadkhah, A.~A. Manaf, Y.~Hori, A.~E. Hassanien, S.~Sadeghi, An effective
  svd-based image tampering detection and self-recovery using active
  watermarking, Signal Processing: Image Communication 29~(10) (2014) 1197 --
  1210.
\newblock \href
  {http://dx.doi.org/http://dx.doi.org/10.1016/j.image.2014.09.001}
  {\path{doi:http://dx.doi.org/10.1016/j.image.2014.09.001}}.

\bibitem{ref16}
X.~Zhang, Y.~Xiao, Z.~Zhao, Self-embedding fragile watermarking based on dct
  and fast fractal coding, Multimedia Tools and Applications 74~(15) (2015)
  5767--5786.

\bibitem{ref17}
D.~Singh, S.~K. Singh, Effective self-embedding watermarking scheme for image
  tampered detection and localization with recovery capability, Journal of
  Visual Communication and Image Representation 38 (2016) 775 -- 789.
\newblock \href
  {http://dx.doi.org/http://dx.doi.org/10.1016/j.jvcir.2016.04.023}
  {\path{doi:http://dx.doi.org/10.1016/j.jvcir.2016.04.023}}.

\bibitem{ref18}
W.-C. Wu, Z.-W. Lin, Svd-based self-embedding image authentication scheme using
  quick response code features, Journal of Visual Communication and Image
  Representation 38 (2016) 18 -- 28.
\newblock \href
  {http://dx.doi.org/http://dx.doi.org/10.1016/j.jvcir.2016.02.005}
  {\path{doi:http://dx.doi.org/10.1016/j.jvcir.2016.02.005}}.

\bibitem{ref19}
C.-S. Hsu, S.-F. Tu, Image tamper detection and recovery using adaptive
  embedding rules, Measurement 88 (2016) 287 -- 296.
\newblock \href
  {http://dx.doi.org/http://dx.doi.org/10.1016/j.measurement.2016.03.053}
  {\path{doi:http://dx.doi.org/10.1016/j.measurement.2016.03.053}}.

\bibitem{ref20}
F.~Cao, B.~An, J.~Wang, D.~Ye, H.~Wang, Hierarchical recovery for tampered
  images based on watermark self-embedding, Displays 46 (2017) 52 -- 60.
\newblock \href
  {http://dx.doi.org/http://dx.doi.org/10.1016/j.displa.2017.01.001}
  {\path{doi:http://dx.doi.org/10.1016/j.displa.2017.01.001}}.

\bibitem{ref21}
K.~Maeno, Q.~Sun, S.-F. Chang, M.~Suto, New semi-fragile image authentication
  watermarking techniques using random bias and nonuniform quantization, IEEE
  Transactions on Multimedia 8~(1) (2006) 32--45.

\bibitem{ref22}
X.~Zhu, A.~T. Ho, P.~Marziliano, A new semi-fragile image watermarking with
  robust tampering restoration using irregular sampling, Signal Processing:
  Image Communication 22~(5) (2007) 515 -- 528.
\newblock \href
  {http://dx.doi.org/http://dx.doi.org/10.1016/j.image.2007.03.004}
  {\path{doi:http://dx.doi.org/10.1016/j.image.2007.03.004}}.

\bibitem{ref23}
R.~Chamlawi, A.~Khan, I.~Usman, Authentication and recovery of images using
  multiple watermarks, Computers and Electrical Engineering 36~(3) (2010) 578
  -- 584.
\newblock \href
  {http://dx.doi.org/http://dx.doi.org/10.1016/j.compeleceng.2009.12.003}
  {\path{doi:http://dx.doi.org/10.1016/j.compeleceng.2009.12.003}}.

\bibitem{ref24}
X.~Qi, X.~Xin, A quantization-based semi-fragile watermarking scheme for image
  content authentication, Journal of Visual Communication and Image
  Representation 22~(2) (2011) 187 -- 200.
\newblock \href
  {http://dx.doi.org/http://dx.doi.org/10.1016/j.jvcir.2010.12.005}
  {\path{doi:http://dx.doi.org/10.1016/j.jvcir.2010.12.005}}.

\bibitem{ref25}
A.~Phadikar, S.~P. Maity, M.~Mandal, Novel wavelet-based \{QIM\} data hiding
  technique for tamper detection and correction of digital images, Journal of
  Visual Communication and Image Representation 23~(3) (2012) 454 -- 466.
\newblock \href
  {http://dx.doi.org/http://dx.doi.org/10.1016/j.jvcir.2012.01.005}
  {\path{doi:http://dx.doi.org/10.1016/j.jvcir.2012.01.005}}.

\bibitem{ref26}
L.~Rosales-Roldan, M.~Cedillo-Hernandez, M.~Nakano-Miyatake, H.~Perez-Meana,
  B.~Kurkoski, Watermarking-based image authentication with recovery capability
  using halftoning technique, Signal Processing: Image Communication 28~(1)
  (2013) 69 -- 83.
\newblock \href
  {http://dx.doi.org/http://dx.doi.org/10.1016/j.image.2012.11.006}
  {\path{doi:http://dx.doi.org/10.1016/j.image.2012.11.006}}.

\bibitem{ref27}
R.~O. Preda, Semi-fragile watermarking for image authentication with sensitive
  tamper localization in the wavelet domain, Measurement 46~(1) (2013) 367 --
  373.
\newblock \href
  {http://dx.doi.org/http://dx.doi.org/10.1016/j.measurement.2012.07.010}
  {\path{doi:http://dx.doi.org/10.1016/j.measurement.2012.07.010}}.

\bibitem{ref28}
Y.~Huo, H.~He, F.~Chen, A semi-fragile image watermarking algorithm with
  two-stage detection, Multimedia tools and applications 72~(1) (2014)
  123--149.

\bibitem{ref29}
H.~M. Al-Otum, Semi-fragile watermarking for grayscale image authentication and
  tamper detection based on an adjusted expanded-bit multiscale
  quantization-based technique, Journal of Visual Communication and Image
  Representation 25~(5) (2014) 1064 -- 1081.
\newblock \href
  {http://dx.doi.org/http://dx.doi.org/10.1016/j.jvcir.2013.12.017}
  {\path{doi:http://dx.doi.org/10.1016/j.jvcir.2013.12.017}}.

\bibitem{ref30}
X.~Qi, X.~Xin, A singular-value-based semi-fragile watermarking scheme for
  image content authentication with tamper localization, Journal of Visual
  Communication and Image Representation 30 (2015) 312 -- 327.
\newblock \href
  {http://dx.doi.org/http://dx.doi.org/10.1016/j.jvcir.2015.05.006}
  {\path{doi:http://dx.doi.org/10.1016/j.jvcir.2015.05.006}}.

\bibitem{ref31}
O.~Benrhouma, H.~Hermassi, S.~Belghith, Tamper detection and self-recovery
  scheme by dwt watermarking, Nonlinear Dynamics 79~(3) (2015) 1817--1833.

\bibitem{ref32}
C.~Li, A.~Zhang, Z.~Liu, L.~Liao, D.~Huang, Semi-fragile self-recoverable
  watermarking algorithm based on wavelet group quantization and double
  authentication, Multimedia tools and applications 74~(23) (2015)
  10581--10604.

\bibitem{ref33}
K.~Chetan, S.~Nirmala, An intelligent blind semi-fragile watermarking scheme
  for effective authentication and tamper detection of digital images using
  curvelet transforms., in: SIRS, 2015, pp. 199--213.

\bibitem{ref39}
A.~H. Taherinia, M.~Jamzad, Blind dewatermarking method based on wavelet
  transform, Optical Engineering 50~(5) (2011) 057006--057006--8.
\newblock \href {http://dx.doi.org/10.1117/1.3581116}
  {\path{doi:10.1117/1.3581116}}.

\bibitem{ref50}
P.~Korus, Digital image integrity – a survey of protection and verification
  techniques, Digital Signal Processing 71 (2017) 1 -- 26.
\newblock \href {http://dx.doi.org/https://doi.org/10.1016/j.dsp.2017.08.009}
  {\path{doi:https://doi.org/10.1016/j.dsp.2017.08.009}}.

\bibitem{ref34}
W.~Sweldens, The lifting scheme: A construction of second generation wavelets,
  SIAM Journal on Mathematical Analysis 29~(2) (1998) 511--546.

\bibitem{ref35}
W.~Sweldens, The lifting scheme: A construction of second generation wavelets,
  SIAM Journal on Mathematical Analysis 29~(2) (1998) 511--546.

\bibitem{ref36}
P.-E. Axelson, Quality measures of halftoned images (a review).

\bibitem{ref37}
J.~F. Jarvis, C.~N. Judice, W.~Ninke, A survey of techniques for the display of
  continuous tone pictures on bilevel displays, Computer Graphics and Image
  Processing 5~(1) (1976) 13--40.

\bibitem{ref38}
R.~Neelamani, R.~D. Nowak, R.~G. Baraniuk, Winhd: Wavelet-based inverse
  halftoning via deconvolution, IEEE Transactions on Image Processing.

\bibitem{ref41}
D.~G. Lowe, Object recognition from local scale-invariant features, in:
  Computer vision, 1999. The proceedings of the seventh IEEE international
  conference on, Vol.~2, Ieee, 1999, pp. 1150--1157.

\bibitem{ref42}
D.~G. Lowe, Distinctive image features from scale-invariant keypoints,
  International Journal of Computer Vision 60~(2) (2004) 91--110.
\newblock \href {http://dx.doi.org/10.1023/B:VISI.0000029664.99615.94}
  {\path{doi:10.1023/B:VISI.0000029664.99615.94}}.

\bibitem{ref43}
H.~Bay, A.~Ess, T.~Tuytelaars, L.~V. Gool, Speeded-up robust features (surf),
  Computer Vision and Image Understanding 110~(3) (2008) 346 -- 359, similarity
  Matching in Computer Vision and Multimedia.
\newblock \href
  {http://dx.doi.org/http://dx.doi.org/10.1016/j.cviu.2007.09.014}
  {\path{doi:http://dx.doi.org/10.1016/j.cviu.2007.09.014}}.

\bibitem{ref46}
D.~Svozil, V.~Kvasnicka, J.~Pospichal, Introduction to multi-layer feed-forward
  neural networks, Chemometrics and Intelligent Laboratory Systems 39~(1)
  (1997) 43 -- 62.
\newblock \href
  {http://dx.doi.org/http://dx.doi.org/10.1016/S0169-7439(97)00061-0}
  {\path{doi:http://dx.doi.org/10.1016/S0169-7439(97)00061-0}}.

\bibitem{ref47}
M.~T. Hagan, M.~B. Menhaj, Training feedforward networks with the marquardt
  algorithm, IEEE transactions on Neural Networks 5~(6) (1994) 989--993.

\bibitem{ref44}
S.~Katzenbeisser, F.~Petitcolas, Information hiding techniques for
  steganography and digital watermarking, Artech house, 2000.

\bibitem{ref45}
Z.~Wang, A.~C. Bovik, H.~R. Sheikh, E.~P. Simoncelli, Image quality assessment:
  from error visibility to structural similarity, IEEE transactions on image
  processing 13~(4) (2004) 600--612.

\end{thebibliography}

\end{document}